\makeatletter 
\declare@file@substitution{revtex4-1.cls}{revtex4-2.cls} 
\makeatother 
\documentclass[twocolumn,times,tighten]{aastex631}

\hypersetup{linkcolor=blue,citecolor=blue,filecolor=cyan,urlcolor=blue}

\def\smallplot#1{\centering \leavevmode
\includegraphics[width=89.5mm]{#1} }

\def\smallplottwo#1#2{\centering \leavevmode
\includegraphics[width=89.5mm]{#1}
 \hfil \includegraphics[width=89.5mm]{#2} }

\def\smallplotsix#1#2#3#4#5#6{\centering \leavevmode
\includegraphics[width=89.5mm]{#1} \includegraphics[width=89.5mm]{#2}
\hfil \includegraphics[width=89.5mm]{#3} \includegraphics[width=89.5mm]{#4}
\hfil \includegraphics[width=89.5mm]{#5} \includegraphics[width=89.5mm]{#6}}

\def\smallploteight#1#2#3#4#5#6#7#8{\centering \leavevmode
\includegraphics[width=89.5mm]{#1} \includegraphics[width=89.5mm]{#2}
\hfil \includegraphics[width=89.5mm]{#3} \includegraphics[width=89.5mm]{#4}
\hfil \includegraphics[width=89.5mm]{#5} \includegraphics[width=89.5mm]{#6}
\hfil \includegraphics[width=89.5mm]{#7} \includegraphics[width=89.5mm]{#8}}

\def\largeplotone#1{\centering \leavevmode
\includegraphics[width=150mm]{#1} }

\def\largeplottwo#1#2{\centering \leavevmode
\includegraphics[width=150mm]{#1}
 \hfil \includegraphics[width=150mm]{#2} }

\graphicspath{{./}{figures/}}

\shorttitle{Infrared properties of carbon stars in our Galaxy}
\shortauthors{Suh} 

\begin{document}

\title{Infrared properties of carbon stars in our Galaxy}

\correspondingauthor{Kyung-Won Suh}
\email{kwsuh@chungbuk.ac.kr}

\author[0000-0001-9104-9763]{Kyung-Won Suh}
\affiliation{Department of Astronomy and Space Science, Chungbuk National University, Cheongju-City, 28644, Republic of Korea}

\begin{abstract}
In this study, we explore the characteristics of carbon stars within our Galaxy 
through a comprehensive analysis of observational data spanning visual and 
infrared (IR) bands. Leveraging datasets from IRAS, ISO, Akari, MSX, 2MASS, 
WISE, Gaia DR3, AAVSO, and the SIMBAD object database, we conduct a detailed 
comparison between the observational data and theoretical models. To facilitate 
this comparison, we introduce various IR two-color diagrams (2CDs), IR 
color-magnitude diagrams (CMDs), and spectral energy distributions (SEDs). We 
find that the CMDs, which utilize the latest distance and extinction data from 
Gaia DR3 for a substantial number of carbon stars, are very useful to 
distinguish carbon-rich asymptotic giant branch (CAGB) stars from extrinsic 
carbon stars that are not in the AGB phase. To enhance the accuracy of our 
analysis, we employ theoretical radiative transfer models for dust shells 
around CAGB stars. These theoretical dust shell models demonstrate a 
commendable ability to approximate the observations of CAGB stars across 
various SEDs, 2CDs, and CMDs. We present the infrared properties of known 
pulsating variables and explore the infrared variability of the sample stars by 
analyzing WISE photometric data spanning the last 14 yr. Additionally, we 
present a novel catalog of CAGB stars, offering enhanced reliability and a 
wealth of additional information. 
\end{abstract}



\section{Introduction} \label{sec:intro}

The asymptotic giant branch (AGB) phase is divided into the early AGB (E-AGB) and 
the thermally pulsing AGB (TP-AGB) phases (e.g., \citealt{iben1983}). Carbon-rich 
AGB (CAGB) stars are generally believed to be the evolutionary successors of 
M-type oxygen-rich AGB (OAGB) stars that are in the E-AGB phase. When the OAGB 
stars of intermediate mass range (1.55 $M_{\odot}$ $\leq$ M $<$ 4 $M_{\odot}$: 
for solar metallicity) go through third dredge-up processes, which occurs due to 
thermal pulses in the AGB phase, the C/O ratio can become larger than one and 
thus O-rich dust formation ceases and the stars become visual CAGB stars 
(\citealt{groenewegen1995}), which are now CAGB stars in the TP-AGB phase. After 
that phase, C-rich dust grains start forming and the stars evolve into infrared 
carbon stars, which are CAGB stars with thick C-rich dust envelopes and high 
mass-loss rates (e.g., \citealt{suh2000}). More evolved or massive OAGB stars 
also undergo the TP-AGB phase. Almost all AGB stars are long-period variables 
(LPVs) with large amplitude pulsations (e.g., \citealt{hofner2018}). 

A carbon star is typically a CAGB star but there are other types of carbon stars. 
Generally, a classical (or intrinsic or type C-N) carbon star is known as a CAGB 
star. And there are non-classical or extrinsic carbon stars that are not in the 
AGB phase: Barium (Ba) stars, CH (or C-H) stars, dwarf carbon (dC) stars, J-type 
(or C-J) stars, and early R (R-hot or RH) type stars (e.g., \citealt{green2013}; 
\citealt{abia2020}). 

Extrinsic carbon stars of the Ba, CH, and dC types are hypothesized to be binary
systems, featuring one giant (or dwarf) star and another in the form of a white
dwarf. The white dwarf was once a CAGB star involved in a mass transfer process,
providing C-rich material (\citealt{lu1991}; \citealt{li2018};
\citealt{abia2022}). While these stars exhibit carbon enhancement, the C/O ratio
in the envelope is not necessarily greater than one. CH stars, identified as
Population II stars, share similarities in evolutionary state, spectral
peculiarities, and orbital statistics with Ba stars. They are considered the
older, metal-poor analogs of the latter (\citealt{escorza2017}).

The nature of J-type or R type extrinsic carbon stars are not yet clearly known.
Late R-type stars are known to be similar to CAGB stars (\citealt{zamora2009}).
On the other hand, early R-type (R-hot or RH) stars are believed to be single
stars that were merged from binary stars that consist of a red giant and a white
dwarf (\citealt{zj2013}; \citealt{abia2020}). The anomalous He-flash after the
red giant star’s merging has been suggested as a cause of the carbon enhancement
for the RH stars (\citealt{zj2013}). J-type carbon stars are suspected to
represent a short and luminous stage in the evolution of a R-hot star
(\citealt{zj2013}; \citealt{abia2020}).

Various IR observational data are available from the Infrared Astronomical 
Satellite (IRAS), Infrared Space Observatory (ISO), Midcourse Space Experiment 
(MSX), AKARI, Two-Micron All-Sky Survey (2MASS), Wide-field Infrared Survey 
Explorer (WISE). These data have been very useful to identify carbon stars and 
understand the nature of them. Additionally, the Near-Earth Object WISE 
Reactivation (NEOWISE-R) mission (\citealt{mainzer2014}) has conducted 
photometric observations at 3.4 and 4.6 $\mu$m since 2013. 

Recently, the Gaia Data Release 3 (DR3) provided useful data at visual bands for 
more than one billion stars (\citealt{rimoldini2023}). The newly obtained 
distance (\citealt{bailer-jones2021}) and extinction (e.g., 
\citealt{lallement2022}) data derived from the Gaia DR3 data can be useful to 
find the absolute luminosity for a large number of carbon stars in our Galaxy. 

In this study, we explore the properties of carbon stars within our Galaxy. In 
Section ~\ref{sec:gcarbon}, we furnish lists of known carbon stars across various 
subclasses. For the sample of carbon stars, we conduct cross-identifications with 
counterparts from IRAS, AKARI, MSX, 2MASS, WISE, Gaia Data Release 3 (DR3), and 
American Association of Variable Star Observers (AAVSO; international variable 
star index; version 2023 December 12; \citealt{watson2023}). 

Section~\ref{sec:irpro} presents diverse infrared two-color diagrams (2CDs) and 
color-magnitude diagrams (CMDs) for distinct subclasses of carbon stars, 
juxtaposed with theoretical models of CAGB stars. This comparison aims to unveil 
potential differences in their infrared properties. Section~\ref{sec:agbmodels} 
outlines the theoretical radiative transfer models for dust shells enveloping 
CAGB stars. 

In Section~\ref{sec:cagb}, we compare observed spectral energy distributions 
(SEDs) of CAGB stars with theoretical models, exploring the properties of CAGB 
stars in depth. Section~\ref{sec:spacial} details the spatial distributions of 
carbon stars across various subclasses throughout our Galaxy. In 
Section~\ref{sec:pul}, we present the infrared properties of known Mira variables 
among the CAGB stars and investigate the infrared variability of all sample 
stars. This analysis is carried out by scrutinizing WISE photometric data at the 
W1 and W2 bands spanning the last 14 yr. In Section~\ref{sec:catalog-cagb}, we 
present a new catalog of CAGB stars. Finally, Section~\ref{sec:sum} consolidates 
and summarizes the key findings and results of this paper.

\section{Carbon stars in our Galaxy \label{sec:gcarbon}}

In this section, we compile lists of known carbon stars spanning various 
subclasses, drawing from diverse literature sources and the Strasbourg 
Astronomical Data Centre (CDS) SIMBAD astronomical database. For the recently 
acquired sample of carbon stars, we perform cross-identifications with 
counterparts from IRAS, AKARI, MSX, 2MASS, WISE, Gaia DR3, and AAVSO. 

\citet{suh2022} presented a catalog of CAGB stars within our Galaxy, divided into 
two parts: one based on the IRAS PSC (4118 objects) and the other on the AllWISE 
source catalog (5366 objects). However, this catalog includes (extrinsic) carbon 
stars not in the AGB phase (refer to Section~\ref{sec:intro}) and objects with 
uncertain chemistry. To enhance the catalog's reliability, we utilize a fresh 
sample of carbon stars in our Galaxy, discerning CAGB stars from other subclasses 
of carbon stars. The inclusion of distance and extinction information from the 
new Gaia DR3 data proves beneficial for this purpose.

\begin{table*}
\centering
\caption{Sample of carbon stars (CS) with IRAS counterparts (CS\_IC) \label{tab:tab1}}
\begin{tabular}{llllllllll}
\hline \hline
Subclass     & Reference & Number & Selected & IRAS & Ba-CH\_IC & RH\_IC  & J-type\_IC &RCB\_IC & CAGB\_IC \\
\hline
CAGB-IRAS  &\citet{suh2022}  & 4118 & 4118 & 4118  & 4  & 3 & 78 &35 & 3523$^a$ \\
CS    &\citet{alksnis2001}  & 6891 & 6817$^{b}$ & 3799 & 6  & 4 & 98 & 26 & 3574  \\
CAGB       &\citet{chen2012} & 974  & 963$^{c}$  & 962  & 0  & 0 & 2   &4  & 938  \\
J-type&\citet{chen2007} & 113  & 112$^{d}$  &103 & 1 & 1  & 97 & 1 & 6 \\
CS    &\citet{abia2022} & 827 & 818$^{b}$    &574 & 0 & 5  & 78 & 0 & 292 \\
CS    &\citet{li2018} & 2650 & 2624$^{b}$    &349 &0  & 0  & 13 & 1  & 328 \\
Ba    &\citet{lu1991} & 389  & 389           &168 &166 & 1  & 0 & 0 & 0 \\
Ba (mostly) &\citet{escorza2017} &437  & 437  &176 &167 & 2 & 2 & 0 & 1\\
dC (mostly) & \citet{green2013}  &1211 & 1211 &1  &0  & 0 & 0 & 0 & 1  \\
CS    & SIMBAD CS  & 21,743 & 7541$^{e}$    &3641 &0  & 4 &95 & 5 & 3511 \\
CS    & Gaia DR3 CS  &30023$^{f}$  &19863$^{g}$  &4498  &0  &0 &89 & 6 & 4132 \\
\hline
Total$^{h}$ &- & -  & -                       &-    &181 &5 &100 &39 & 4909  \\
\hline
\end{tabular}
\begin{flushleft}
$^a$color-selected objects without any other evidences are excluded
$^b$sources of duplicate SIMBAD identifers are excluded.
$^c$wrongly classified objects are excluded.
$^d$IRAS 18006-3213 is an OH/IR star
$^e$CDS SIMBAD main type C* in our Galaxy.
$^f$C-rich stars from Gaia DR3 spectra for LPVs (\citealt{lebzelter2023}).
$^g$objects in our Galaxy.
$^h$duplicate objects are excluded.
\end{flushleft}
\end{table*}

\begin{table*}
\centering
\caption{Sample of carbon stars (CS) without IRAS counterparts (CS\_NI; AllWISE or Gaia DR3 sources)\label{tab:tab2}}
\begin{tabular}{llllllllll}
\hline \hline
Subclass     & Reference & Number & Selected & non-IRAS & Ba-CH\_NI & dC\_NI & RH\_NI & J-type\_NI & CAGB\_NI \\
\hline
CAGB-WISE  &\citet{suh2022} & 5366 & 2440$^a$   & 2440 & 24  & 2 & 28 & 69 & 1663 \\
CS    &\citet{alksnis2001} & 6891  & 6817$^{b}$ & 3018 &28 & 2 & 31 & 75 & 2022 \\
CAGB   &\citet{chen2012} & 974  & 963$^{c}$   &1    & 0  & 0 & 0 & 0 & 1\\
J-type &\citet{chen2007} & 113  & 112$^{d}$   &9    & 0  & 0 & 0 & 8 & 0\\
CS        &\citet{abia2022} & 827 & 818$^{b}$ &244  & 0  & 0 & 230 &1  &3 \\
CS        &\citet{li2018} & 2650 & 2624$^{b}$ &2275 &1442 & 6 &204  &124  &75 \\
Ba    &\citet{lu1991} & 389 & 389              & 221 &221 & 0 & 0  & 0 & 0      \\
Ba (mostly) &\citet{escorza2017} & 437 & 437   & 261 &244 & 0 & 1  & 0 & 0 \\
dC (mostly) & \citet{green2013}  & 1211 & 1211 &1210 &25  &1086 &0 & 2 & 2  \\
CS   & SIMBAD CS   & 21,743 & 7541$^{e}$      &3901 &80  &448 &203 &78 &1789  \\
CS   & Gaia DR3 CS &30023$^{f}$  &19863$^{g}$    &15365$^{h}$ &2  &2 & 0 & 66 & 1857 + (7470)$^{i}$   \\ 
\hline
Total$^j$ &- & -  & -                            &-   &1707 &1087 &230 &131 &2254 + (7470)$^{i}$ \\
\hline
\end{tabular}
\begin{flushleft}
$^a$color-selected objects without any other evidences are excluded
$^b$sources of duplicate SIMBAD identifers are excluded.
$^c$wrongly classified objects are excluded.
$^d$IRAS 18006-3213 is an OH/IR star.
$^e$CDS SIMBAD main type C* in our Galaxy.
$^f$C-rich stars from Gaia DR3 spectra for LPVs (\citealt{lebzelter2023}).
$^g$objects in our Galaxy.
$^h$denoted by GC(all)\_NI.
$^i$candidate objects for new CAGB stars solely identified from Gaia DR3 spectra and absolute magnitudes
(denoted by GC-CAGB\_NI; see Section~\ref{sec:gaiacarbon}).
$^j$duplicate objects are excluded.
\end{flushleft}
\end{table*}

\subsection{Sample stars \label{sec:carbon-gs}}

We have compiled lists of known carbon stars in our Galaxy from both the 
literature and CDS SIMBAD astronomical database, as detailed in 
Tables~\ref{tab:tab1} and \ref{tab:tab2}. Table~\ref{tab:tab1} comprises the 
sample stars with corresponding IRAS counterparts (CS\_IC), while 
Table~\ref{tab:tab2} lists stars without IRAS counterparts (CS\_NI). The tables 
provide references and the numbers of selected objects for each subclass of 
carbon stars. 

We present the sample of carbon stars in two parts: one (CS\_IC) based on the 
IRAS source catalog for brighter or more isolated objects (Table~\ref{tab:tab1}), 
and the other (CS\_NI) based on the AllWISE or Gaia DR3 source catalog for less 
bright objects or those in crowded regions (Table~\ref{tab:tab2}). While most 
CS\_IC objects have AllWISE or Gaia DR3 counterparts, CS\_NI objects lack IRAS 
counterparts. 

For each carbon star outlined in Tables~\ref{tab:tab1} and \ref{tab:tab2}, we 
meticulously assigned specific subclasses by referencing pertinent literature and 
the CDS SIMBAD database. The subclasses include Ba or CH stars (Ba-CH\_IC and 
Ba-CH\_NI), dC stars (dC\_NI; relevant only for non-IRAS objects), R-hot stars 
(RH\_IC and RH\_NI), J-type stars (J-type\_IC and J-type\_NI), R Coronae Borealis 
(RCB) stars (RCB\_IC; considered only for IRAS objects), and CAGB stars (CAGB\_IC 
and CAGB\_NI). Ambiguous cases, where the carbon-rich nature was unclear due to 
conflicting information (e.g., certain C-type objects identified as OAGB stars), 
were omitted from the selection. 

R Coronae Borealis (RCB) variables, which are a class of Hydrogen-deficient
Carbon (HdC) stars, are believed to be formed via the merger of two white dwarf
(WD) stars. Some RCB stars were known to be CAGB stars or J-type carbon stars in
previous studies.

While the IR color-selection method using a 2CD is beneficial for distinguishing 
between CAGB and OAGB stars (e.g., \citealt{sh2017}; \citealt{suh2022}), we 
acknowledge its potential unreliability for a significant proportion of AGB stars 
compared to spectroscopic or spectrophotometric methods. To enhance the 
reliability of our carbon star sample, we excluded objects solely classified 
based on the IR color-selection method without additional supporting evidence. 

In this study, we have identified CAGB stars from a substantial sample of carbon 
stars spanning various subclasses (refer to Tables~\ref{tab:tab1} and 
\ref{tab:tab2}), leveraging information from diverse literature sources, the CDS 
SIMBAD database, and examining their IR properties through 2CDs, CMDs, and SEDs 
(refer to Sections~\ref{sec:irpro} and \ref{sec:cagb}). 

The criteria for classifying CAGB stars can be summarized as follows: 1) they 
exhibit characteristics indicative of being carbon-rich stars from the 
spectroscopic information (see Section~\ref{sec:specdata}); 2) their IR 
properties (2CDs, CMDs, and SEDs) align with those of known CAGB stars; 3) their 
absolute magnitudes are brighter than -6 mag at IR[12] or W3[12] (refer to 
Section~\ref{sec:irpro}); 4) there is no conclusive evidence of J-type or complex 
binary star characteristics; 5) Mira variables can be considered as AGB stars 
(refer to Section~\ref{sec:pul}). 

For all the sample stars, we carefully cross-identified available counterparts 
from IRAS, Akari, MSX, 2MASS, WISE, Gaia DR3, AAVSO (version 2023-12-12; 
\citealt{watson2023}), and the CDS SIMBAD object database.

\subsection{Infrared Photometric Data \label{sec:photdata}}

2MASS \citep{cutri2003} provided fluxes at J (1.25 $\mu$m), H (1.65 $\mu$m), and 
K (2.16 $\mu$m) bands. The WISE \citep{jarrett2011} mapped the sky at four bands 
(3.4, 4.6, 12, and 22 $\mu$m; W1, W2, W3, and W4). The FOV pixel sizes of the 
IRAS, MSX, AKARI PSC, AKARI BSC, 2MASS, and WISE images are 
0$\farcm$75x(4$\farcm$5-4$\farcm$6), 18$\farcs$3, 10$\arcsec$, 30$\arcsec$, 
2$\arcsec$, and 2$\farcs$75, respectively. 

Table~\ref{tab:tab3} list IR bands used in this work for 2CDs and CMDs. For each 
band, the reference wavelength ($\lambda_{ref}$) and zero magnitude flux (ZMF) 
value, which are useful to obtain theoretical models colors (see 
Section~\ref{sec:agbmodels}) and SEDs, are also shown. For more detailed 
description of IR photometric data and ZMF, refer to Section 2.1 in 
\citet{suh2021}. 

In this work, we use only good quality observational data at all wavelength bands
for the  photometric data (quality better than 1 for the IRAS, AKARI, and MSX
data; quality better than U for the WISE data).

\begin{table}
\scriptsize
\caption{IR bands and zero magnitude flux values\label{tab:tab3}}
\centering
\begin{tabular}{lllll}
\hline \hline
Band &$\lambda_{ref}$ ($\mu$m)	&ZMF (Jy) &Telescope &Reference 	\\
\hline
K[2.2]  &2.159	&	666.7	&	2MASS	& \citet{cohen2003}\\
W1[3.4]	&3.35	&	306.682	&	WISE    & \citet{jarrett2011}	\\
W2[4.6]	&4.60	&	170.663	&	WISE    & \citet{jarrett2011}	\\
IR[12]	&12     &  28.3 	&	IRAS     & \citet{beichman1988}	\\
W3[12]$^a$	&12	    &  28.3 	&	WISE & \citet{jarrett2011}\\
W4[22]	&22.08	&	8.284	&	WISE    & \citet{jarrett2011}	\\
IR[25]	&25 & 6.73 	&	IRAS      & \citet{beichman1988}	\\
IR[60]	&60 & 1.19 	&	IRAS      & \citet{beichman1988}	\\
\hline
\end{tabular}
\begin{flushleft}
$^a$For W3[12], we use a new reference wavelength and zero magnitude flux for theoretical models
(see section 4.2 in ~\citealt{suh2020}).
\end{flushleft}
\end{table}

\subsection{The IRAS Data \label{sec:iras}}

Given the considerably low angular resolution of IRAS, utilizing AllWISE or 2MASS 
counterparts derived from the positions of cross-matched AKARI or MSX sources 
(which possess higher angular resolution, as discussed in 
Section~\ref{sec:photdata}) improves the reliability of the data (e.g., 
\citealt{suh2021}). 

In the IRAS Point Source Catalog (PSC), there are 245,889 sources, and in the 
Faint Source Catalog (FSC), there are 173,044 sources. This study utilizes the 
combined source catalog of IRAS, as presented by \citet{abrahamyan2015}, 
consisting of 345,163 sources with 73,770 associated sources from the PSC and 
FSC. A correction is applied, acknowledging that IRAS PSC 17501-0333 is distinct 
from IRAS FSC F17499-0334 due to differing Akari PSC counterparts with distinct 
flux values. Consequently, a combined source catalog of 345,164 sources with 
73,769 associated sources is employed.

In contrast to the approach taken by \citet{abrahamyan2015}, this study utilizes 
FSC data (position and flux) only for the 32,079 associated sources with FSC flux 
at 12 $\mu$m fainter than the average FSC flux. For the brighter 41,690 
associated sources, PSC data is preferred, as it yields higher-quality data. 

For all 345,164 IRAS sources, AKARI, MSX, AllWISE, 2MASS, OGLE4, Gaia DR3, and 
AAVSO counterparts are identified using the following method. The AKARI PSC, MSX, 
and AKARI BSC counterparts are cross-identified by selecting the nearest source 
within 60$\arcsec$, utilizing the position provided in the IRAS PSC or FSC. 
Subsequently, the AllWISE, 2MASS, Gaia DR3, and AAVSO counterparts are 
cross-identified using the best position among the available AKARI PSC, MSX, or 
AKARI BSC counterparts. The position of the IRAS source is employed only in cases 
where there is no AKARI or MSX counterpart. When a Gaia DR3 counterpart is 
present, the Gaia position can be used as the best position for the IRAS source. 

Due to the substantial beam size of IRAS, some sources represent multiple 
objects. Notably, certain IRAS objects recognized as CAGB stars (e.g., IRAS 
17209-3318 and 10151-6008) are now identified as multiple objects. These 
particular objects are excluded from the CS\_IC list. The majority of these 
objects find inclusion in the CS\_NI list, with consideration given to SIMBAD 
information and their counterparts in AKARI, WISE, and Gaia DR3. It is important 
to note that there are no duplicate objects in the CS\_IC and CS\_NI lists.

\subsection{Spectroscopic data \label{sec:specdata}}

The SIMBAD astronomical database provides useful information on the optical and 
IR spectra for the sample stars. AAVSO (\citealt{watson2023}) also provides 
spectrocopic information for a major portion of sample stars. 

In identifying important dust features for AGB stars, IRAS Low Resolution
Spectrograph (LRS; $\lambda$ = 8$-$22 $\mu$m) data are useful
(\citealt{kwok1997}). The class E (the 10 $\mu$m silicate feature in emission)
and class A (the 10 $\mu$m silicate feature in absorption) objects are typically
OAGB stars. Class C (the 11.3 $\mu$m SiC feature in emission) objects are
generally carbon stars.

Among the 715 IRAS LRS sources identified as class C, 713 were initially 
categorized as CAGB stars by \citet{suh2022}, with only two exceptions (IRAS 
13136-4426: S-type star; IRAS 22306+5918: composite object). However, further 
examination reveals that 23 objects (9 J-type stars, 3 OAGB stars, 11 others) 
deviate from the CAGB classification, leaving a total of 690 confirmed CAGB 
stars. 

Additionally, the ISO Short Wavelength Spectrometer (SWS: $\lambda$=2.4–45.4 
$\mu$m) data can be employed for analysis (see Section~\ref{sec:modelsed}). The 
ISO SWS catalogue by \citet{sloan2003} encompasses high-resolution spectral data 
from 1271 observations. 

Carbon stars exhibit diverse gas-phase emission or absorption features at the NIR 
and MIR bands, attributable to circumstellar molecules like C$_2$, CO, CN, HCN, 
and C$_2$H$_2$ (e.g., \citealt{lancon2000}). Identification of 139 cool carbon 
stars in the NIR spectrophotometric survey of the Infrared Telescope in Space 
(IRTS) was conducted by \citet{LeBertre2005}, based on the conspicuous presence 
of molecular absorption bands at 1.8, 3.1, and 3.8 $\mu$m. These objects were 
duly considered in the CAGB lists presented by \citet{sk2011} and 
\citet{suh2022}. 

The Gaia DR3 LPV (\citealt{lebzelter2023}) data provided the low-resolution RP 
spectra at visual wavelength bands to identify carbon stars for the 1,720,558 LPV 
candidates. \citet{li2018} presented a catalog of 2651 carbon stars from the 
optical spectra of the Large Sky Area Multi-Object Fiber Spectroscopy Telescope 
(LAMOST). While these spectra in the visual bands may exhibit limitations for 
objects with thick dust envelopes or those experiencing significant extinction 
(e.g., \citealt{suh2022}), they remain valuable for identifying visual carbon 
stars within a substantial sample of observed objects. 

Using Gaia RP spectral data for the LPV candidates, 30,023 objects can be 
classified as carbon stars (\citealt{suh2022}) using the selection criteria (7 
$<$ median\_delta\_wl\_rp $<$ 11 and G$_{BP}$ $<$ 19 mag) outlined in Section 2.4 
of \citet{lebzelter2023}. Among these, 19,863 objects are located in our Galaxy. 
From these 19,863 objects, 4498 have IRAS counterparts, and 15,365 do not. It is 
noteworthy that most of the 4498 objects with IRAS counterparts are already 
well-known as carbon stars in various literature sources and SIMBAD. Among them, 
we identify 4132 objects as CAGB\_IC objects. 

The 15,365 objects without IRAS counterparts are considered candidate carbon 
stars identified from Gaia DR3 spectra (denoted as GC(all)\_NI; refer to 
Table~\ref{tab:tab2}) without any additional evidence. From this group of 15,365 
GC(all)\_NI objects, we select candidates for CAGB stars (GC-CAGB\_NI objects) 
based on their IR properties (refer to Section~\ref{sec:gaiacarbon}).

\begin{figure*}
\centering
\largeplottwo{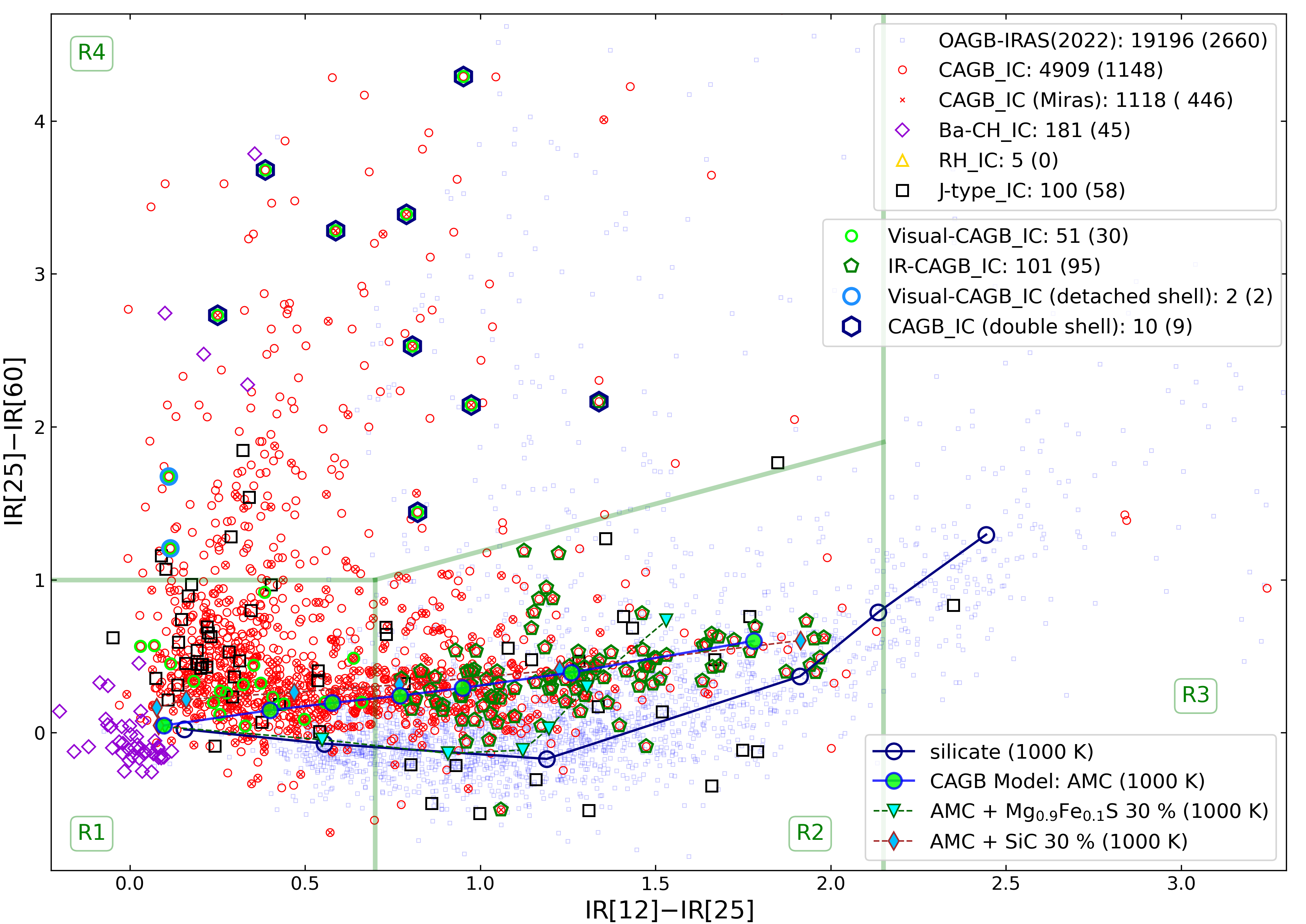}{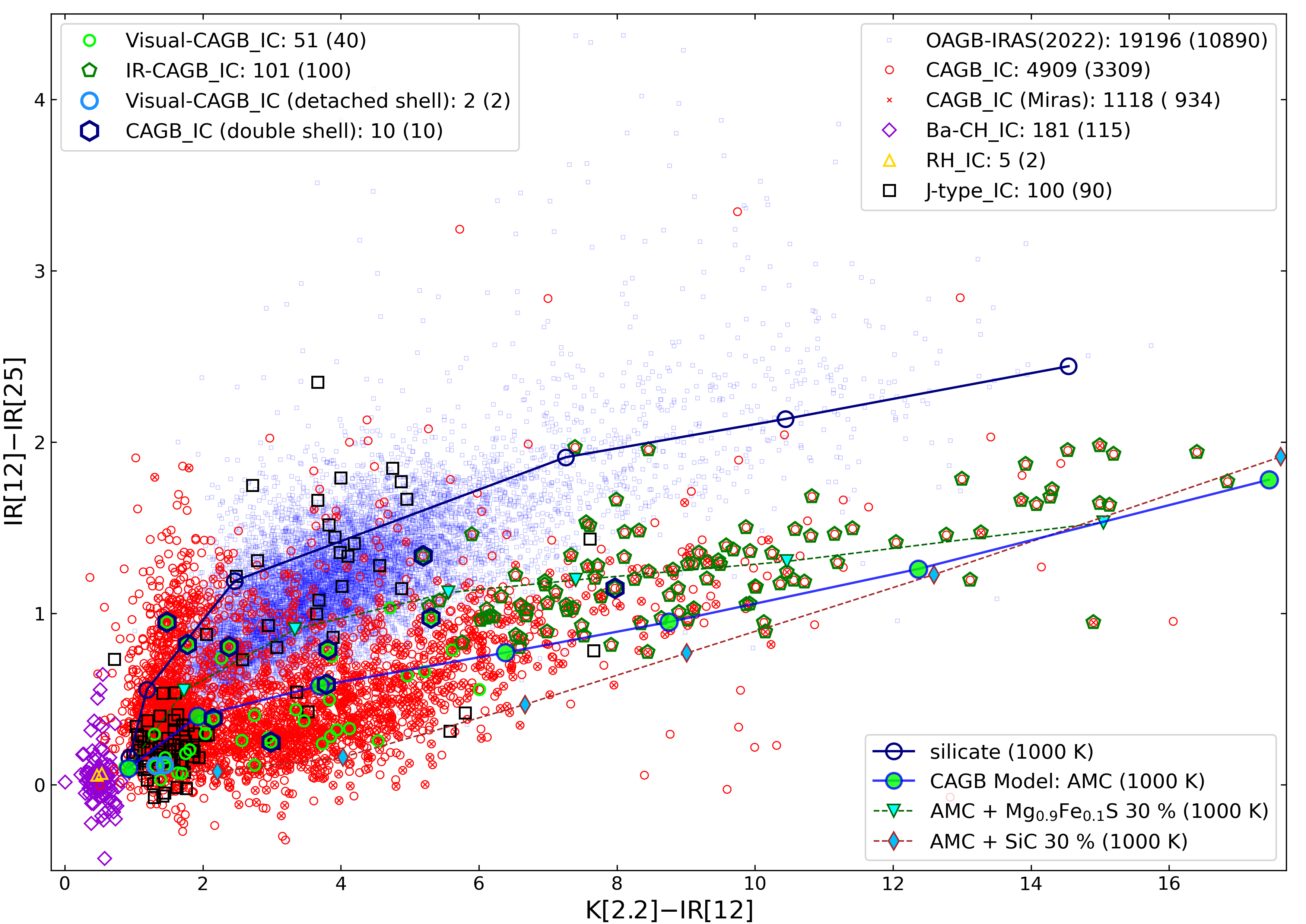}\caption{IRAS-2MASS 2CDs for various subclasses of CS\_IC objects (see Table~\ref{tab:tab1})
and special subgroups of CAGB stars (see Section~\ref{sec:irpro}) compared with theoretical models (see Section~\ref{sec:agbmodels}).
For CAGB models (AMC $T_c$ = 1000 K): $\tau_{10}$ = 0.0001, 0.01, 0.1, 0.5, 1, 2, and 4 from left to right.
For each subclass, the number of objects is shown.
The number in parentheses denotes the number of the plotted objects on the 2CD with good-quality observed colors.
See Section~\ref{sec:ircd}.} \label{f1}
\end{figure*}

\begin{figure*}
\centering
\largeplottwo{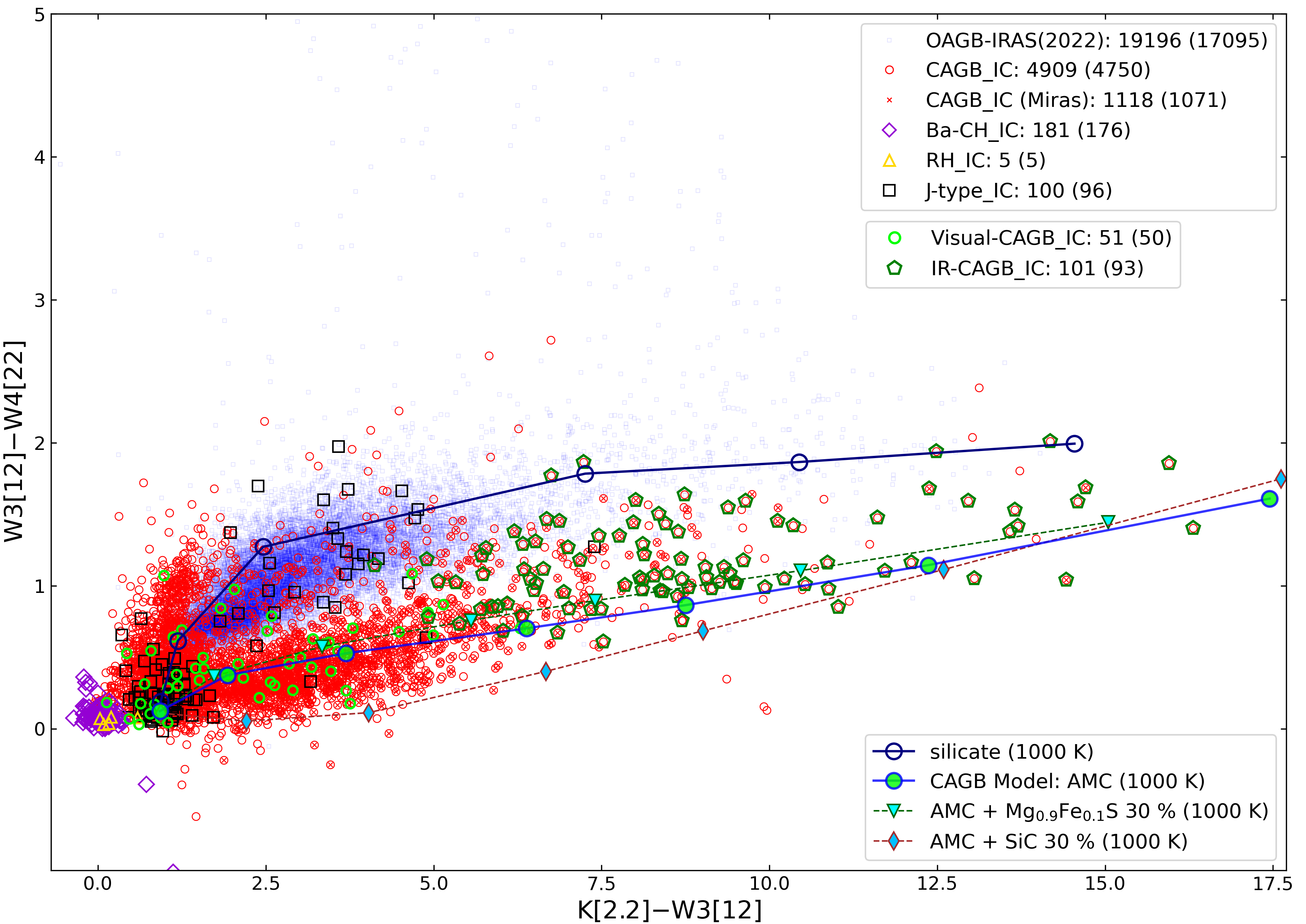}{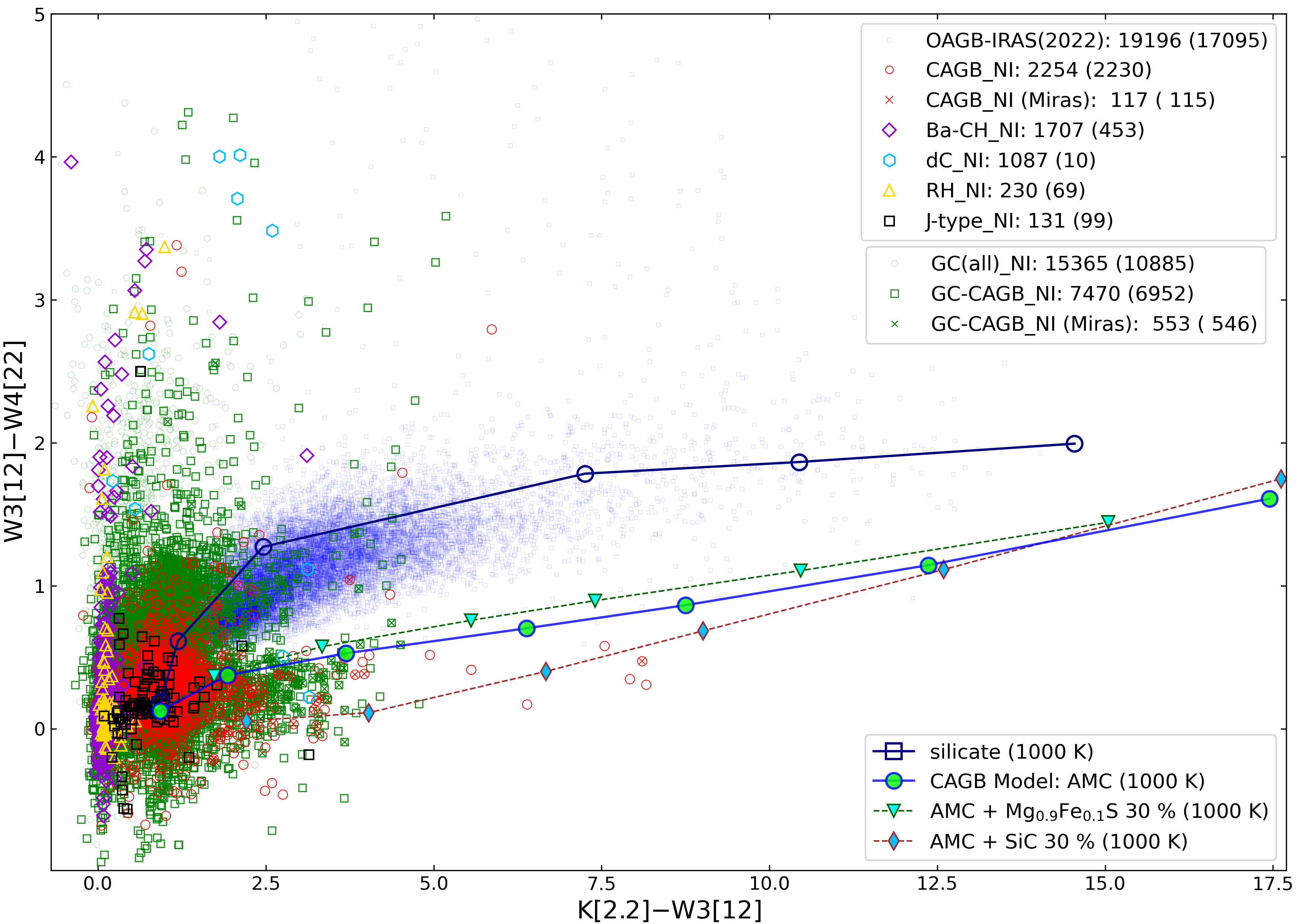}\caption{WISE-2MASS 2CDs for CS\_IC and CS\_NI objects
(see Tables~\ref{tab:tab1} and \ref{tab:tab2}) compared with theoretical models (see Section~\ref{sec:agbmodels}).
For CAGB models (AMC $T_c$ = 1000 K): $\tau_{10}$ = 0.0001, 0.01, 0.1, 0.5, 1, 2, and 4 from left to right.
For each subclass, the number of objects is shown.
The number in parentheses denotes the number of the plotted objects on the 2CD with good-quality observed colors.
See Section~\ref{sec:wise}.} \label{f2}
\end{figure*}

\begin{figure}
\centering
\smallplot{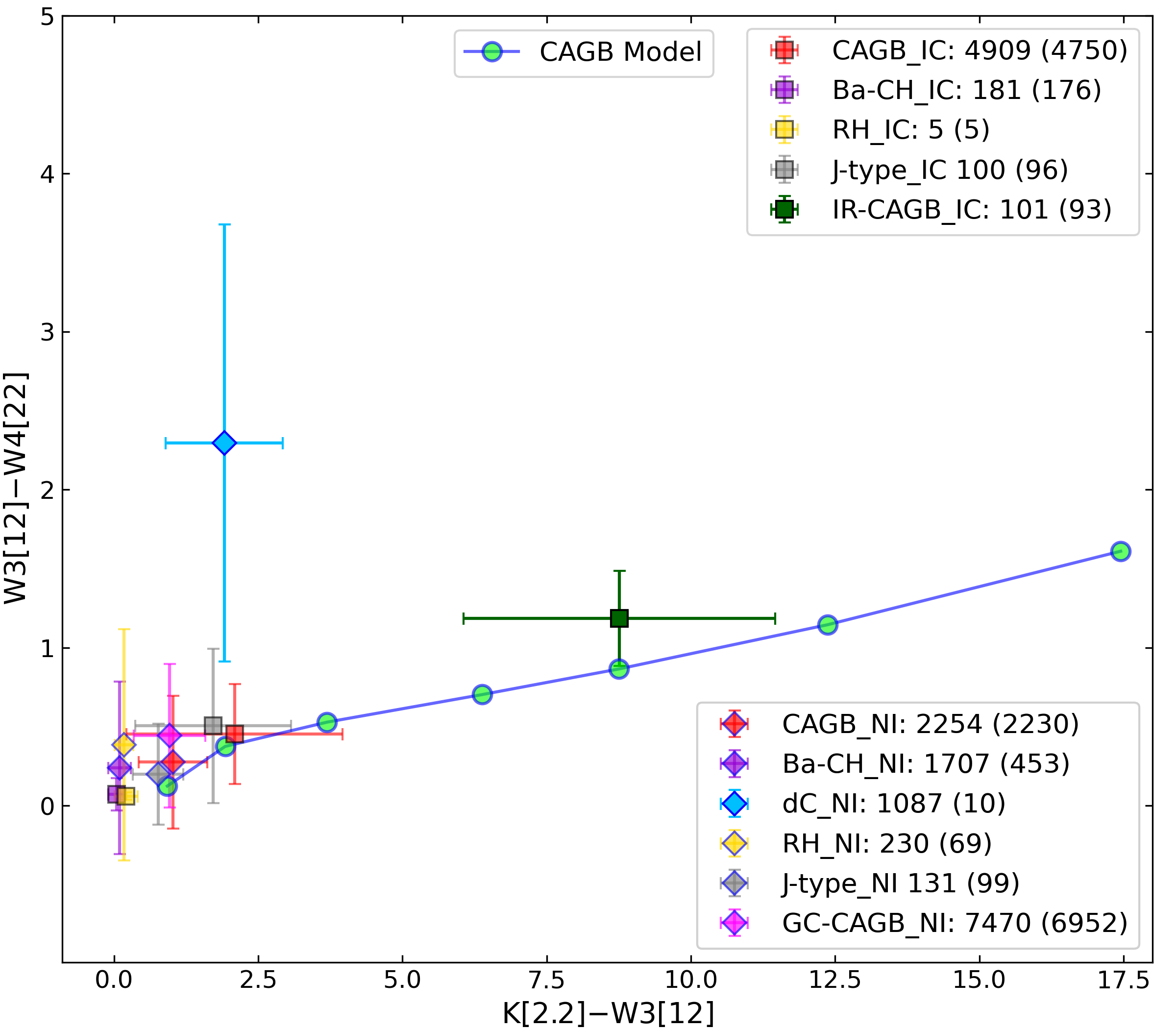}\caption{Error bar plot of the WISE-2MASS 2CD for all carbon stars
(CS\_IC and CS\_NI objects; see Figure~\ref{f4}).
See Section~\ref{sec:wise}.}
\label{f3}
\end{figure}

\begin{figure}
\centering
\smallplot{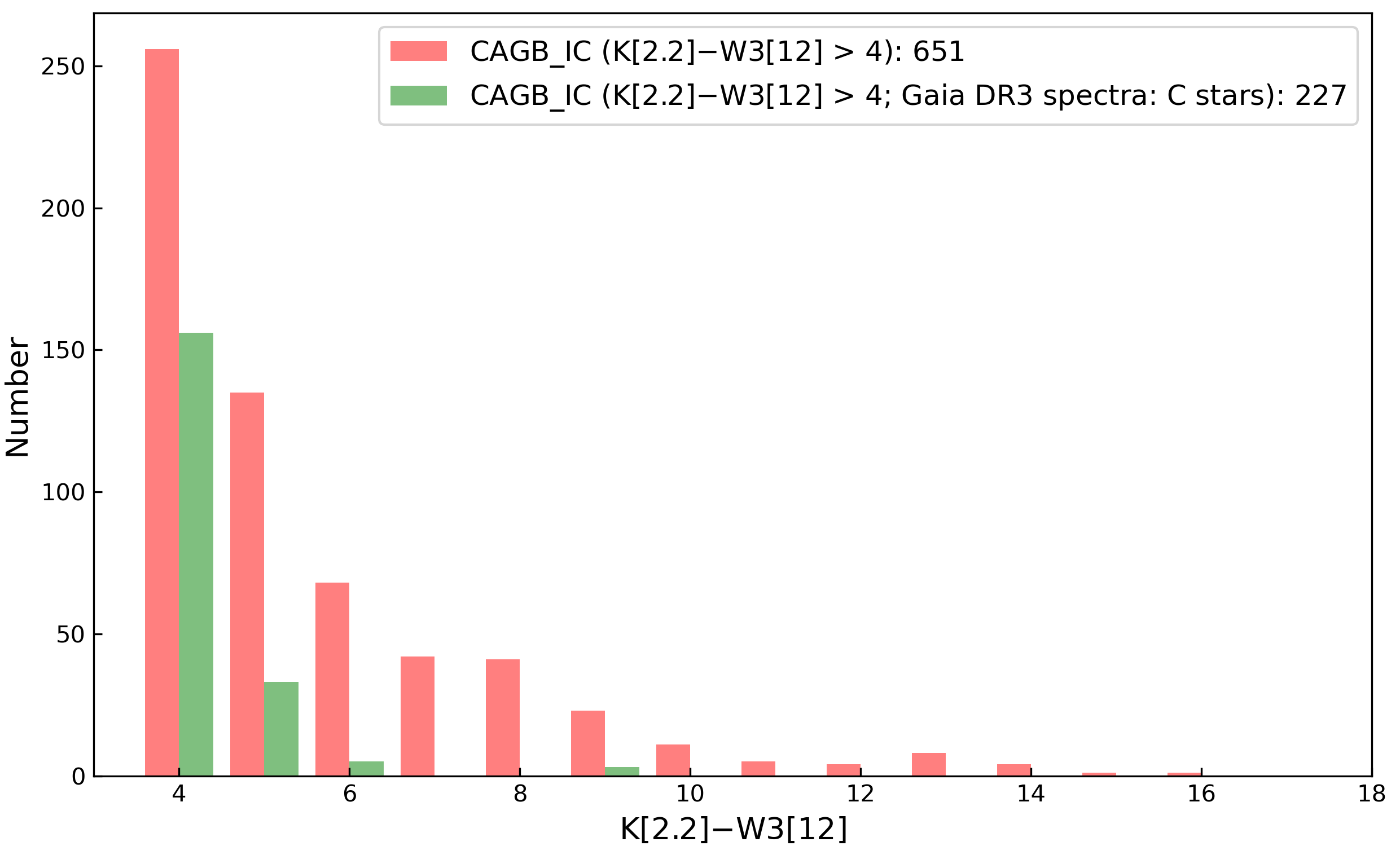}\caption{Histograms of the K[2.2]$-$W3[12] colors for CAGB stars (CAGB\_IC objects) with thick dust shells.
CAGB stars that can be classified as C-rich stars based on the Gaia DR3 spectra are compared with all CAGB stars.
See Section~\ref{sec:wise}.}
\label{f4}
\end{figure}

\begin{figure*}
\centering
\smallplottwo{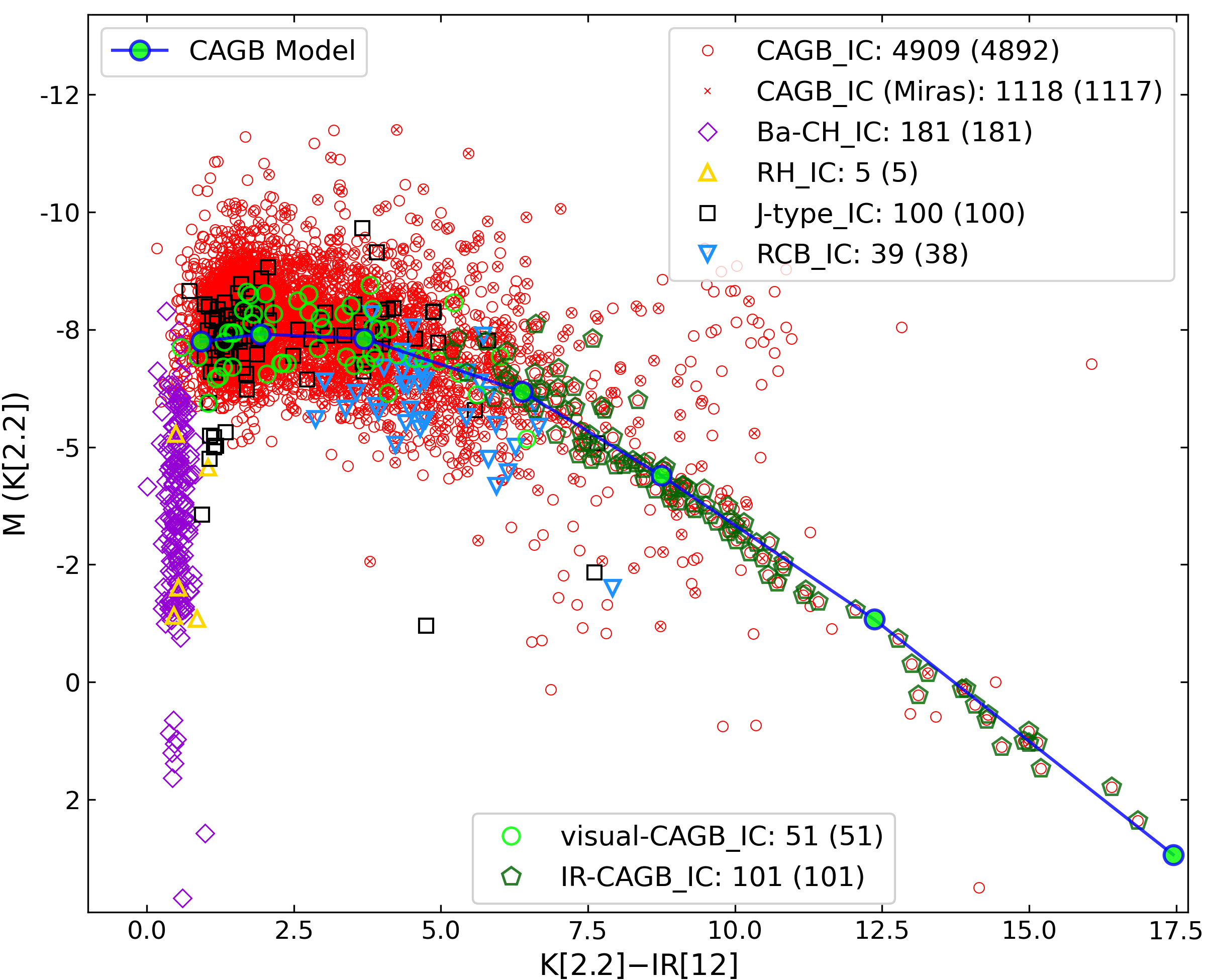}{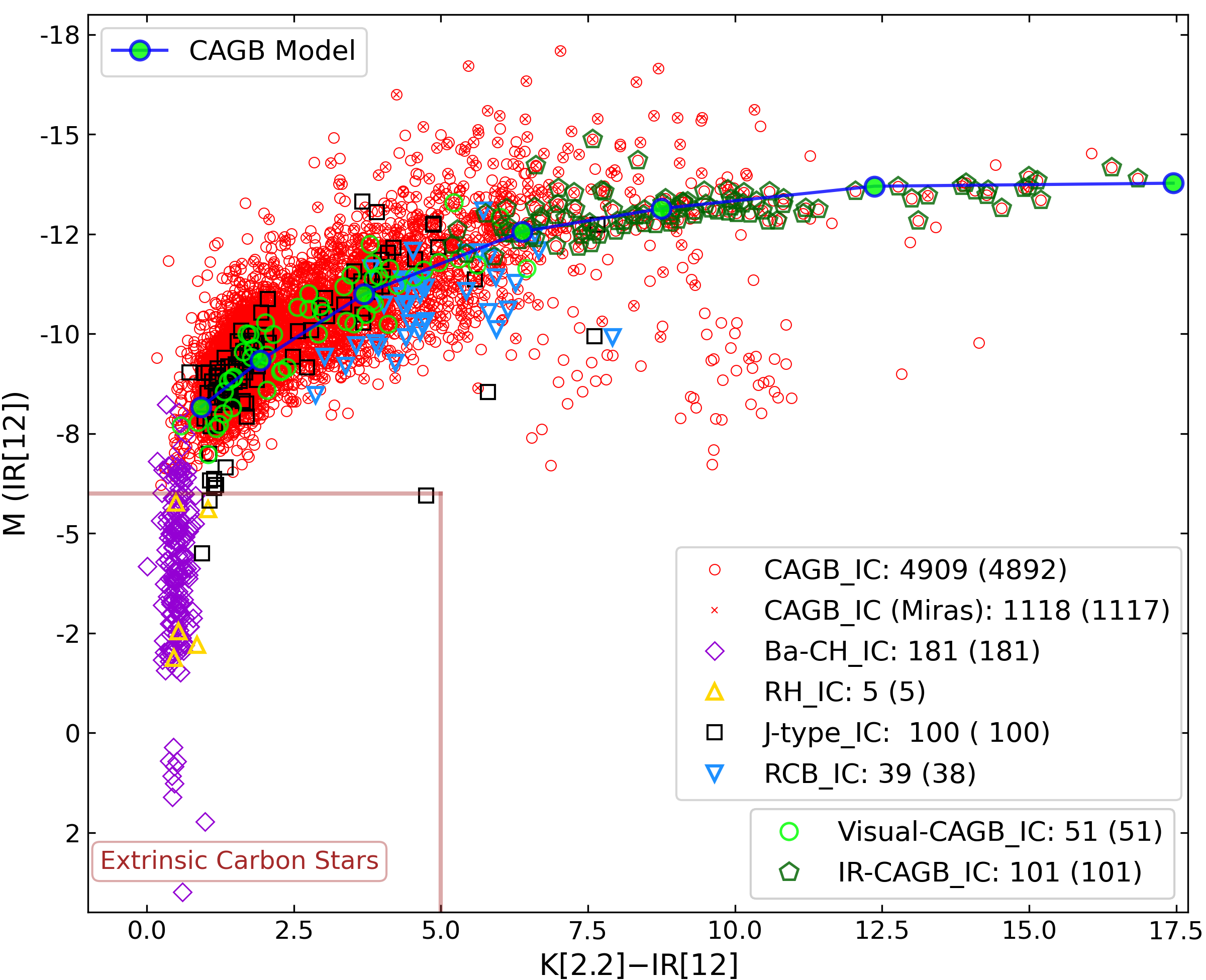}\caption{IRAS-2MASS CMDs for carbon stars with IRAS counterparts
spanning various subclasses (see Table~\ref{tab:tab1}). 
For CAGB models (AMC $T_c$ = 1000 K): $\tau_{10}$ = 0.0001, 0.01, 0.1, 0.5, 1, 2, and 4 from left to right (see Section~\ref{sec:agbmodels}).
For each subclass, the number of objects is shown.
The number in parenthesis denotes the number of the plotted objects with good quality observed data.
See Section~\ref{sec:CMD}.} \label{f5}
\end{figure*}

\begin{figure*}
\centering
\smallplotsix{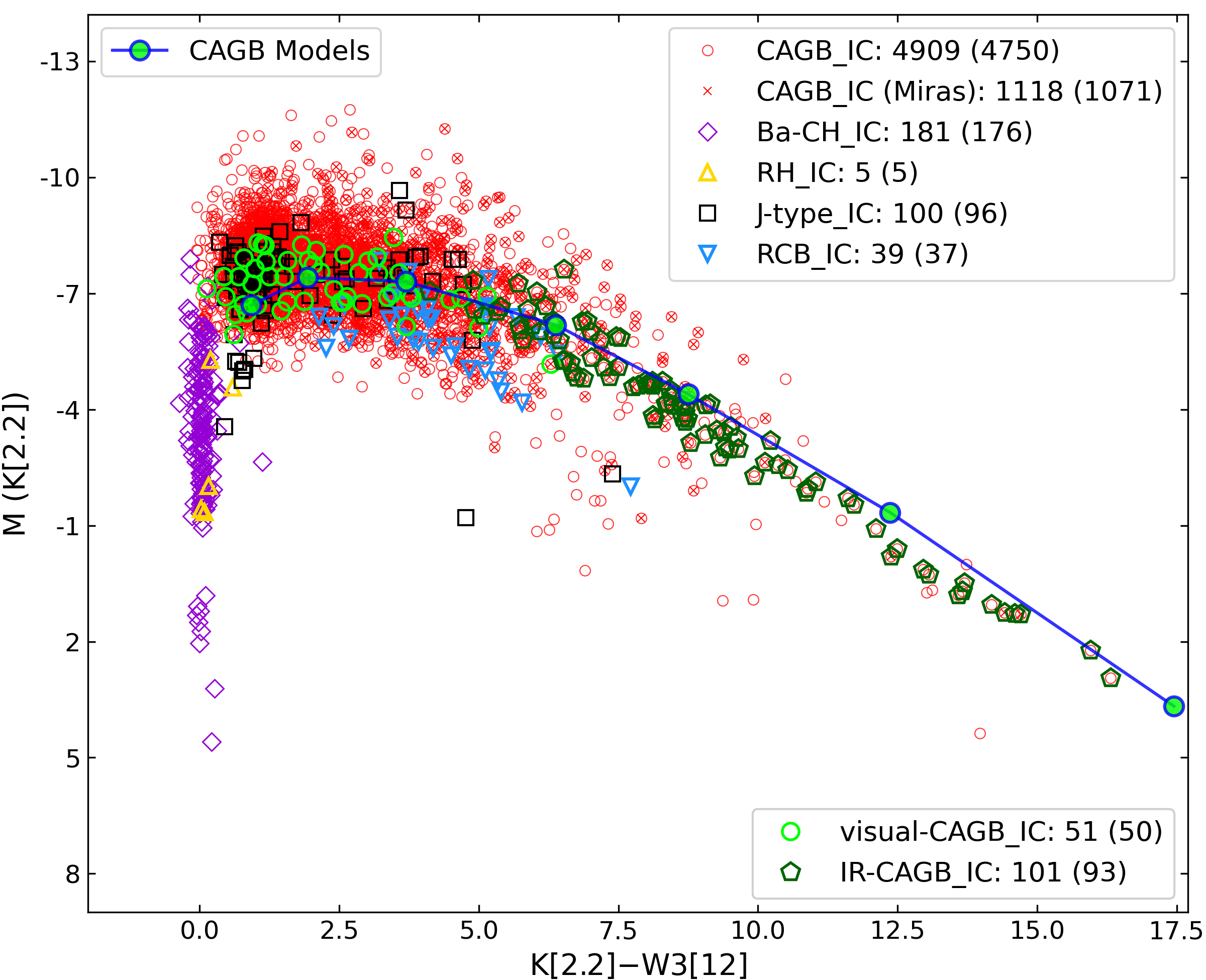}{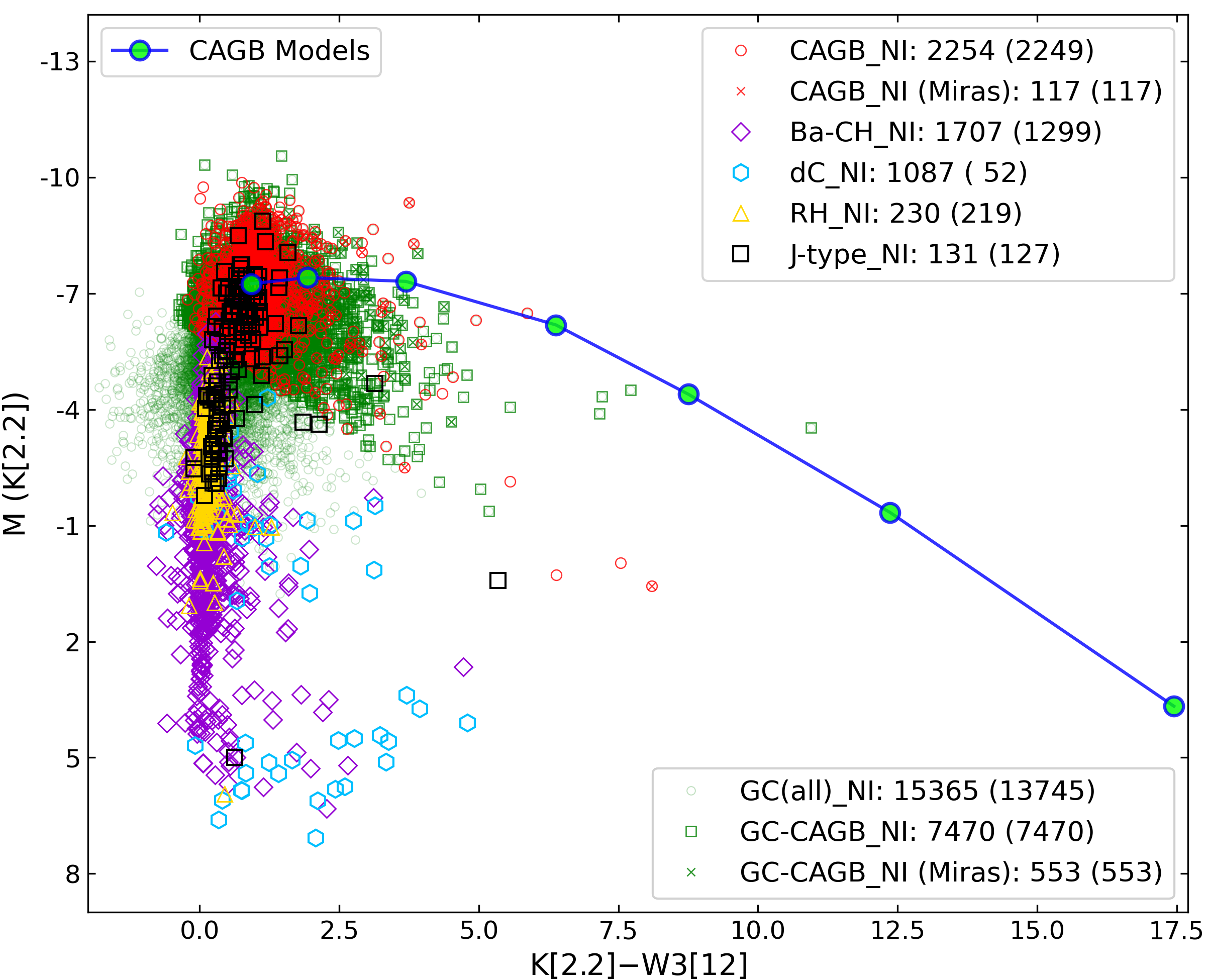}{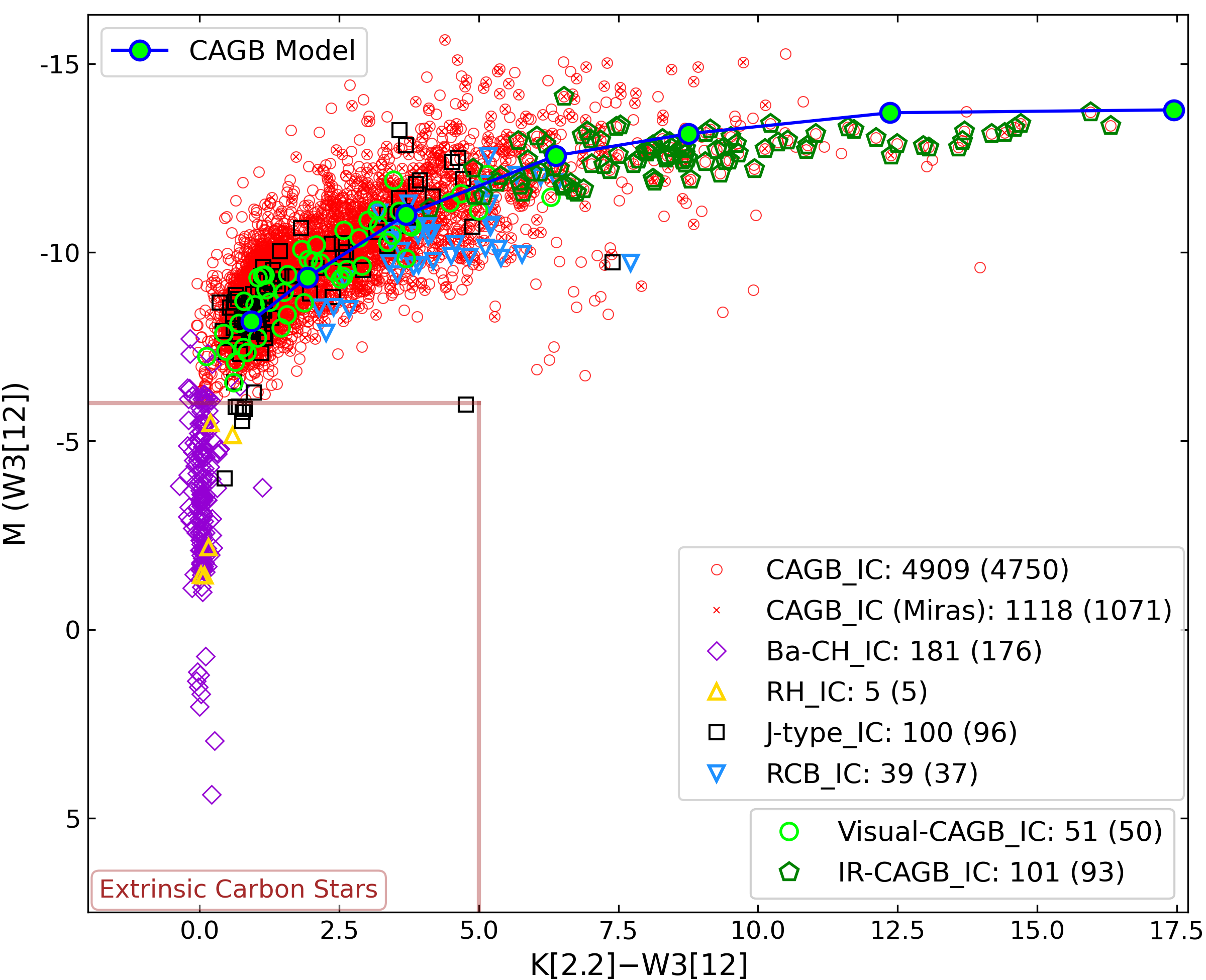}{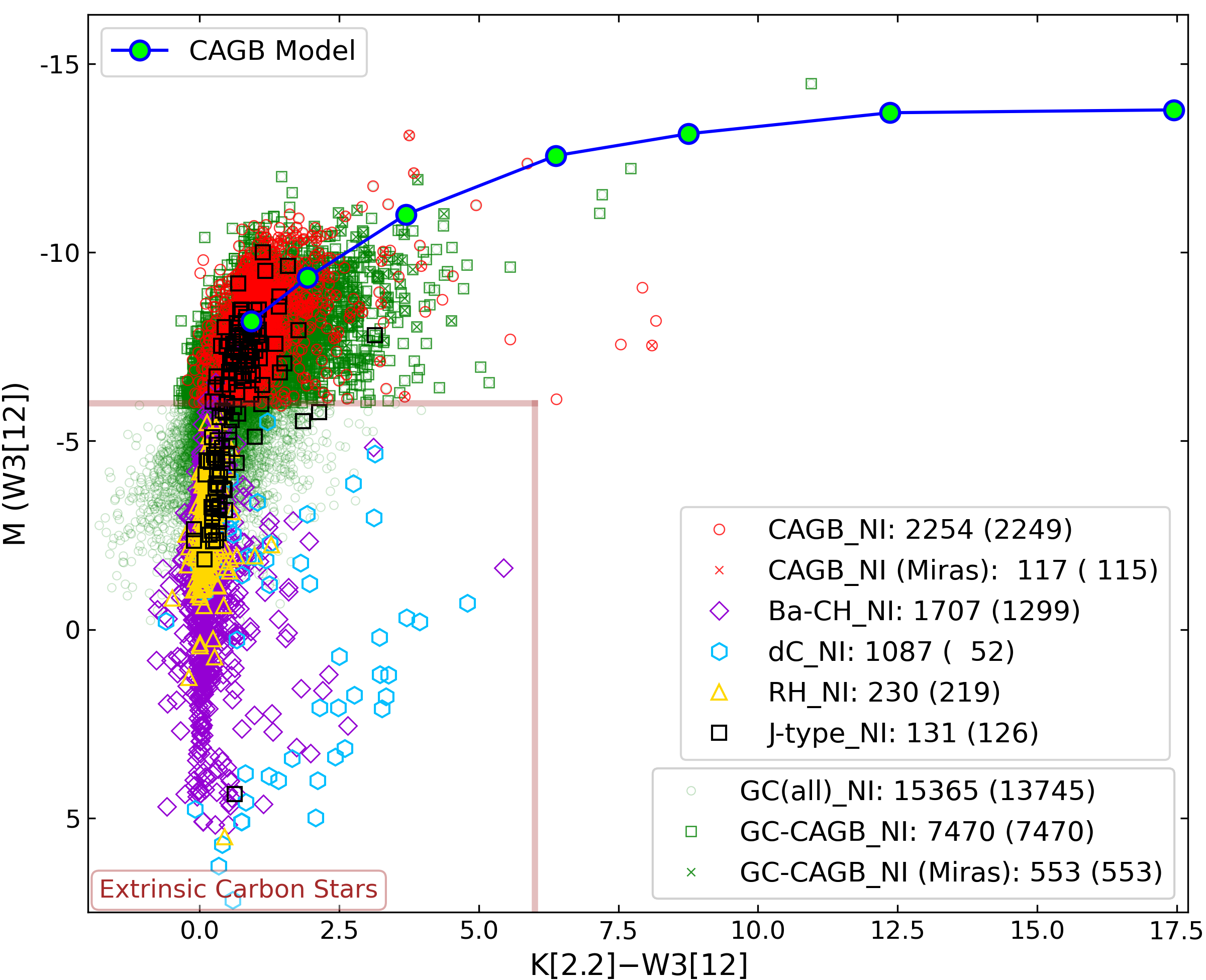}{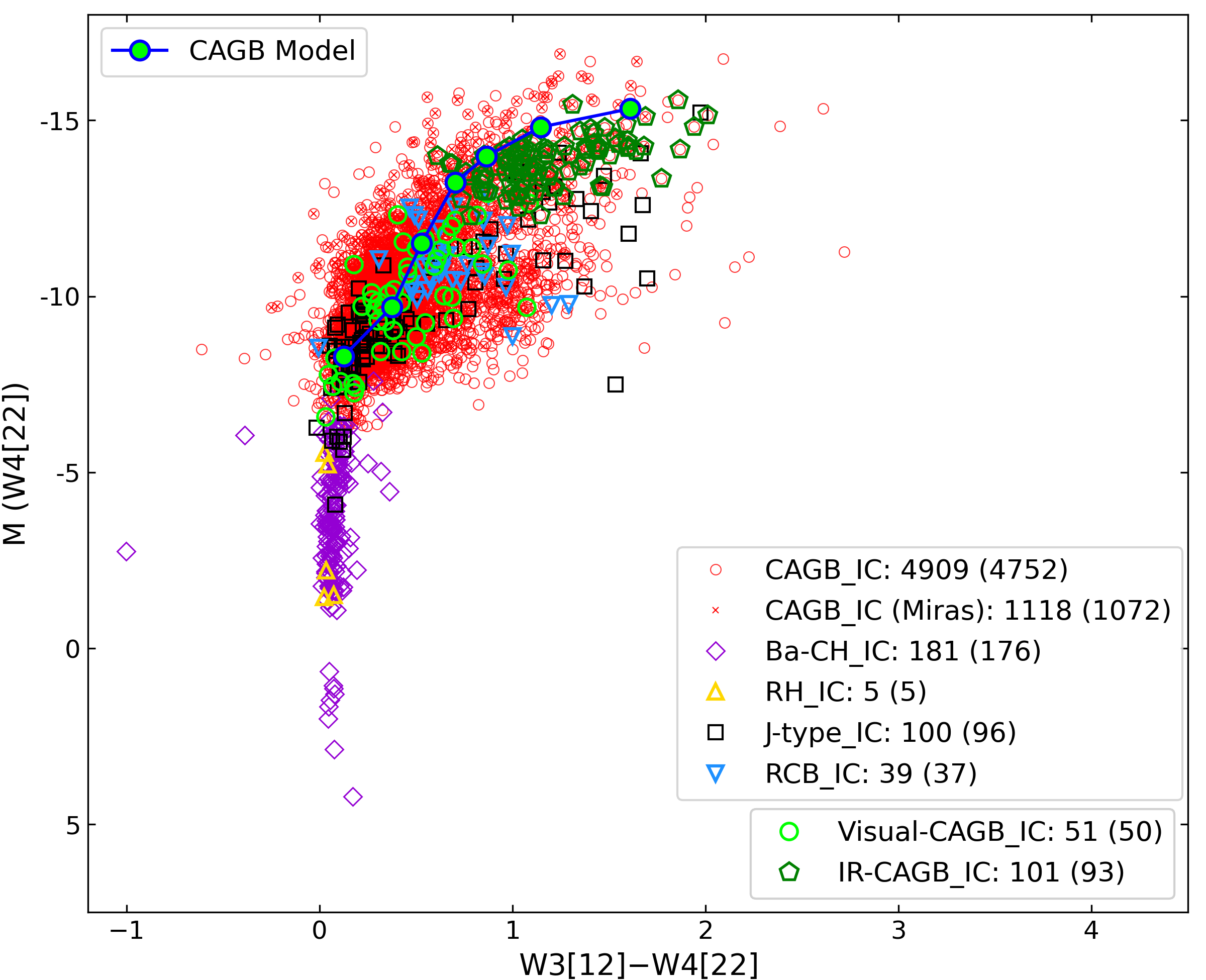}{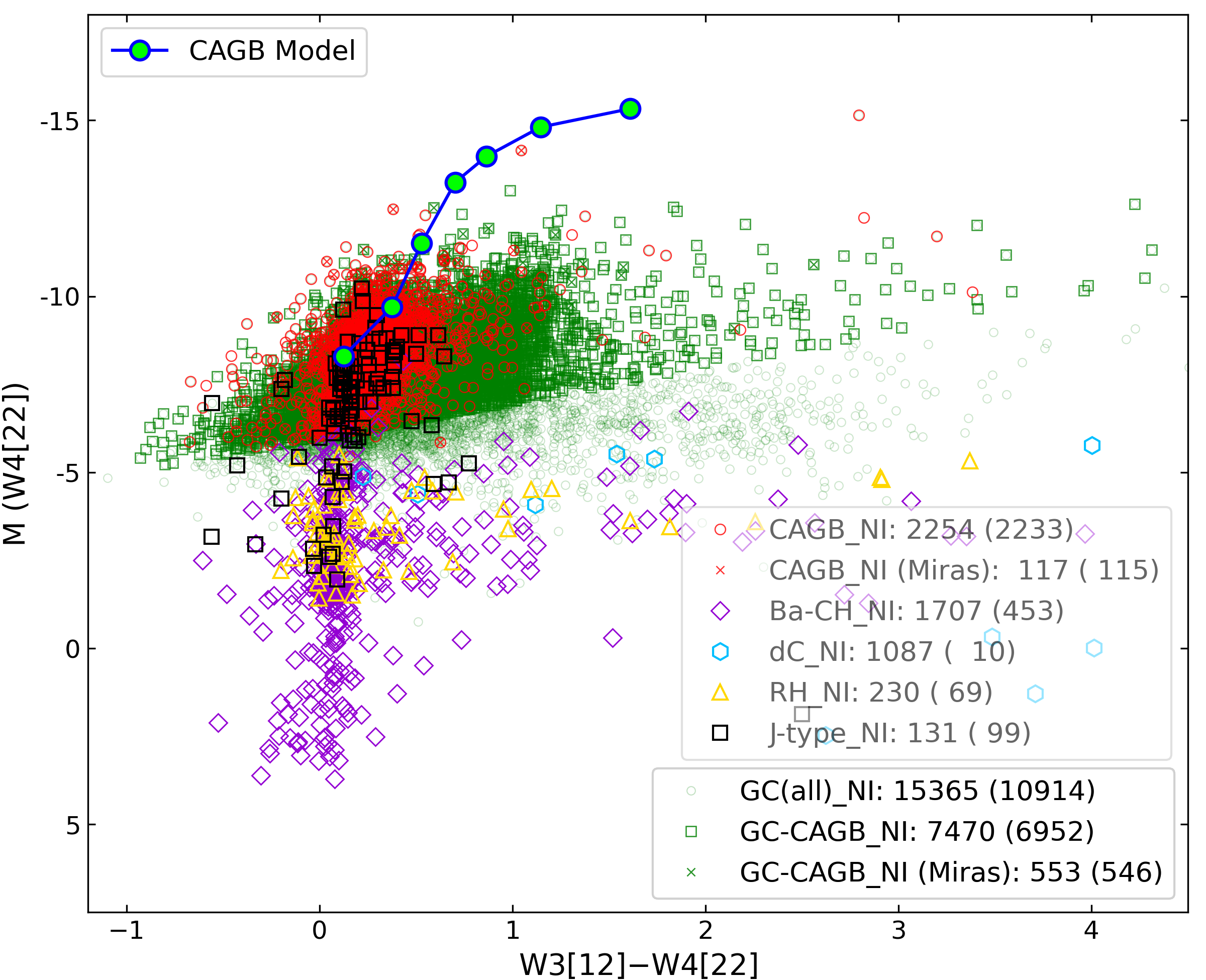}\caption{WISE-2MASS CMDs for carbon stars spanning
various subclasses (see Tables~\ref{tab:tab1} and ~\ref{tab:tab2}).
For CAGB models (AMC $T_c$ = 1000 K): $\tau_{10}$ = 0.0001, 0.01, 0.1, 0.5, 1, 2, and 4 from left to right (see Section~\ref{sec:agbmodels}).
For each subclass, the number of objects is shown.
The number in parenthesis denotes the number of the plotted objects with good quality observed data.
See Section~\ref{sec:CMD}.} \label{f6}
\end{figure*}

\subsection{Distances and Galactic extinction \label{sec:ext}}

The recently obtained distance information (\citealt{bailer-jones2021}) and 
extinction data from \citet{lallement2022}, derived from Gaia DR3 data, play a 
crucial role in determining the absolute luminosity of a significant number of 
carbon stars in our Galaxy. 

For the majority of carbon stars, distances derived from Gaia DR3 data are 
readily available. In instances where Gaia DR3 distances are absent, we turn to 
distances from SIMBAD (mainly obtained from Hipparcos) when accessible. In cases 
where such information is unavailable for CAGB stars, we determine the distance 
through a comparative analysis between the observed SED and the theoretical CAGB 
model SED, as elaborated in Section~\ref{sec:modeldistance}. 

Having established the distance and position of each object, we obtained 
extinction values ($A_V$) at the visual band from the data server provided by 
\citet{lallement2022}. Expanding our analysis, we can calculate extinction values 
at Gaia, 2MASS, and other bands, employing the optical to MIR extinction law 
presented by \citet{wang2019}. 

The obtained distance and extinction data are used in the construction of various 
IR 2CDs and CMDs, as detailed in Section~\ref{sec:irpro}. Note that the 
extinction data are applied to the colors and magnitudes using the K[2.2] band 
data for the IR 2CDs and CMDs presented in Section~\ref{sec:irpro}.

\section{Infrared properties of carbon stars\label{sec:irpro}}

In this section, we investigate IR properties of known carbon stars in our Galaxy 
presented in Section~\ref{sec:gcarbon}. Using various observational data, we 
present useful IR 2CDs and CMDs for the large sample of the carbon stars compared 
with the theoretical models for CAGB stars (see Section~\ref{sec:agbmodels}) and 
distinguish CAGB stars from other subclasses of carbon stars. 
Table~\ref{tab:tab3} lists the IR bands used for the IR 2CDs and CMDs presented 
in this work (see Section~\ref{sec:photdata}). 

Although the photometric fluxes are less useful than a full SED, the large number 
of observations at various wavelength bands can be used to form a 2CD, which can 
be compared with theoretical models. IR 2CDs are useful to statistically 
distinguish various properties of AGB stars and post-AGB stars (e.g., 
\citealt{epchtein1990}; \citealt{sk2011}; \citealt{suh2015}). 

Figures~\ref{f1} and~\ref{f2} show various IR 2CDs using different combinations 
of observed IR colors for various subclasses of carbon stars (see 
Tables~\ref{tab:tab1}and~\ref{tab:tab2}). For comparison, we also plot OAGB stars 
from \citet{suh2022}. We juxtapose the observations on the IR 2CDs with the 
theoretical dust shell models for CAGB stars (refer to 
Section~\ref{sec:agbmodels}). Generally, the theoretical dust shell models for 
both OAGB and CAGB stars demonstrate a commendable ability to reproduce the 
observed points. However, certain objects deviate significantly from the 
theoretical models. This discrepancy may arise from AGB stars with non-spherical 
dust envelopes, which are not accurately represented by the theoretical dust 
shell models. Additionally, the substantial effects on IR colors caused by the 
large amplitude pulsations of AGB stars, along with instances of thermal pulses 
(superwinds), may contribute to the observed deviations.

Figures~\ref{f5} and~\ref{f6} show CMDs using IR bands. We find that the CMDs, 
which utilize the latest distance and extinction data from Gaia DR3 for a 
substantial number of carbon stars, are very useful to distinguish CAGB stars 
from extrinsic carbon stars that are not in the AGB phase. 

A specific subset of well-known 152 CAGB stars are notably highlighted on all the 
2CDs and CMDs, encompassing 51 visual-CAGB\_IC objects and 101 IR-CAGB\_IC 
objects, for which detailed SEDs are compared with theoretical models (see 
Section~\ref{sec:modelsed}). Among the 152 CAGB stars, a simple single dust shell 
model incorporating amorphous carbon (AMC) and SiC dust effectively reproduces 
the observed SEDs for the majority of objects. However, certain objects require 
more intricate models, such as detached or double shells, or non-spherical dust 
envelopes (refer to Section~\ref{sec:cmodel}). The twelve objects necessitating 
more complex models, as indicated exclusively in Figure~\ref{f1}, consist of two 
visual-CAGB\_IC (detached shell) objects and ten CAGB\_IC (double shell) objects.

\subsection{The IRAS 2CD \label{sec:ircd}}

The upper panel of Figure~\ref{f1} illustrates an IRAS 2CD using IR[25]$-$IR[60] 
versus IR[12]$-$IR[25]. We observe that the basic theoretical model tracks can 
provide a rough explanation for the observed points. This 2CD has been 
extensively utilized since \citet{van der Veen1988} (note that the authors did 
not perform zero-magnitude calibrations for their 2CD), who divided it into eight 
regions representing different classes of celestial bodies. Building on 
theoretical dust shell models for AGB stars and post-AGB stars, \citet{suh2015} 
presented potential evolutionary tracks from AGB stars to post-AGB stars and 
ultimately to planetary nebulae on this 2CD. In our work, we categorize this 2CD 
into four regions (from R1 to R4) to enhance the classification of carbon stars. 

On the IRAS 2CD, CAGB stars trace a 'C'-shaped curve, with visual CAGB stars and 
extrinsic carbon stars primarily located in the lower left region (R1) and 
extending to the right side (R2). IR CAGB stars, or infrared carbon stars, 
situated on the right side (R2), exhibit thicker dust envelopes with larger dust 
optical depths (see Section~\ref{sec:intro} for the evolution of CAGB stars). 
Among them, 51 visual-CAGB\_IC objects are positioned in R1 and R4, while all 101 
IR-CAGB\_IC objects are in R2 (with only one object located in R4 requiring a 
complex model). 

A group of stars in the upper-left region (R4) comprises visual carbon stars, 
including the early phase of CAGB stars, exhibiting excessive flux at 60 $\mu$m 
due to remnants of an earlier phase when the stars were OAGB stars (see e.g., 
\citealt{ck1990}). Notably, the twelve objects requiring more complex models, 
such as detached or double shell models, are located in region R4 on the 2CD. 
This subset includes two visual-CAGB\_IC (detached shell) objects and ten 
CAGB\_IC (double shell) objects (refer to Section~\ref{sec:cmodel}). 

Furthermore, the region R4 hosts various subclasses of extrinsic carbon stars, 
primarily binary stars, and some CAGB stars with intricate dust envelopes, 
potentially arising from binary interactions or the presence of substantial 
planets. 

Most extrinsic carbon stars (Ba, CH, and RH stars) in regions R1 and R4 are not 
found in R2, while J-type extrinsic carbon stars that may exhibit CAGB-like 
nature are scattered across all regions (R1, R2, and R4). Especially, extrinsic 
carbon stars of Ba and CH types are notably concentrated in the far-left region. 

On the far right side (R3), where most post-AGB stars and planetary nebulae are 
distributed (e.g., \citealt{suh2015}), there are no carbon stars in our sample.

\subsection{IRAS and 2MASS 2CDs\label{sec:iras2mass}}

The lower panel of Figure~\ref{f1} displays an IRAS-2MASS 2CD using 
IR[12]$-$IR[60] versus K[2.2]$-$IR[12]. Galactic extinction is taken into account 
for the K[2.2]$-$IR[12] color, as explained in Section~\ref{sec:ext}. On this 
2CD, the differentiation between OAGB and CAGB stars becomes more distinct. 
Visual carbon stars occupy the lower-left region of this 2CD due to their 
relatively bluer K[2.2]$-$IR[12] colors compared to others. Specifically, 
extrinsic carbon stars of Ba and CH types are notably concentrated in the 
far-left region.

\subsection{WISE and 2MASS 2CDs\label{sec:wise}}

Figure~\ref{f2} shows 2CDs using W3[12]$-$W4[22] versus K[2.2]$-$W3[12]. Galactic 
extinction is taken into account for the K[2.2]$-$W3[12] color. The upper panel 
shows carbon stars with IRAS counterparts (CS\_IC) while the lower panel shows 
carbon stars without IRAS counterparts (CS\_NI). In general, the theoretical dust 
shell models for both OAGB and CAGB stars demonstrate a satisfactory ability to 
replicate the observed points on these IR 2CDs. Noticeable differences emerge 
between CS\_IC and CS\_NI objects. CS\_IC objects, being more evolved and 
brighter at the MIR bands, exhibit much redder K[2.2]$-$W3[12] colors. 

A distinctive group of CS\_NI objects deviates from other carbon stars in the 
upper-left region of the 2CD. These objects display bluer K[2.2]$-$W3[12] colors 
but redder W3[12]$-$W4[22] colors, and the reason for this phenomenon remains 
unclear. Given that many known extrinsic carbon stars (binary stars) are situated 
in this region, there could be a potential connection with the binary nature of 
extrinsic carbon stars, such as radiation from white dwarfs. 

Figure~\ref{f3} displays error bar plots for W3[12]$-$W4[22] versus 
K[2.2]$-$W3[12] for all carbon stars. In this diagram, we present averaged colors 
and their errors (standard deviations) for all sample stars across various 
subclasses. We find that CS\_NI objects consistently exhibit redder 
W3[12]$-$W4[22] colors for extrinsic carbon stars of Ba, CH, dC, and RH types. 

Out of the 4909 CAGB stars, 4132 (84 \%) can be classified as C-rich stars based 
on the Gaia DR3 spectra (See Section~\ref{sec:specdata}). However, when focusing 
on  CAGB stars with thick dust shells (K[2.2]$-$W3[12] $>$ 4), the percentage 
drops to 34 \%. Figure~\ref{f4} illustrates histograms of the K[2.2]$-$W3[12] 
colors for CAGB stars (CAGB\_IC objects) with thick dust shells. This affirms 
that the Gaia DR3 spectral data can be more effective for stars characterized by 
thinner dust shells.

\begin{figure*}
\centering
\smallplottwo{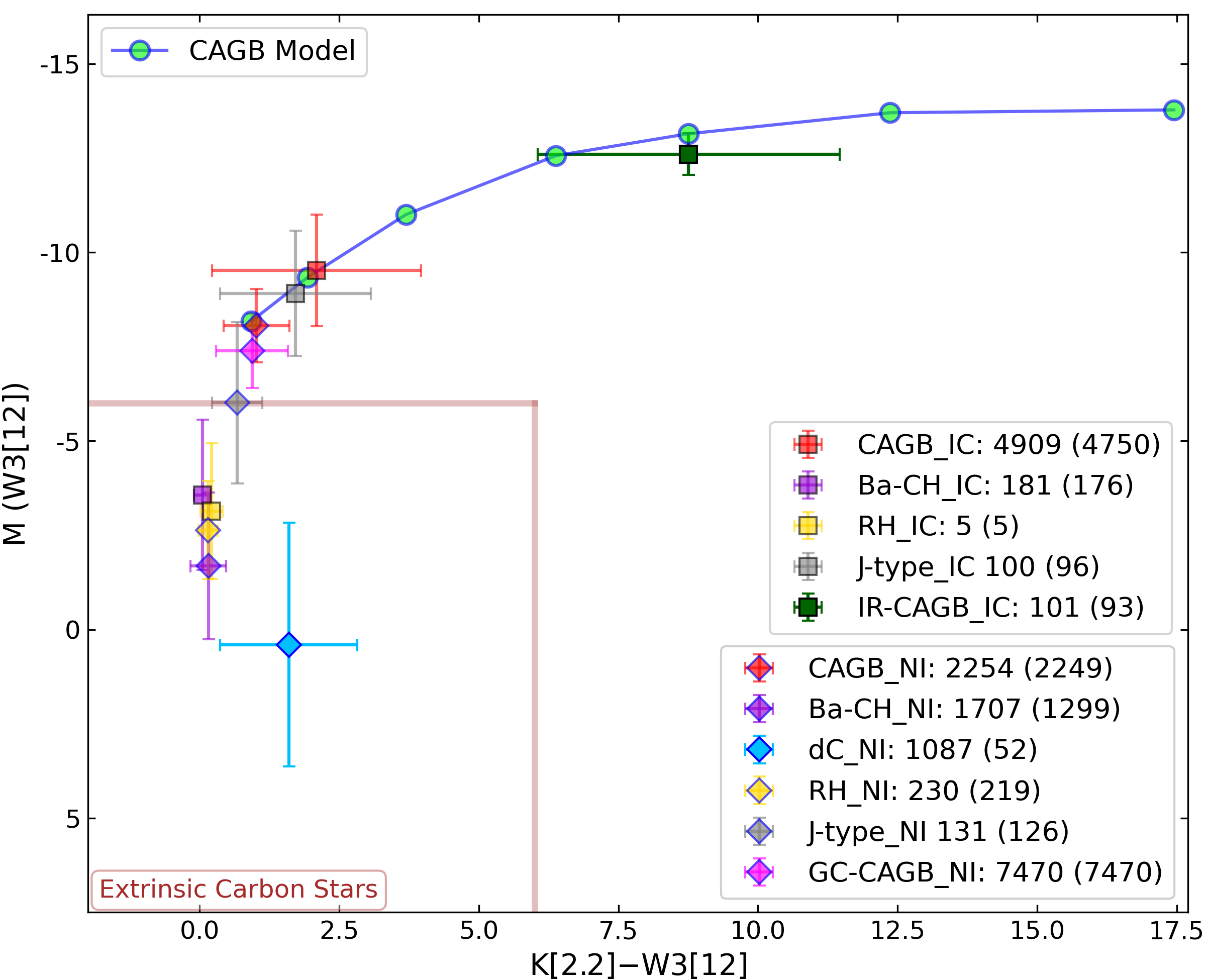}{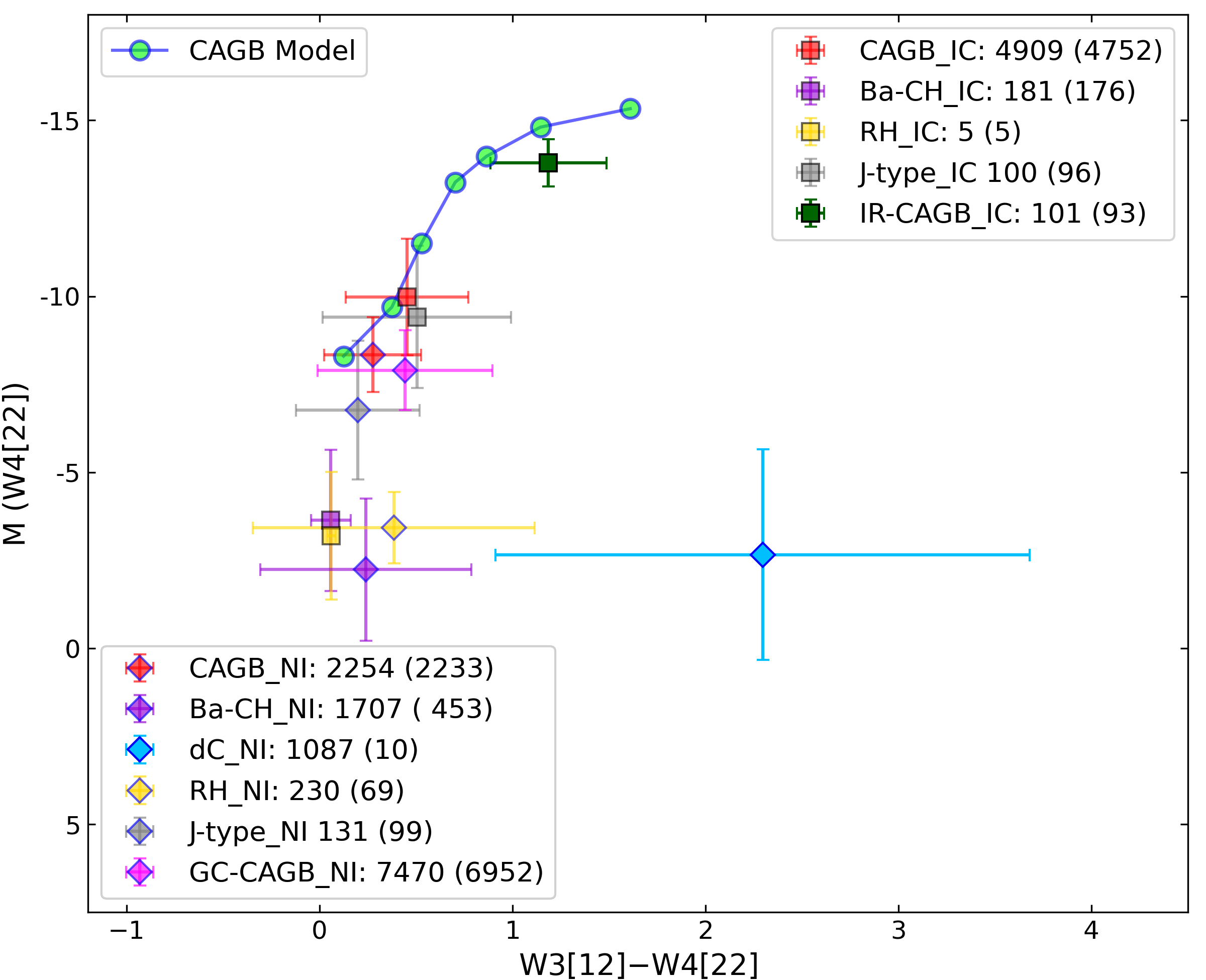}\caption{Error-bar plots of averaged colors and magnitudes 
for various subclasses of carbon stars.
For each subclass, the number of objects is shown.
The number in parenthesis denotes the number of the plotted objects with good quality observed data.
See Section~\ref{sec:class}.} \label{f7}
\end{figure*}

\begin{figure*}
\centering
\smallplottwo{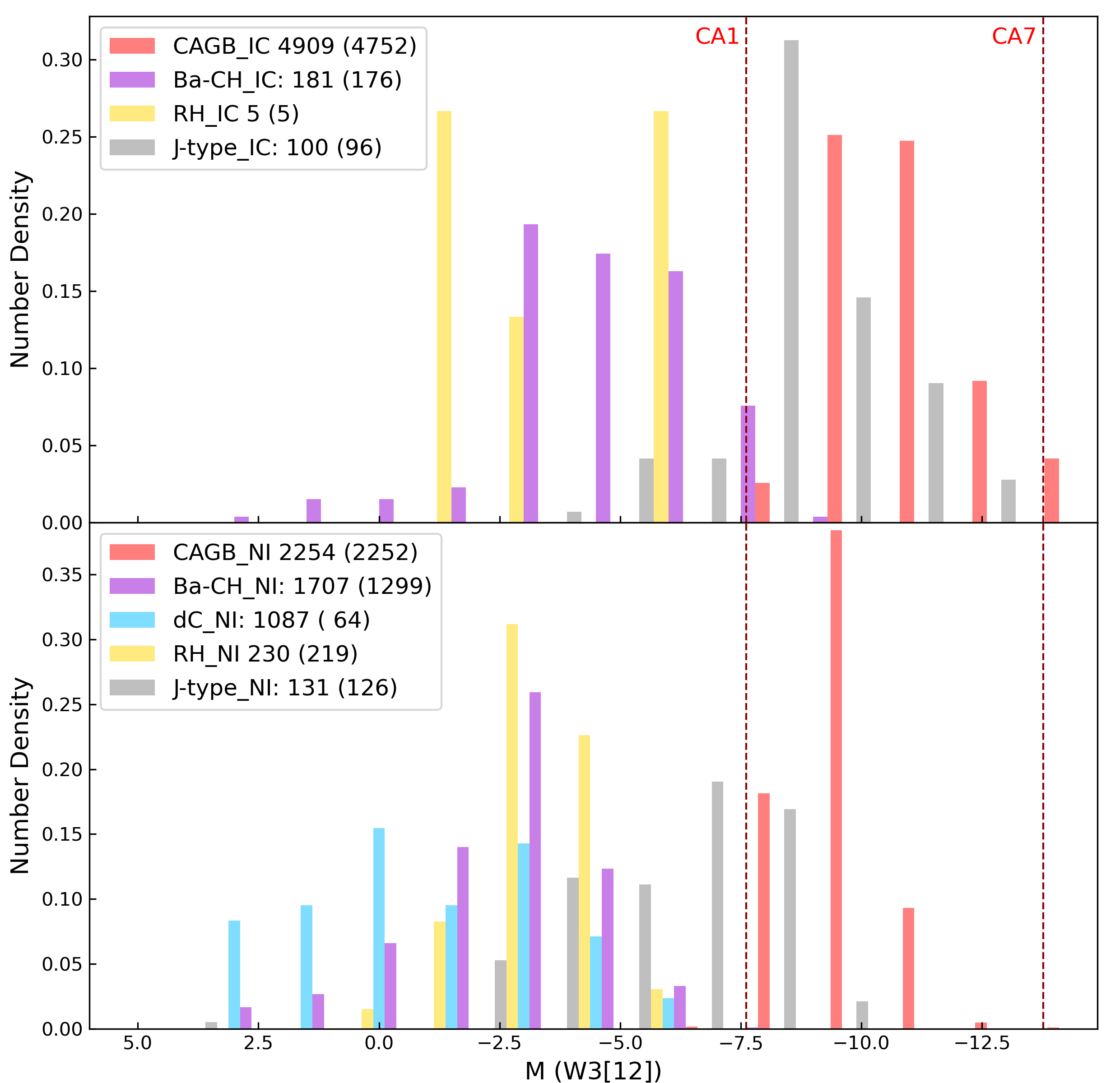}{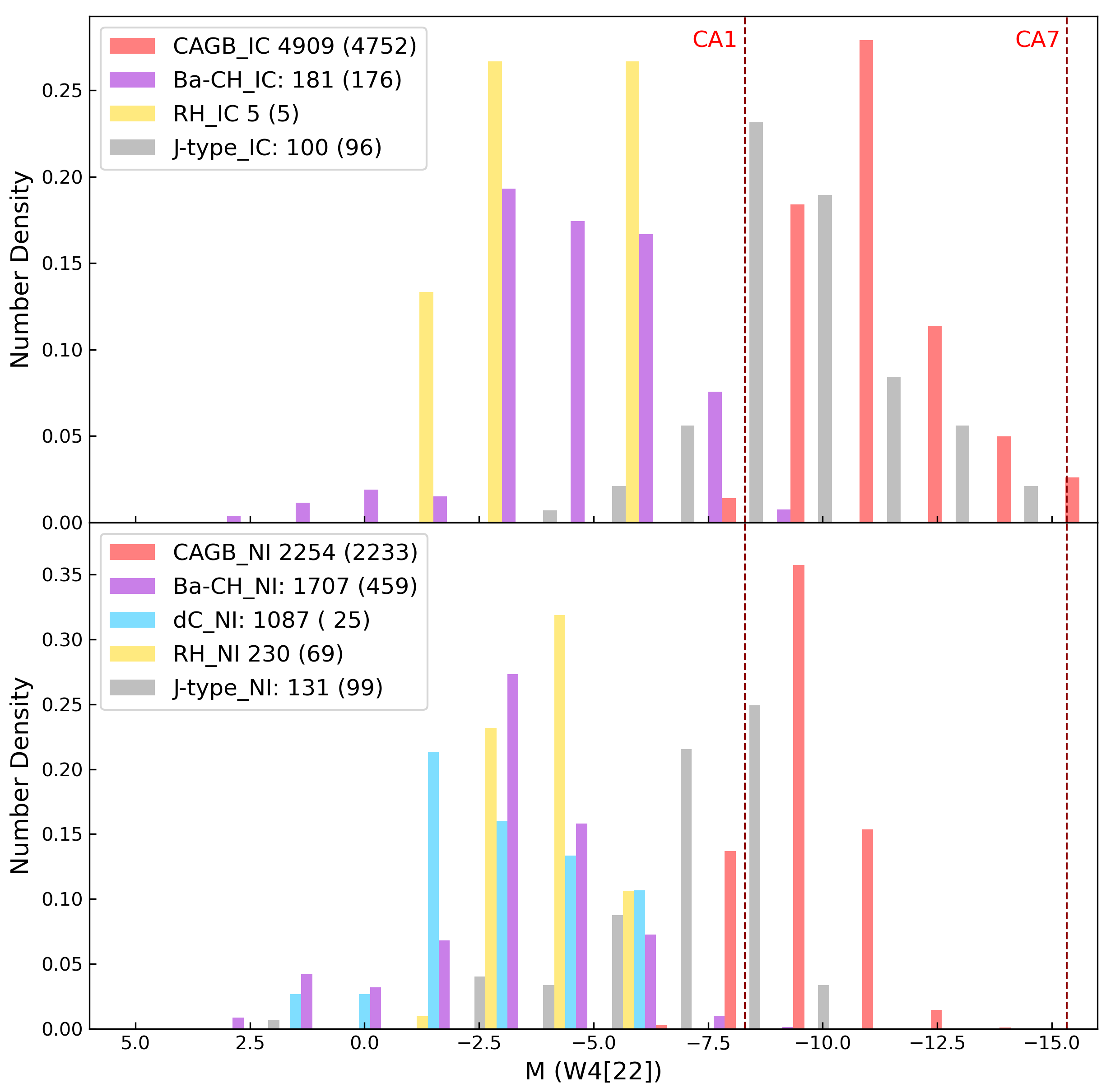}\caption{Histograms of observed absolute magnitudes for various 
subclasses of carbon stars. 
The vertical lines indicate theoretical magnitudes for CAGB models CA1 and CA7 (see Table~\ref{tab:tab4}).
See Section~\ref{sec:class}.} \label{f8}
\end{figure*}

\subsection{IR CMDs\label{sec:CMD}}

Utilizing the recently acquired distance (\citealt{bailer-jones2021}) and 
extinction data from \citealt{lallement2022} derived from the Gaia DR3 data, 
facilitates the determination of absolute magnitudes and unreddened colors across 
a broad spectrum. Leveraging this information for a significant number of carbon 
stars in our Galaxy, we present various CMDs that prove instrumental in 
distinguishing between CAGB (or intrinsic carbon) stars and extrinsic carbon 
stars. Galactic extinction is taken into account when determining the 
K[2.2]$-$IR[12] and K[2.2]$-$IR[12] colors, as well as the absolute magnitude at 
the K[2.2] band. 

Figure~\ref{f5} displays CMDs using K[2.2]$-$IR[12] color and absolute magnitudes 
at K[2.2] and IR[12] bands for various subclasses of carbon stars with IRAS 
counterparts (CS\_IC objects). Since objects with thicker dust shells exhibit 
redder K[2.2]$-$IR[12] colors, most CAGB stars are characterized by colors that 
are redder than those of extrinsic carbon stars. The distinction between CAGB 
stars and extrinsic carbon stars is apparent on these IR CMDs. 

Extrinsic carbon stars of subclasses Ba, CH, and RH are notably distinguishable 
from CAGB stars on these CMDs, displaying dimmer absolute magnitudes at the 
IR[12] and K[2.2] bands and bluer K[2.2]-W3[12] colors. However, J-type stars, 
representing another subclass of extrinsic carbon stars, demonstrate properties 
shared with both CAGB stars and other extrinsic carbon stars. Remarkably, we find 
that RCB stars exhibit very similar properties to CAGB stars, making it 
challenging to distinguish them on these CMDs. 

The left panel of Figure~\ref{f5} displays a CMD using K[2.2]$-$IR[12] color and 
magnitudes at K[2.2]. CAGB stars exhibit brighter absolute magnitudes at K[2.2] 
only in bluer K[2.2]$-$W3[12] color ranges, because more evolved CAGB stars with 
redder K[2.2]$-$IR[12] colors have thicker circumstellar dust shells. These dense 
dust shells absorb radiation at shorter wavelengths (e.g., K[2.2]) and emit at 
longer wavelengths. 

The right panel of Figure~\ref{f5} illustrates a CMD using K[2.2]$-$IR[12] color 
and magnitudes at IR[12]. CAGB stars exhibit brighter absolute magnitudes at 
IR[12] across all K[2.2]-W3[12] color ranges. We find that the absolute 
magnitudes at the IR[12] band are brighter than -6 for all CAGB stars. 

Figure~\ref{f6} presents CMDs utilizing K[2.2]$-$W3[12] and W3[12]$-$W4[22] 
colors, along with absolute magnitudes at K[2.2], W3[12], and W4[22] bands for 
various subclasses of carbon stars (CS\_IC and CS\_NI) objects. The left panels 
display CMDs for CS\_IC objects, exhibiting similar characteristics as depicted 
in Figure~\ref{f5}, while the right panels feature CS\_NI objects. 

In the upper four panels of Figure~\ref{f6}, CMDs are presented using the 
K[2.2]$-$W3[12] color. Given that objects with thicker dust shells manifest 
redder K[2.2]$-$W3[12] colors, the majority of CAGB stars exhibit colors that are 
redder than those of extrinsic carbon stars. More evolved CAGB stars exhibit 
redder K[2.2]$-$W3[12] colors but dimmer absolute magnitudes at K[2.2] due to 
their thicker circumstellar dust shells. 

Notably, CAGB\_IC objects are distributed across a wide range of K[2.2]$-$W3[12] 
colors, whereas CAGB\_NI objects tend to concentrate in bluer K[2.2]$-$W3[12] 
colors. This distinction arises from the general trend of CAGB\_IC objects being 
brighter or more evolved compared to CAGB\_NI objects. Conversely, extrinsic 
carbon stars, whether CS\_IC or CS\_NI objects, tend to cluster in bluer 
K[2.2]$-$W3[12] colors and dimmer absolute magnitudes. 

For all CAGB stars, we consistently observe absolute magnitudes at the W3[12] 
band that are brighter than -6, mirroring our findings in the case of the IR[12] 
band.

In the lower two panels of Figure~\ref{f6}, CMDs are shown using the 
W3[12]$-$W4[22] color and absolute magnitudes at W4[22]. For nearly all CAGB 
stars, the absolute magnitudes at the W4[22] band are brighter than -5. CS\_IC 
objects are distributed across a limited range W3[12]$-$W4[22] colors, while 
CS\_NI objects are distributed across a much wider range into the bluer and 
redder spectrum. The reason for this phenomenon remains unclear (see 
Section~\ref{sec:wise}).

\subsection{Different subclasses of carbon stars\label{sec:class}}

The CMDs (see Figures~\ref{f5} and \ref{f6}) prove valuable in distinguishing 
CAGB (intrinsic carbon) stars from extrinsic carbon stars, particularly when 
comparing magnitudes at MIR bands (IR[12], W3[12], and W3[22]). We find that the 
absolute magnitudes at the IR[12] or W3[12] band are brighter than -6 for all 
CAGB stars. In contrast, extrinsic carbon stars of Ba, CH, dC, and RH types 
consistently exhibit dimmer magnitudes than CAGB stars at the MIR bands. J-type 
stars, constituting another subclass of extrinsic carbon stars, are distributed 
across a broader range of magnitudes demonstrating properties that are shared 
with both CAGB stars and other extrinsic carbon stars.

Figure~\ref{f7} displays error bar plots for CMDs using W3[12] versus 
K[2.2]$-$W3[12] and W4[22] versus W3[12]$-$W4[22]. These diagrams present 
averaged colors and magnitudes and their errors (standard deviations) for all 
sample stars across various subclasses.

Figure~\ref{f8} presents histograms depicting the observed absolute magnitudes at 
W3[12] and W3[22] bands for different subclasses of carbon stars. The plots 
clearly delineate the distinction between CAGB stars and extrinsic carbon stars, 
with CAGB stars consistently brighter than -6 mag at the W3[12] band (-5 mag at 
the W4[22] band). Conversely, most Ba, CH, dC, and RH type extrinsic carbon stars 
exhibit dimmer magnitudes. Some J-type carbon stars are as faint as other 
extrinsic carbon stars, while others are as bright as CAGB stars.

In contrast to typical extrinsic carbon stars, certain J-type carbon stars, such 
as IRAS 20166+3717 and IRAS 21566+5309, are recognized as Mira variables (AGB 
stars). This observation may be attributed to their binary nature, wherein one 
star is an AGB (not a RGB or a red dwarf) star, and the mass transfer process 
enhances the $^{13}$C content.

\subsection{Carbon stars from Gaia DR3 spectra\label{sec:gaiacarbon}}

As discussed in Section~\ref{sec:specdata}, there are 15,365 candidate objects 
for carbon stars identified from Gaia DR3 spectra (denoted by GC(all)\_NI; refer 
to Table~\ref{tab:tab2}) without any other evidence. We selected candidate 
objects for new CAGB stars when their absolute magnitudes at the W3[12] band are 
brighter than -6. There are 7470 candidate objects for new CAGB stars solely 
identified from Gaia DR3 spectra and absolute magnitudes (denoted by GC-CAGB\_NI; 
see Table~\ref{tab:tab2}). 

When selecting the 7470 GC-CAGB\_NI objects, special considerations were given to 
Mira variables (597 Miras are identified through AAVSO). Among these 597 Miras, 
the absolute magnitudes at the W3[12] band were dimmer than -6.5 for 43 objects 
according to the Gaia DR3 distances. Since Miras can be considered AGB stars, we 
calculated theoretical distances assuming that the absolute magnitudes at the 
W3[12] band are -6. We found that the Gaia DR3 distances were too small compared 
to the theoretical distances for the 43 objects. By using the mean distances from 
Gaia DR3 and theoretical distances, the absolute magnitudes for the 43 objects 
became brighter than -6. Therefore, we have used the mean distances for the 43 
Mira variables.

\begin{table}
\caption{Models for typical AGB stars\label{tab:tab4}}
\centering
\begin{tabular}{llllll}
\hline \hline
Model &Class &Dust$^1$ &$\tau_{10}$ & $T_*$ (K) & $L_* (10^{3} L_{\odot}$) \\
\hline
LM1  & OAGB  &silicate  &0.001 & 3000  & 1 \\
LM2  & OAGB  &silicate  &0.01 & 3000  & 2 \\
LM3  & OAGB  &silicate  &0.1  & 3000  & 3 \\
\hline
CA1  & CAGB  &AMC       &0.0001  & 3000  & 5 \\
CA2  & CAGB  &AMC       &0.01  & 2700  & 5 \\
CA3  & CAGB  &AMC       &0.1  & 2500  & 5 \\
CA4  & CAGB  &AMC       &0.5  & 2500  & 7 \\
CA5  & CAGB  &AMC       &1    & 2200  & 8 \\
CA6  & CAGB  &AMC       &2    & 2000  & 10 \\
CA7  & CAGB  &AMC       &4    & 2000  & 10 \\
\hline
HM4  & OAGB  &silicate  &7    & 2000  & 10 \\
HM5  & OAGB  &silicate  &15   & 2000  & 20 \\
HM6  & OAGB  &silicate  &30   & 2000  & 20 \\
\hline
\end{tabular}
\begin{flushleft}
\scriptsize
$^1$See Section~\ref{sec:agbmodels} for details. For all models, $T_c$ = 1000 K.
\end{flushleft}
\end{table}

\section{Theoretical Dust Shell Models\label{sec:agbmodels}}

On all of the 2CDs and CMDs in Figures~\ref{f1} through \ref{f8}, theoretical 
models for AGB stars are plotted to be compared with the observations. To compute 
theoretical model SEDs for CAGB stars, we employ radiative transfer models for 
spherically symmetric dust shells around central stars. The radiative transfer 
code RADMC-3D 
(\url{http://www.ita.uni-heidelberg.de/~dullemond/software/radmc-3d/}), following 
the same methodologies as utilized by \citet{sk2013a}, \citet{suh2015}, and 
\citet{suh2020}, is employed. Refer to \citet{suh2020} for details about the 
theoretical models and their limitations. 

Table~\ref{tab:tab4} details the parameters for seven models representing typical 
CAGB stars and six models representing typical OAGB stars. We assume a continuous 
power-law ($\rho \propto r^{-2}$) dust density distribution with a dust formation 
temperature ($T_c$) of 1000 K. The inner radius of the dust shell is determined 
by $T_c$, and the outer radius is set at the point where the dust temperature 
reaches 30 K. Spherical dust grains with a uniform radius of 0.1 $\mu$m are 
assumed. The dust optical depth ($\tau_{10}$) is taken at 10 $\mu$m as the 
fiducial wavelength. We assume a stellar blackbody temperature ($T_*$) in the 
range of 2000-3000 K and a stellar luminosity ($L_*$) in the range of 1x$10^{3}$ 
- 2x$10^{4}$ $L_{\odot}$. 

For OAGB stars, we use optical constants of warm and cold silicate dust from 
\citet{suh1999}. We use warm silicate for OAGB stars with thin dust shells (3 
models with $\tau_{10}$ $\leq 3$) and cold silicate for OAGB stars with thick 
dust shells (3 models $\tau_{10}$ $> 3$). 

For CAGB stars, the optical constants of amorphous carbon (AMC) and SiC dust 
grains are sourced from \citet{suh2000} and \citet{pegouri1988}, respectively. 
Additionally, for MgS dust, we utilize the optical constants of 
Mg$_{0.9}$Fe$_{0.1}$S, a composition close to pure MgS, as provided by 
\citet{begemann1994}. We adopt three distinct dust opacity models: a simple 
mixture of AMC and Mg$_{0.9}$Fe$_{0.1}$S (30\% by mass), a simple mixture of AMC 
and SiC (30\% by mass), and pure AMC.

\section{CAGB stars in our Galaxy\label{sec:cagb}}

We have selected CAGB stars from the sample of carbon stars spanning various 
subclasses (see Section~\ref{sec:irpro}). In this section, we compare observed 
SEDs of the CAGB stars with theoretical models (see Section~\ref{sec:agbmodels}), 
exploring the properties of CAGB stars in depth.

\begin{figure*}
\centering
\smallploteight{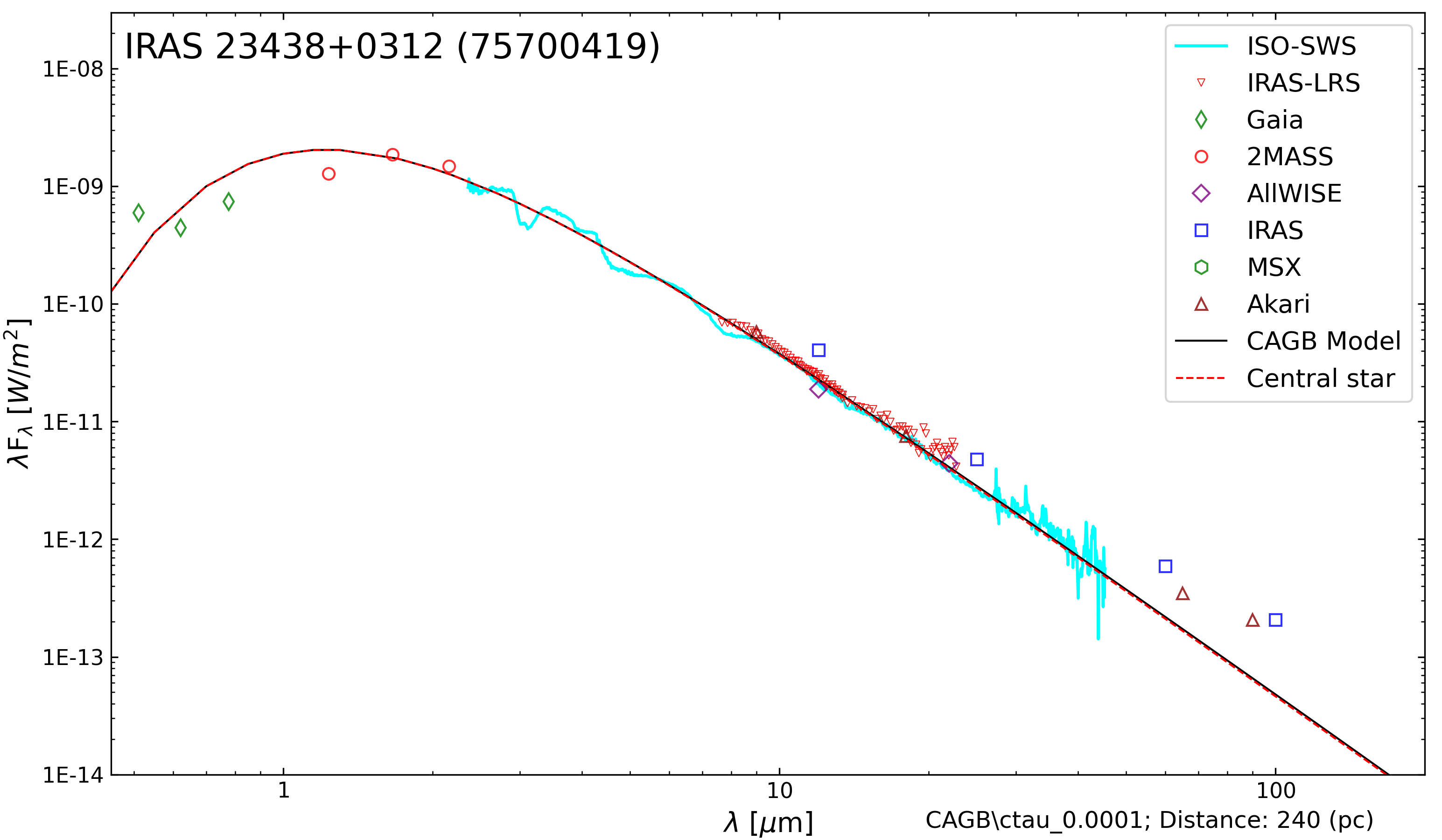}{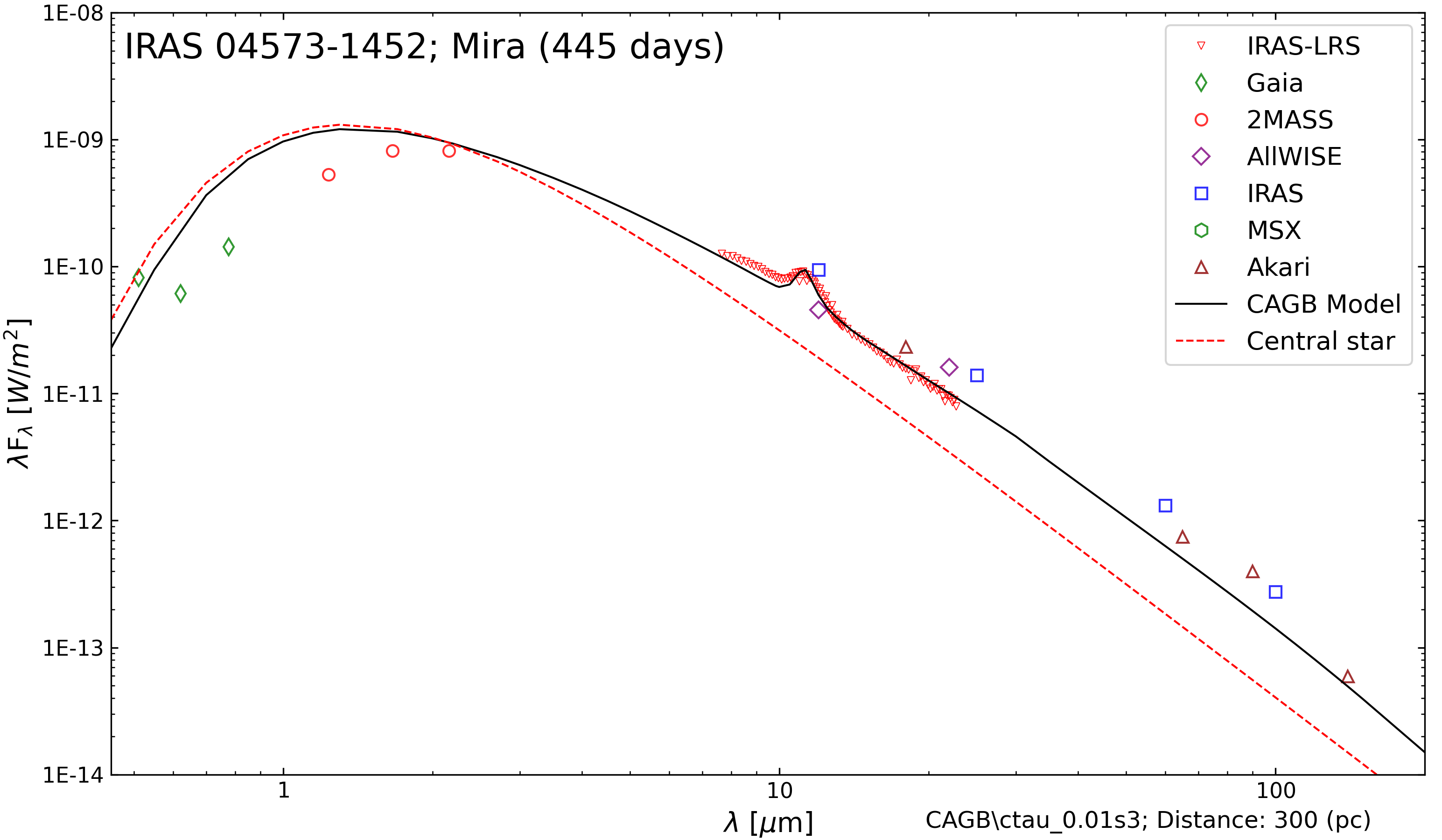}{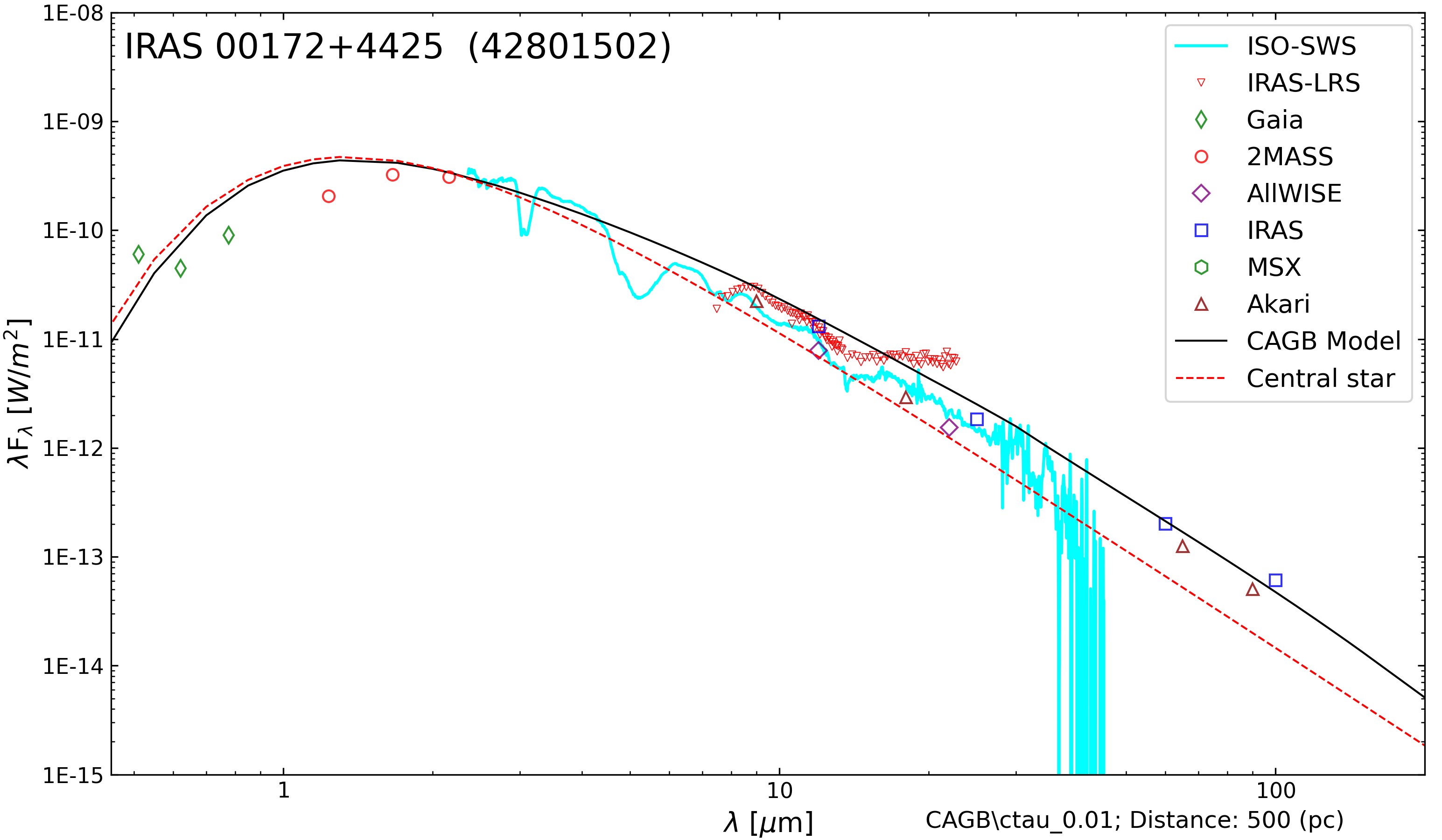}{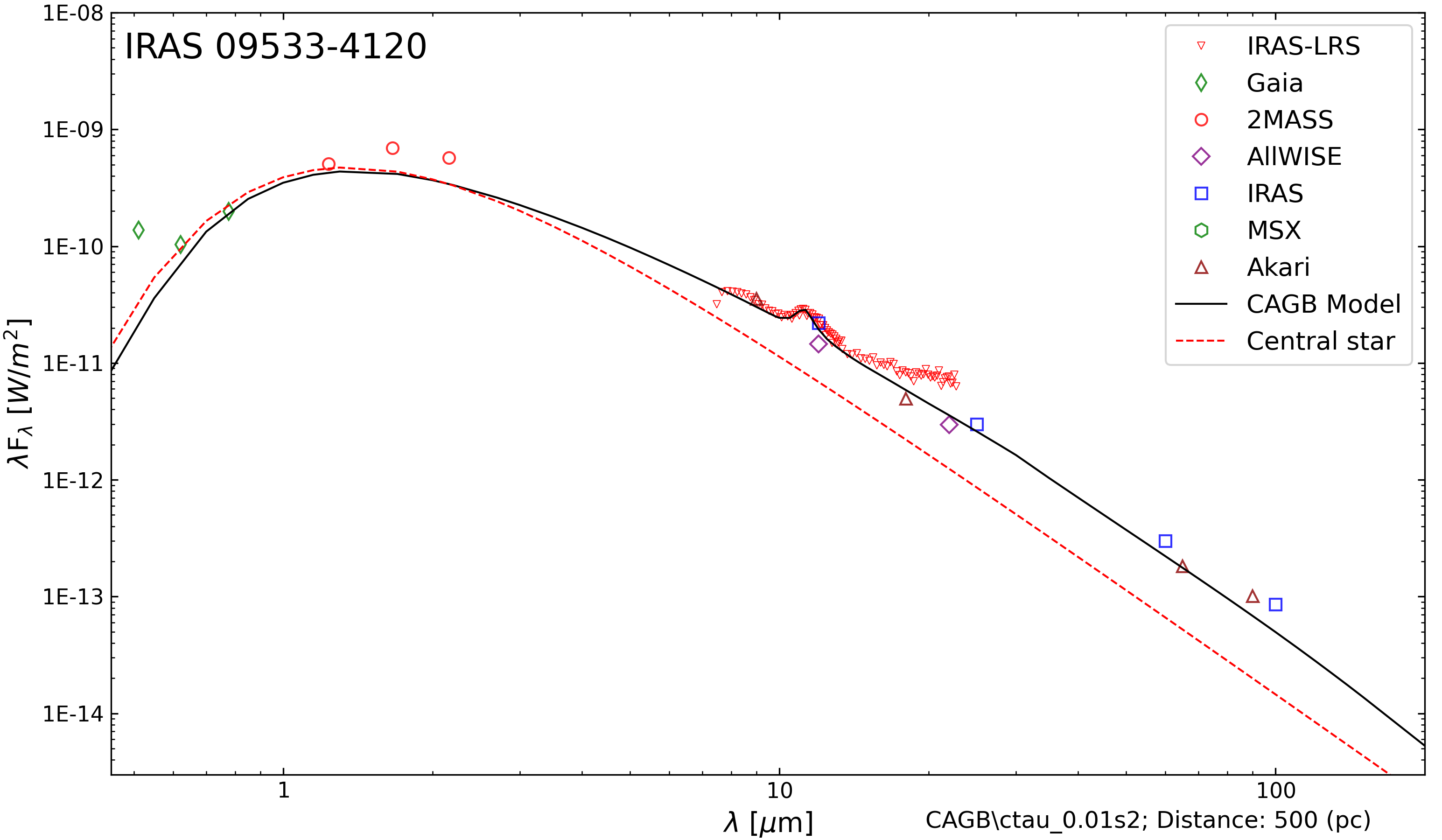}{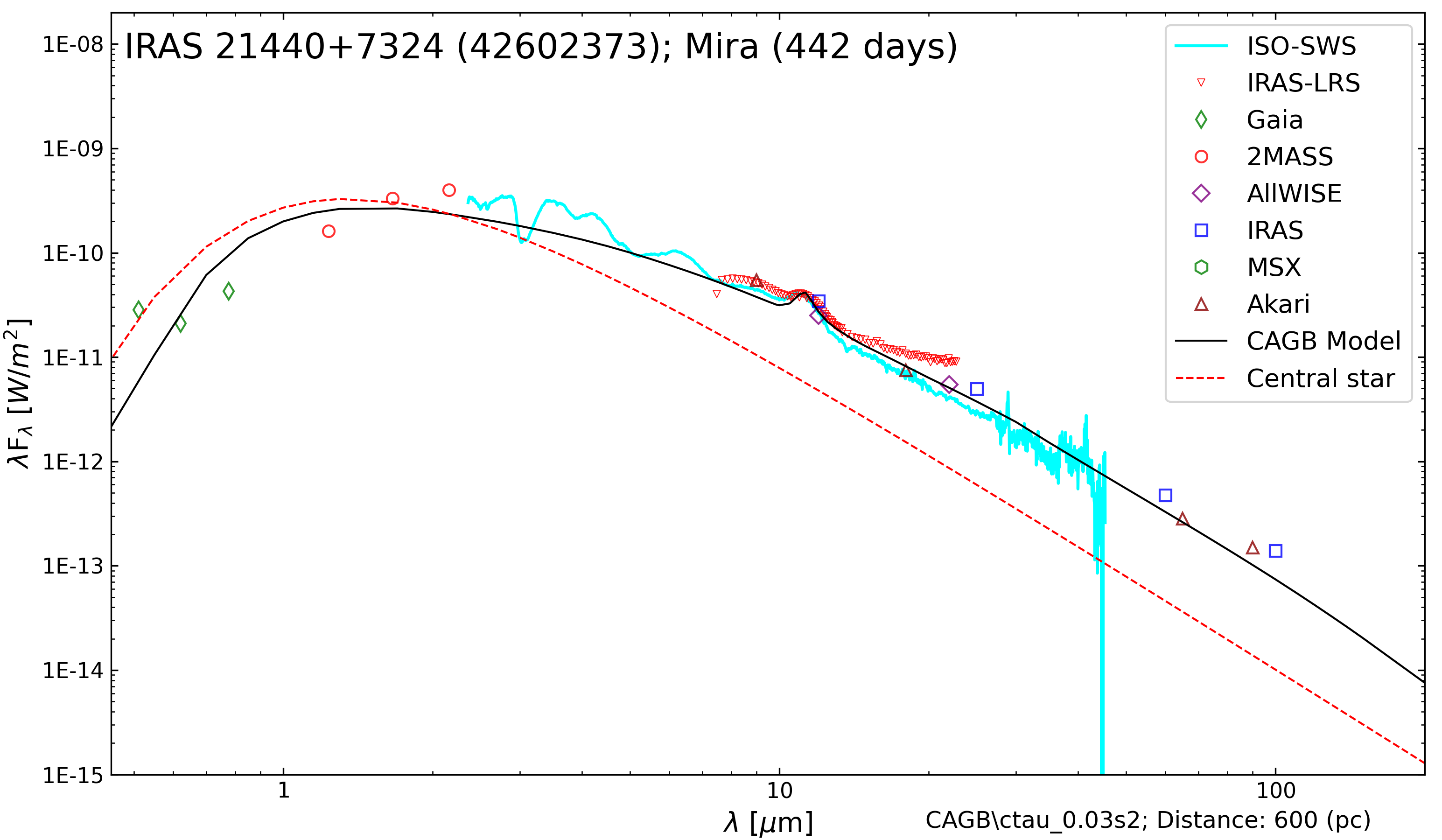}{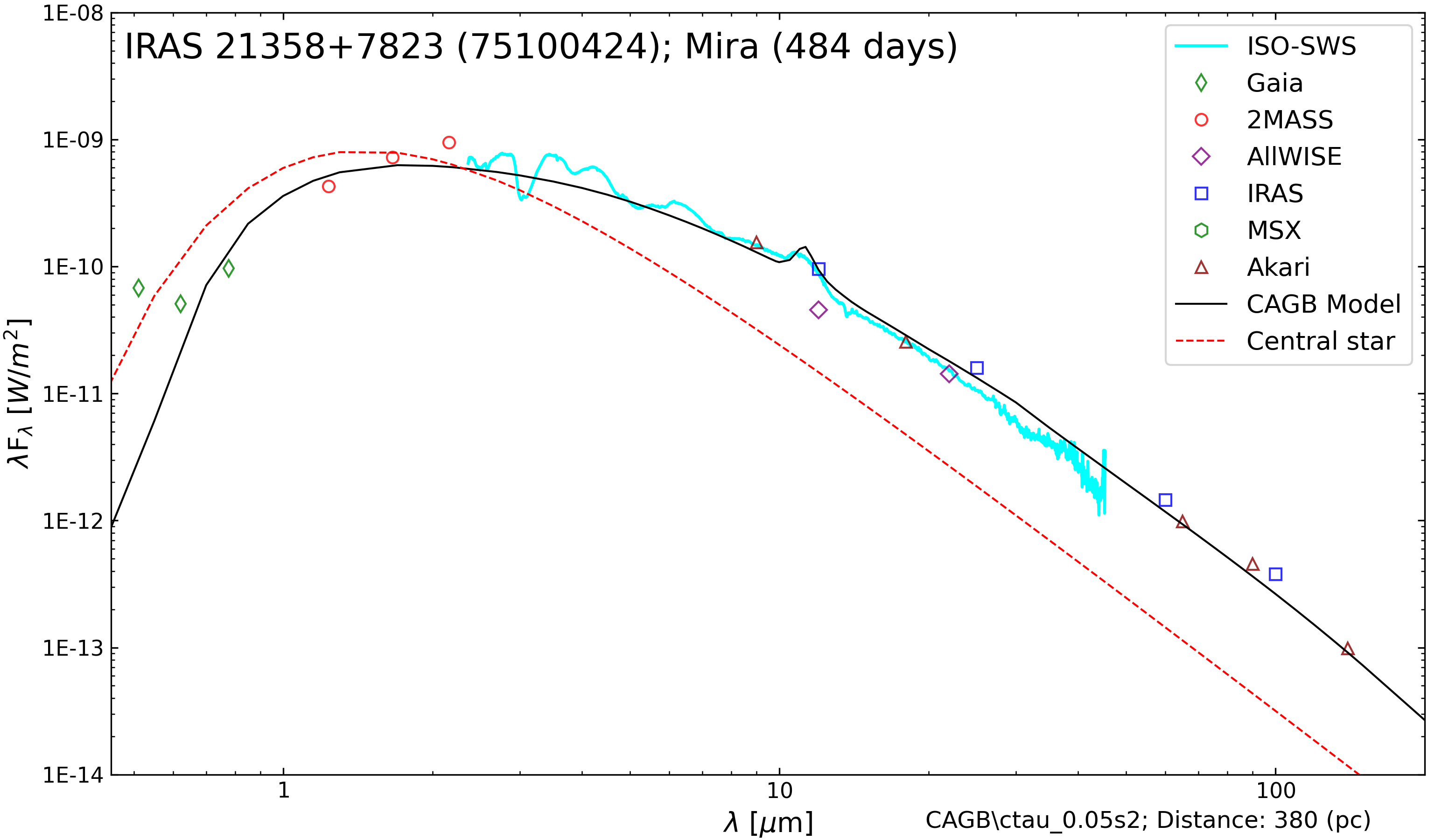}{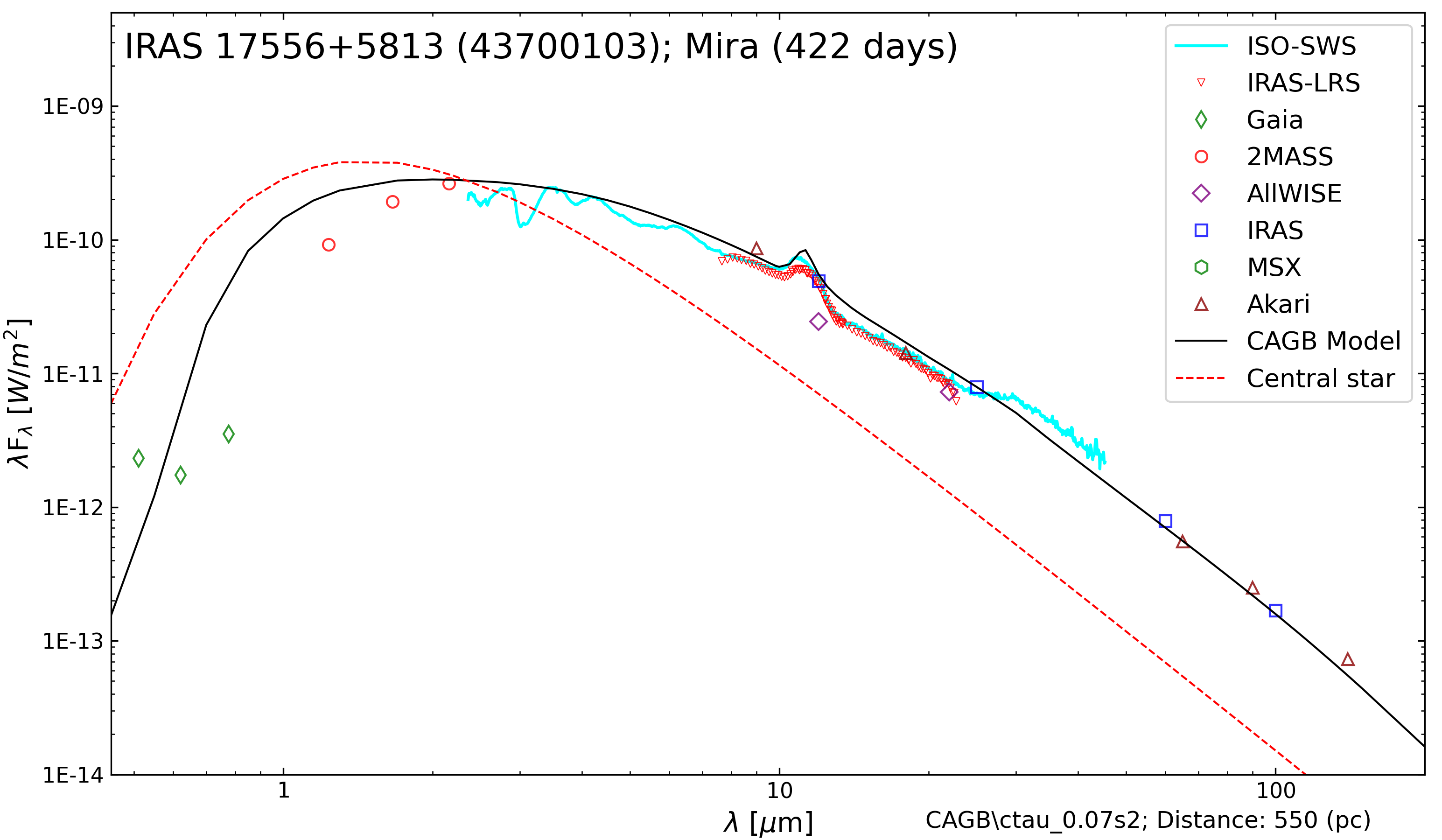}{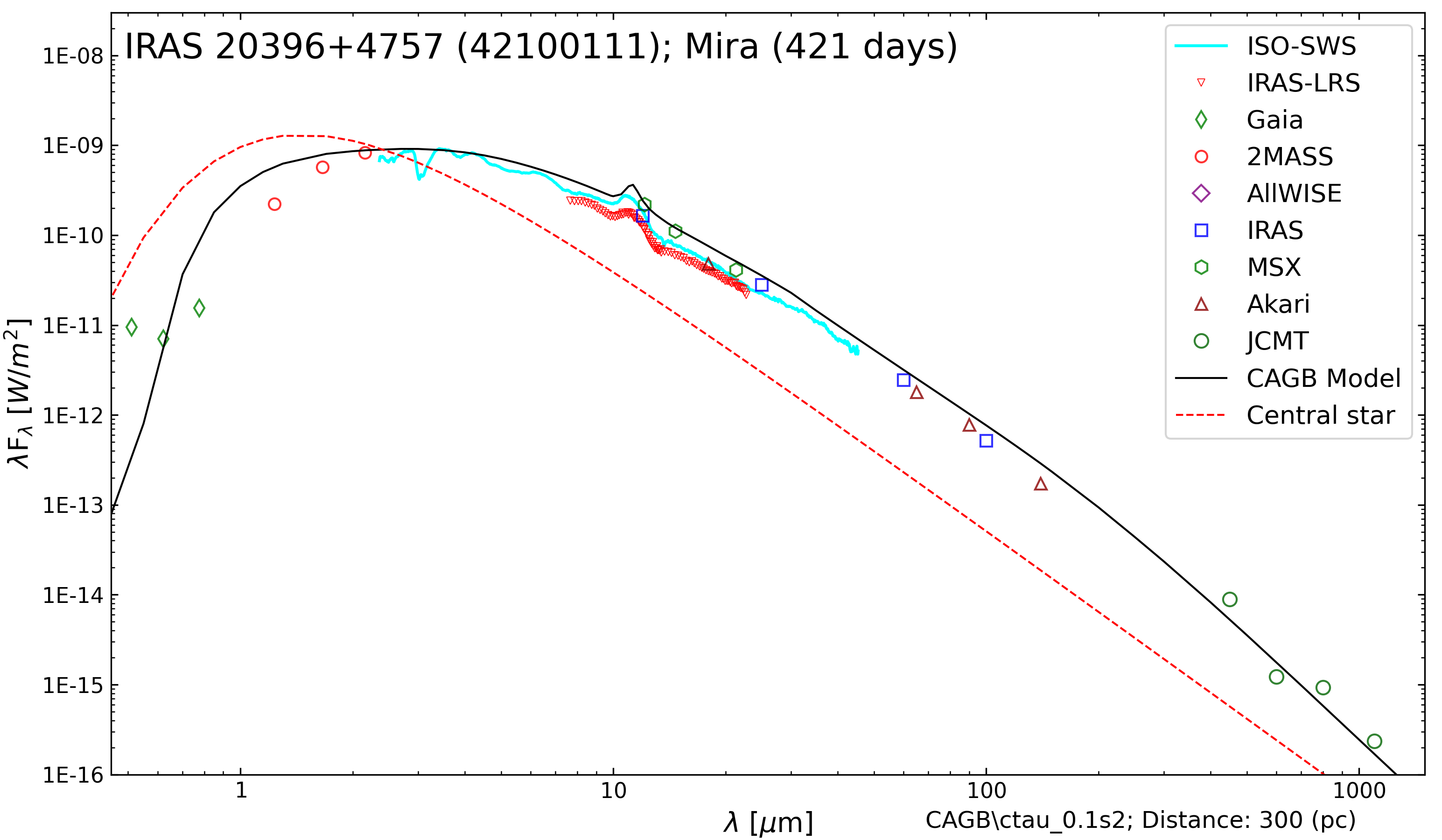}
\caption{Observed SEDs compared with simple dust shell models:
eight visual CAGB stars (visual-CAGB\_IC).
See Section~\ref{sec:modelsed}.} \label{f9}
\end{figure*}

\begin{figure*}
\centering
\smallploteight{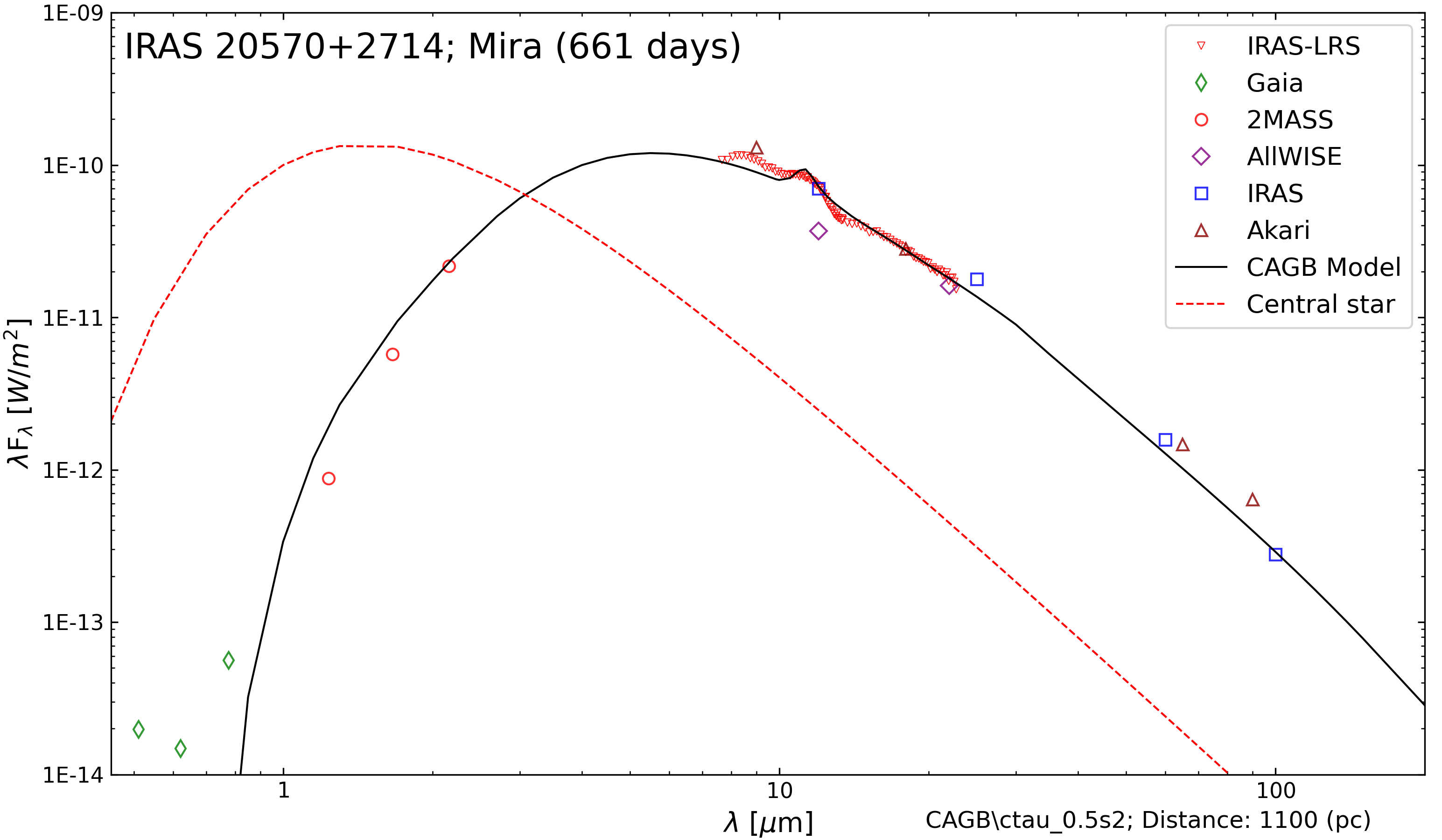}{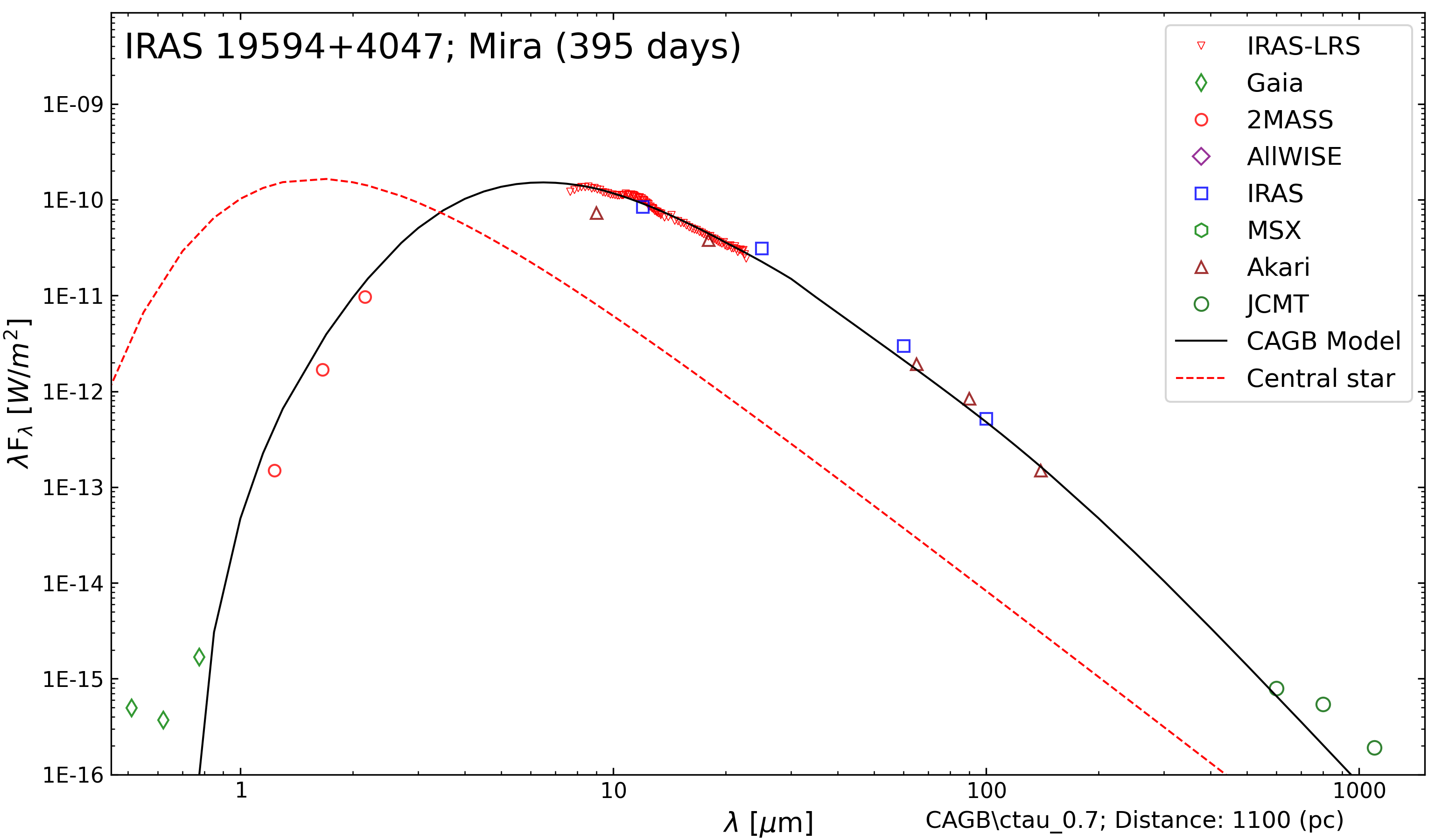}{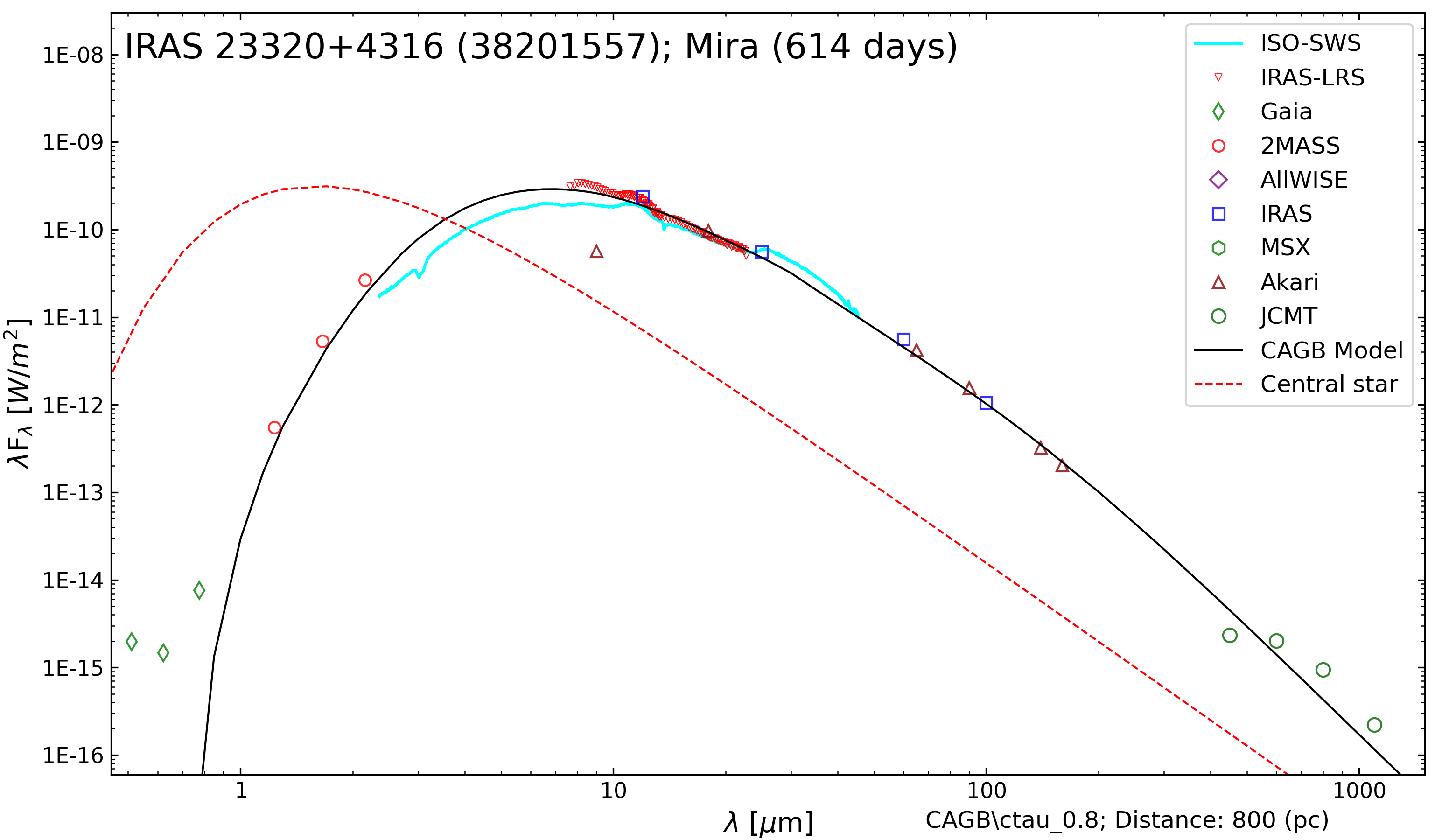}{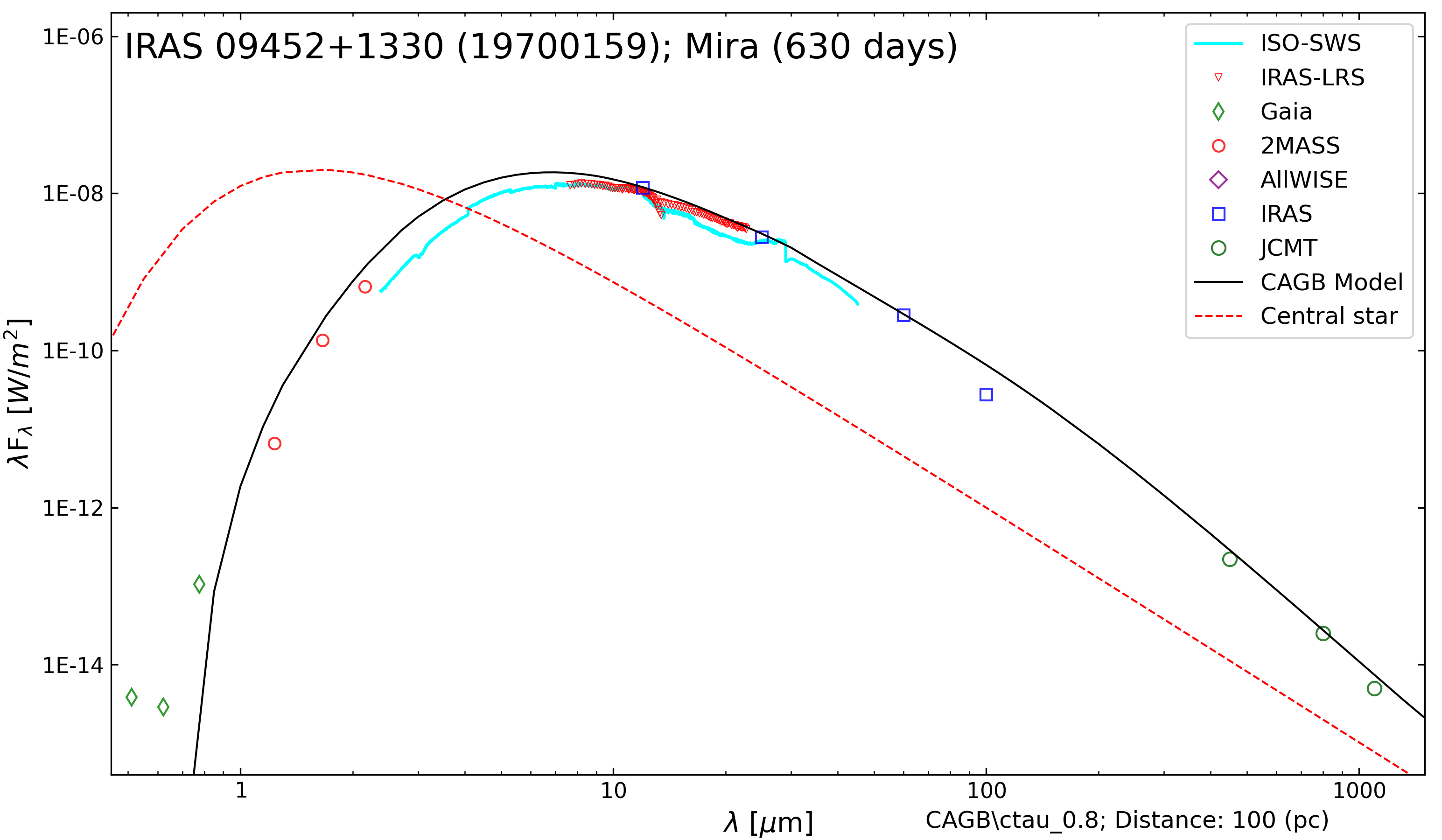}{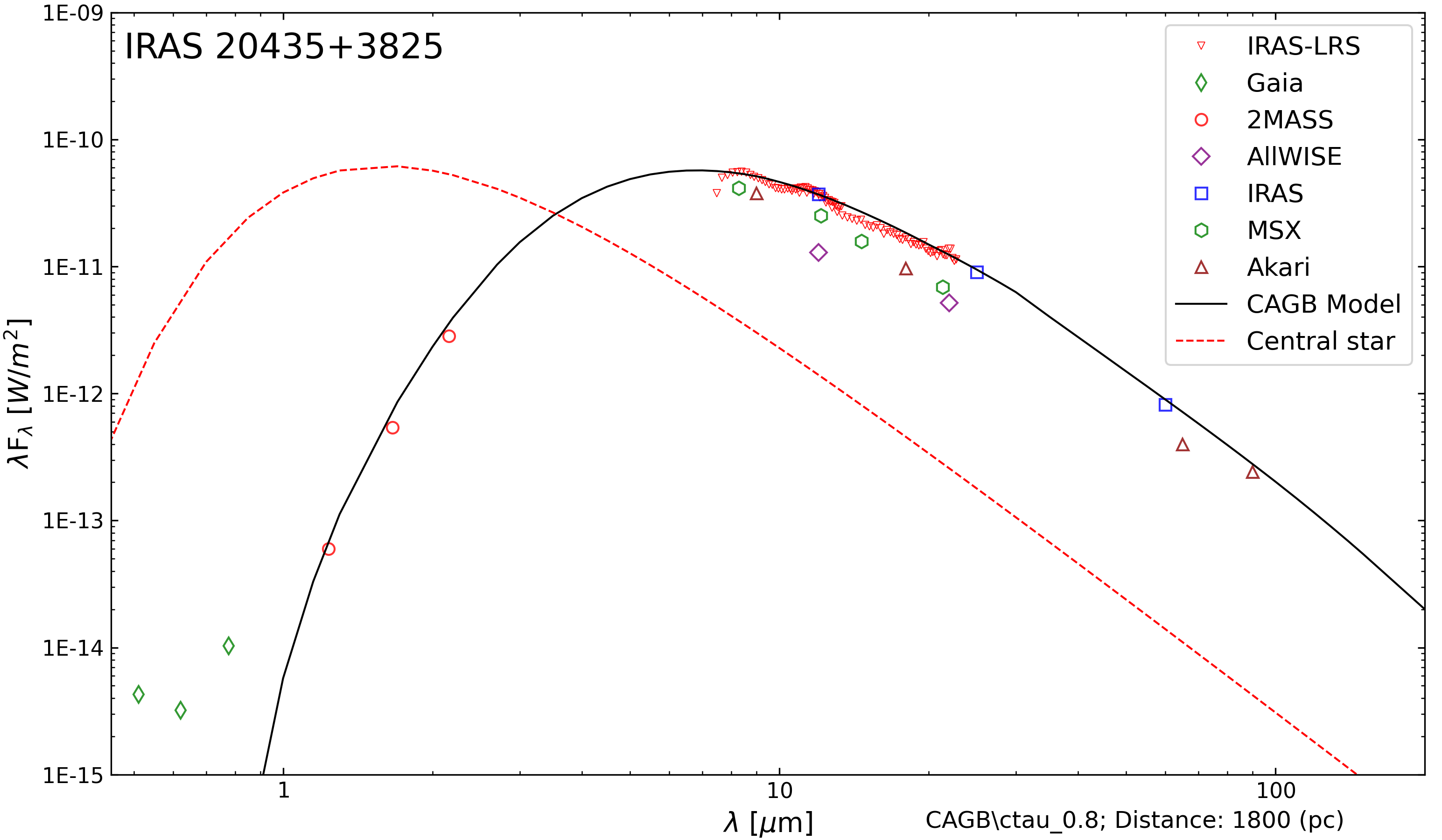}{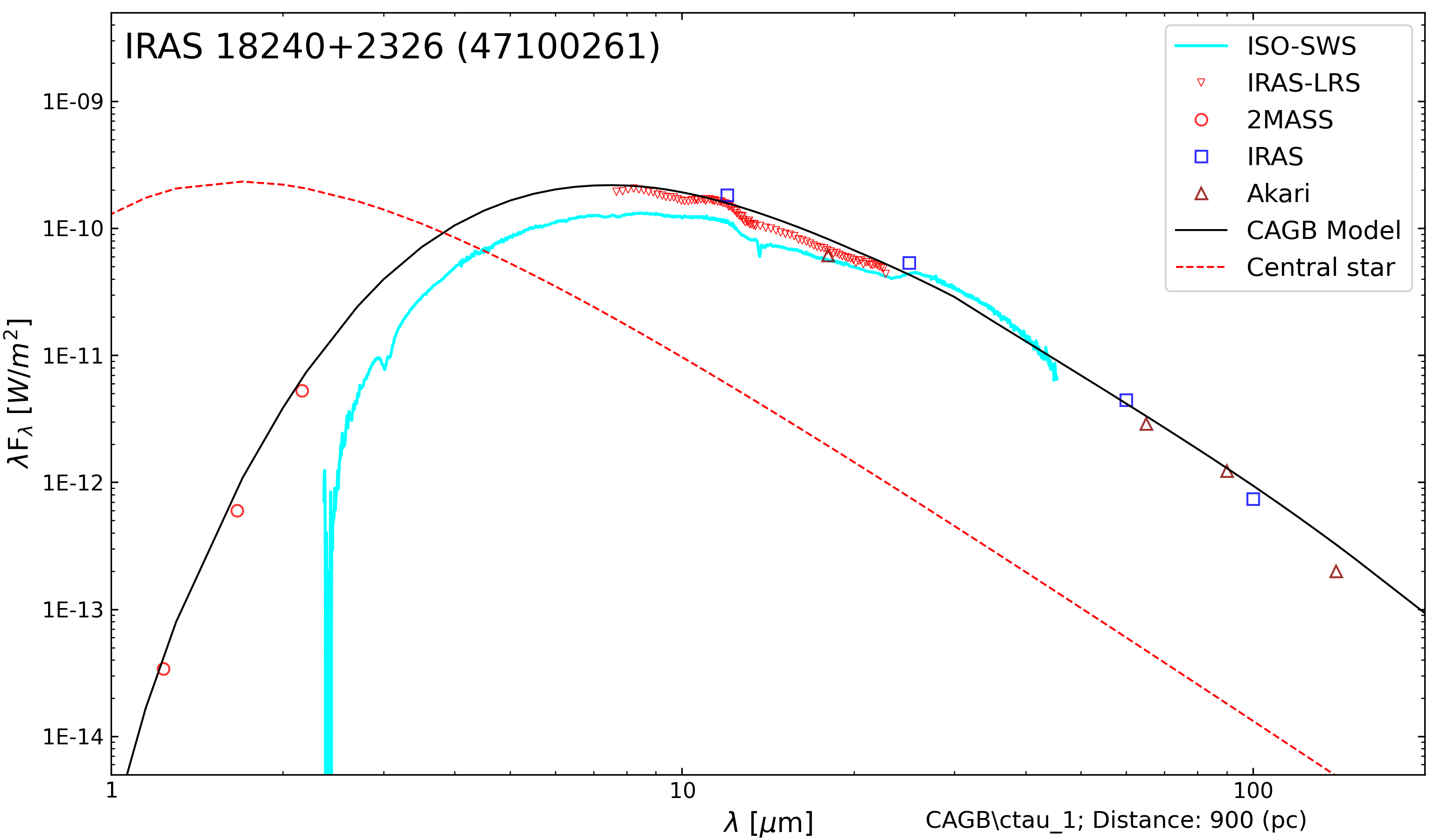}{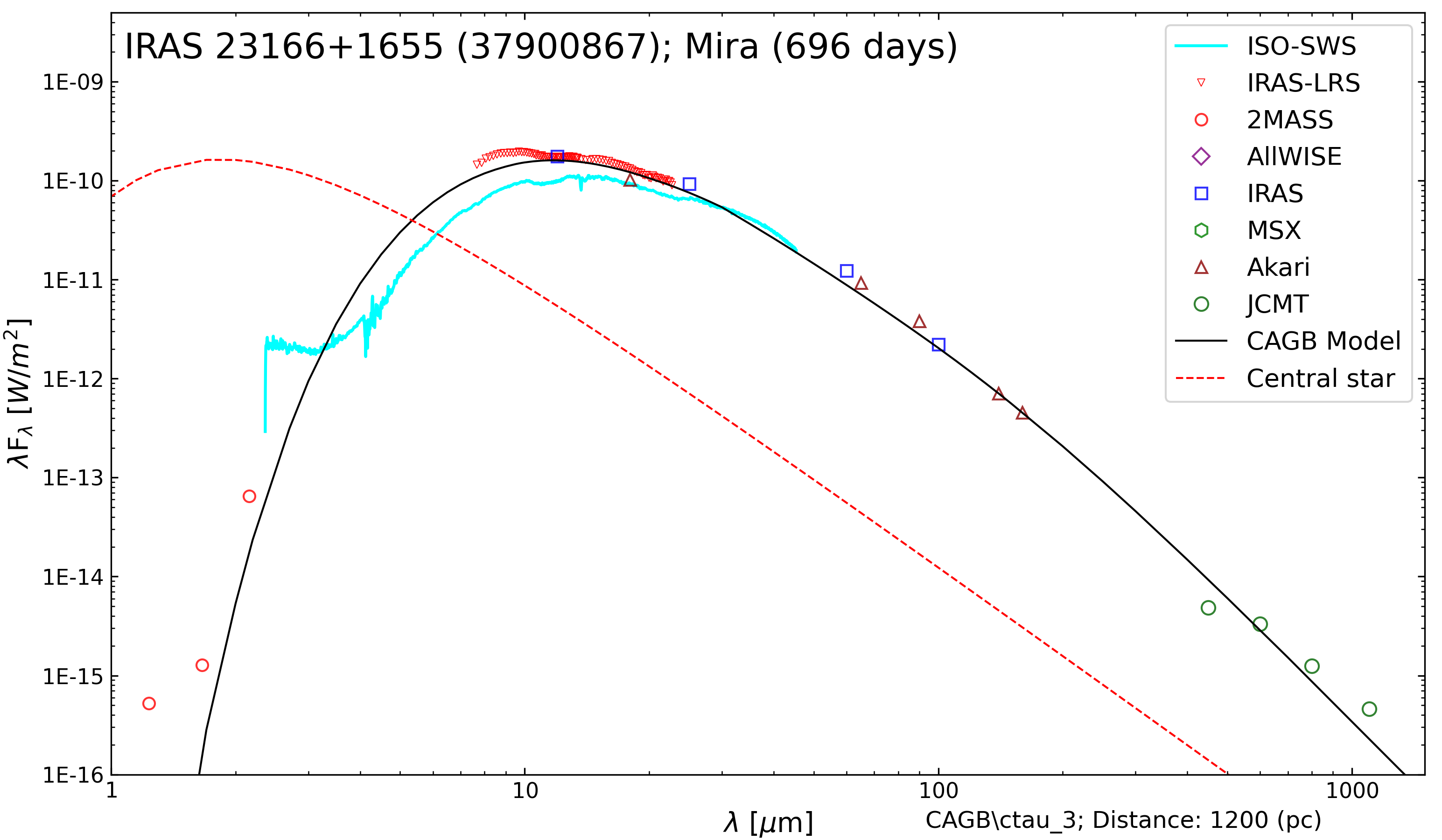}{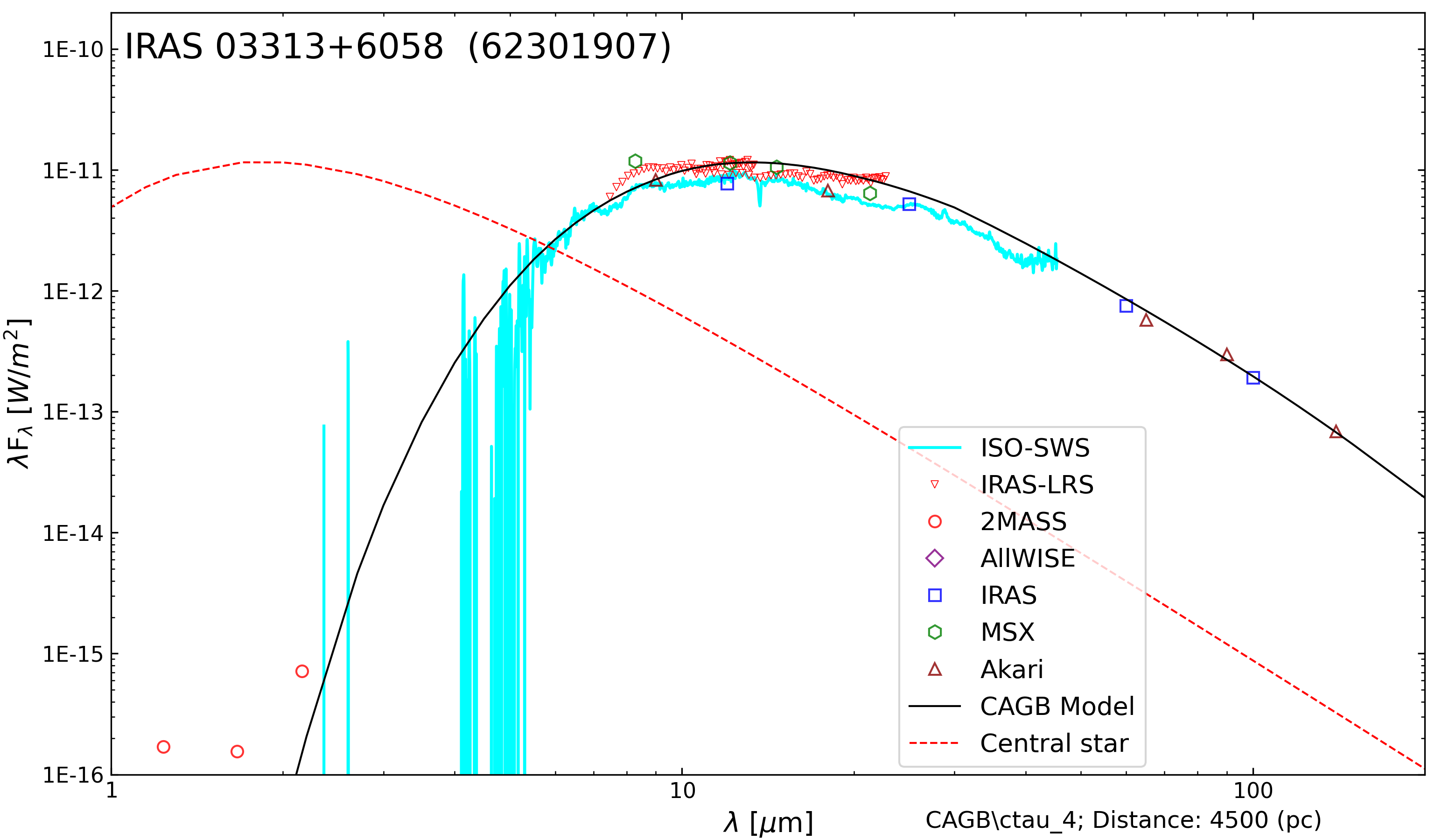}
\caption{Observed SEDs compared with simple dust shell models:
eight IR CAGB stars (IR-CAGB\_IC).
See Section~\ref{sec:modelsed}.} \label{f10}
\end{figure*}

\begin{figure*}
\centering
\smallplotsix{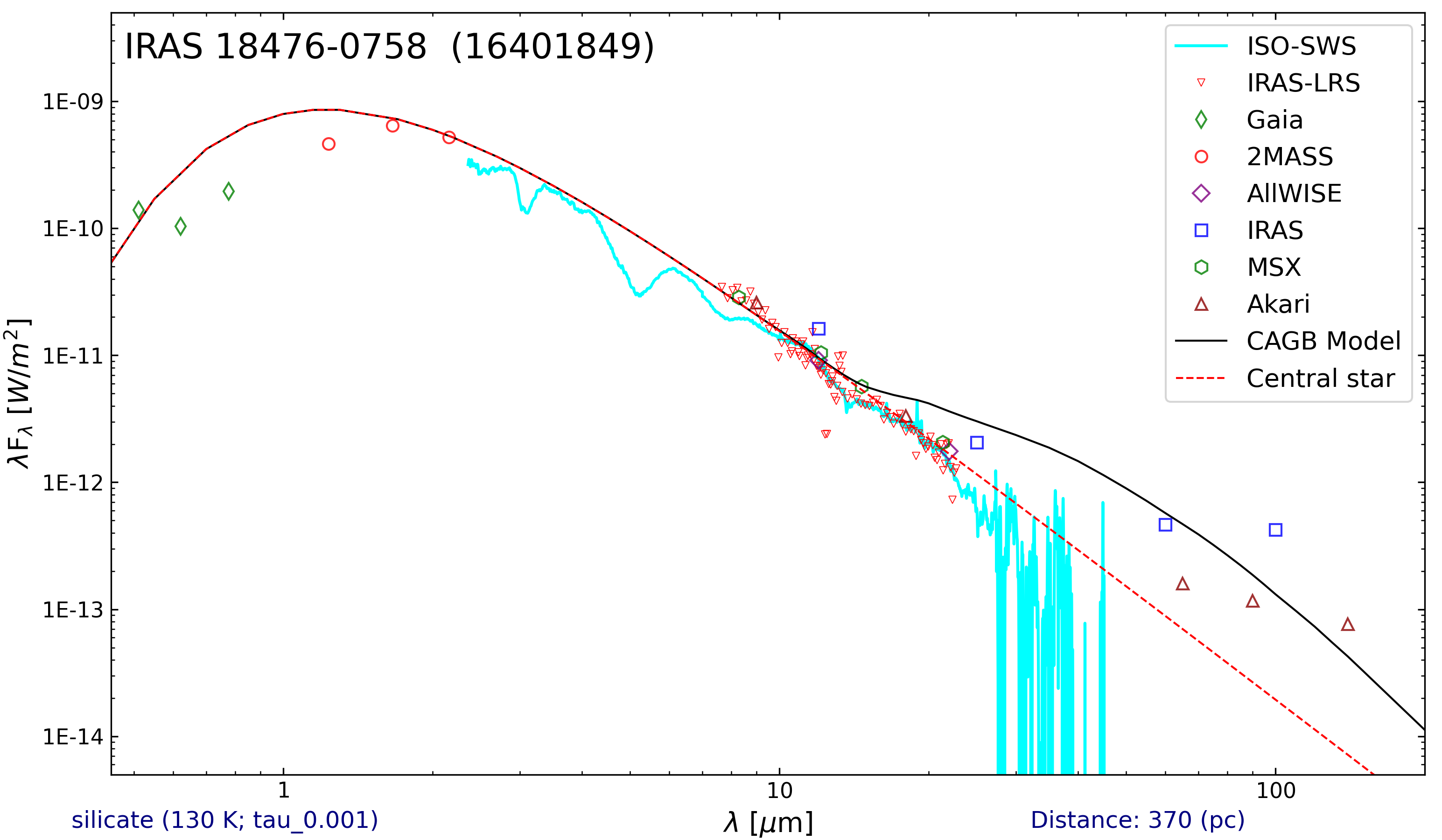}{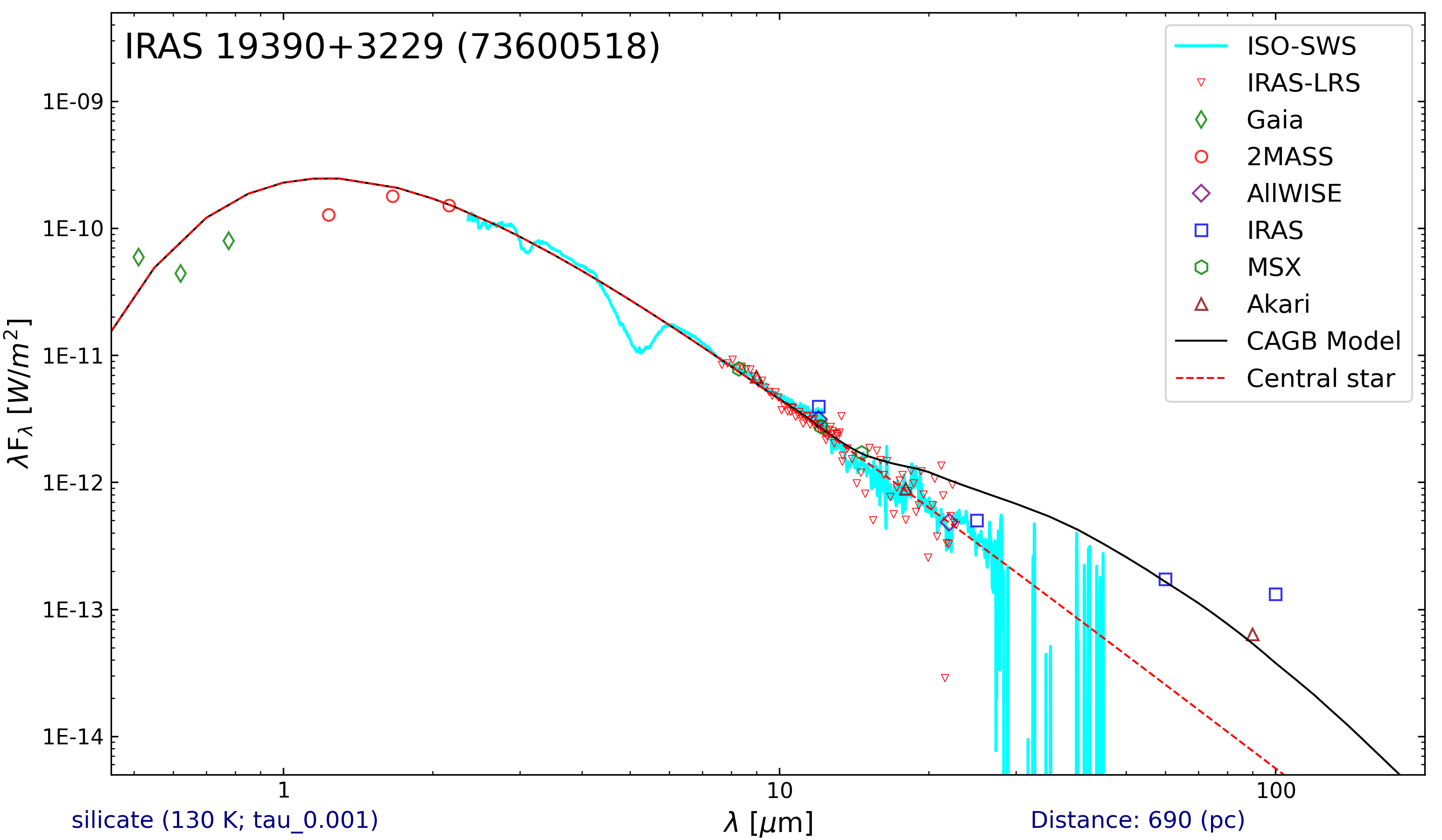}{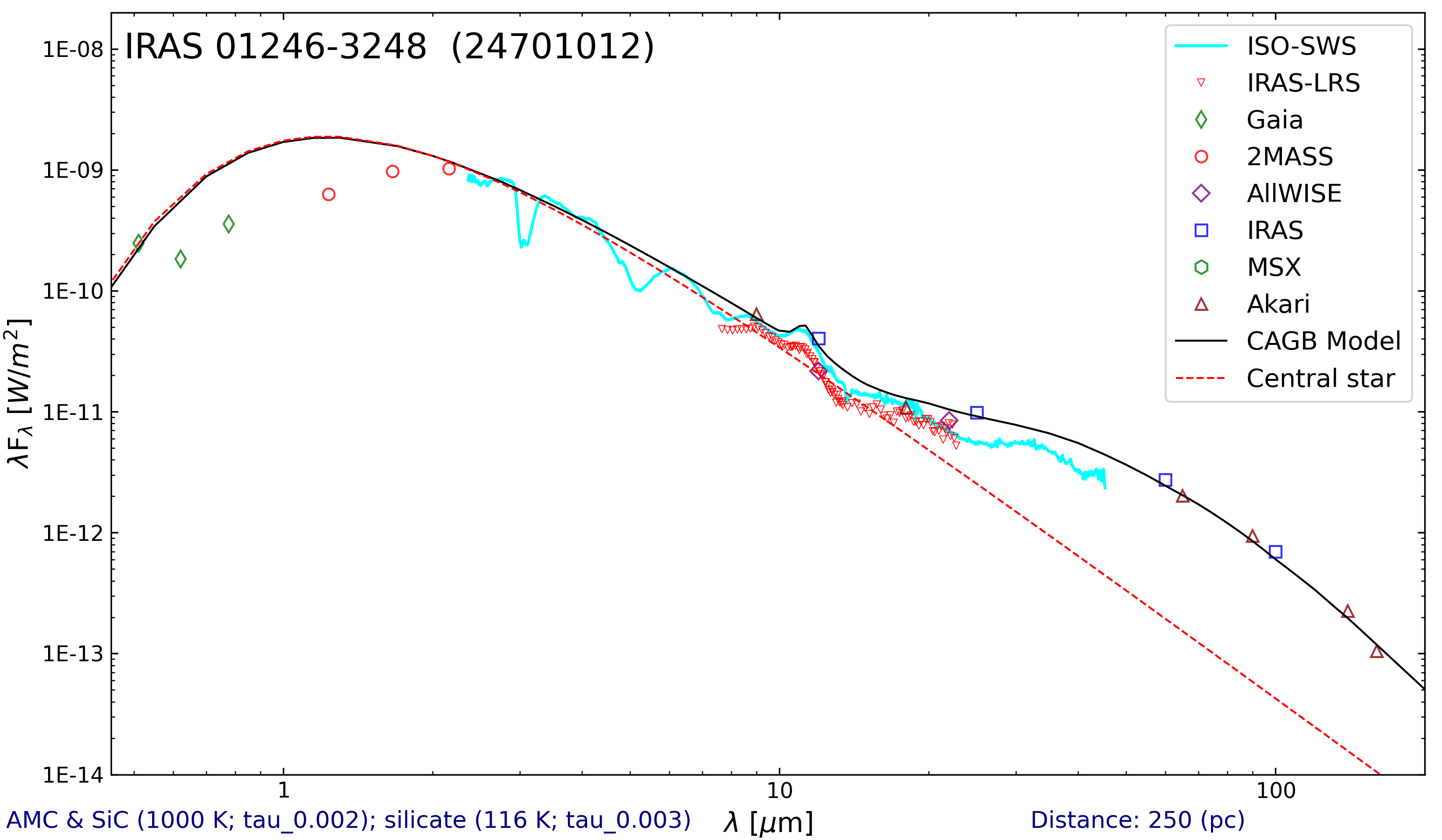}{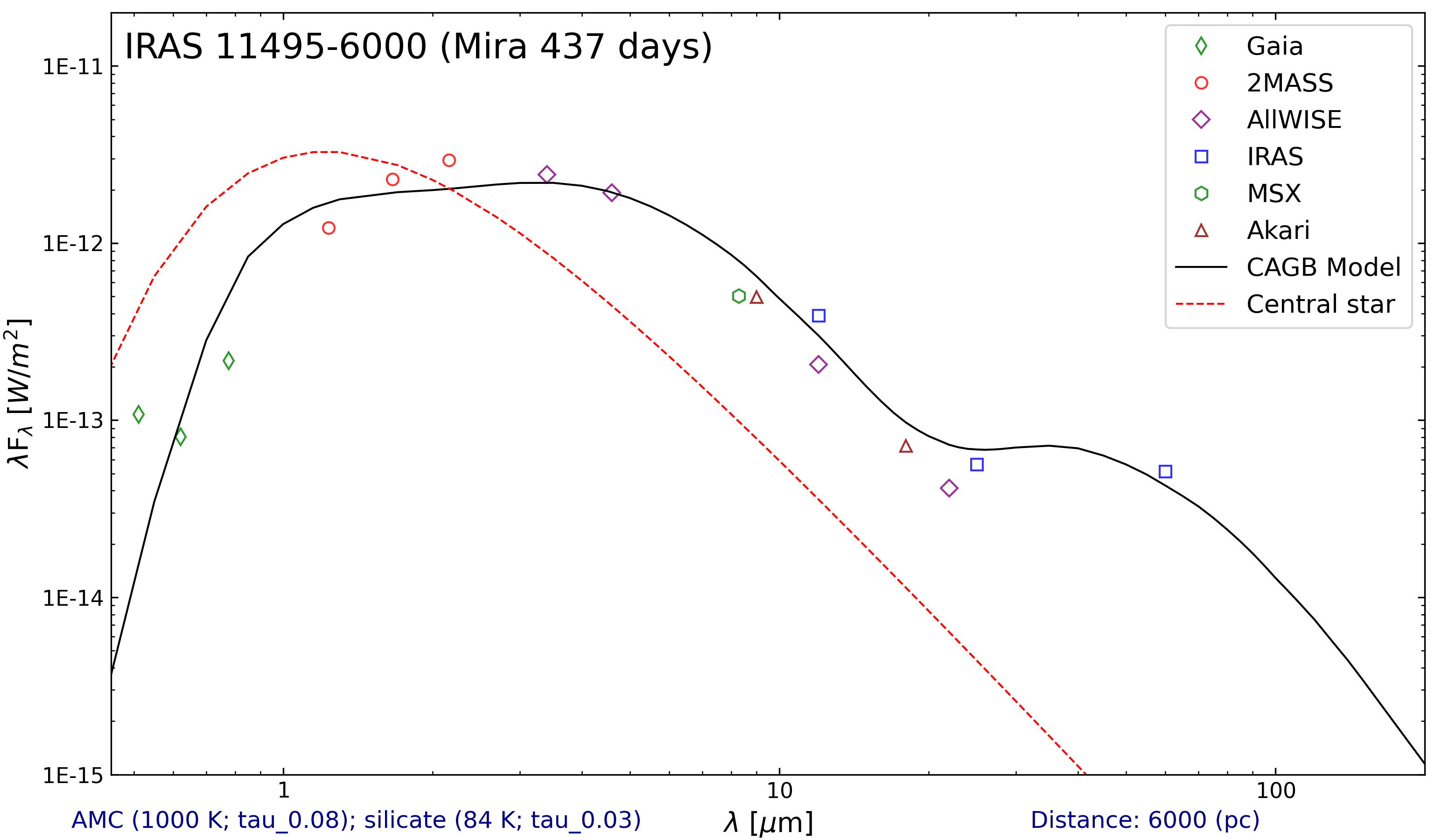}{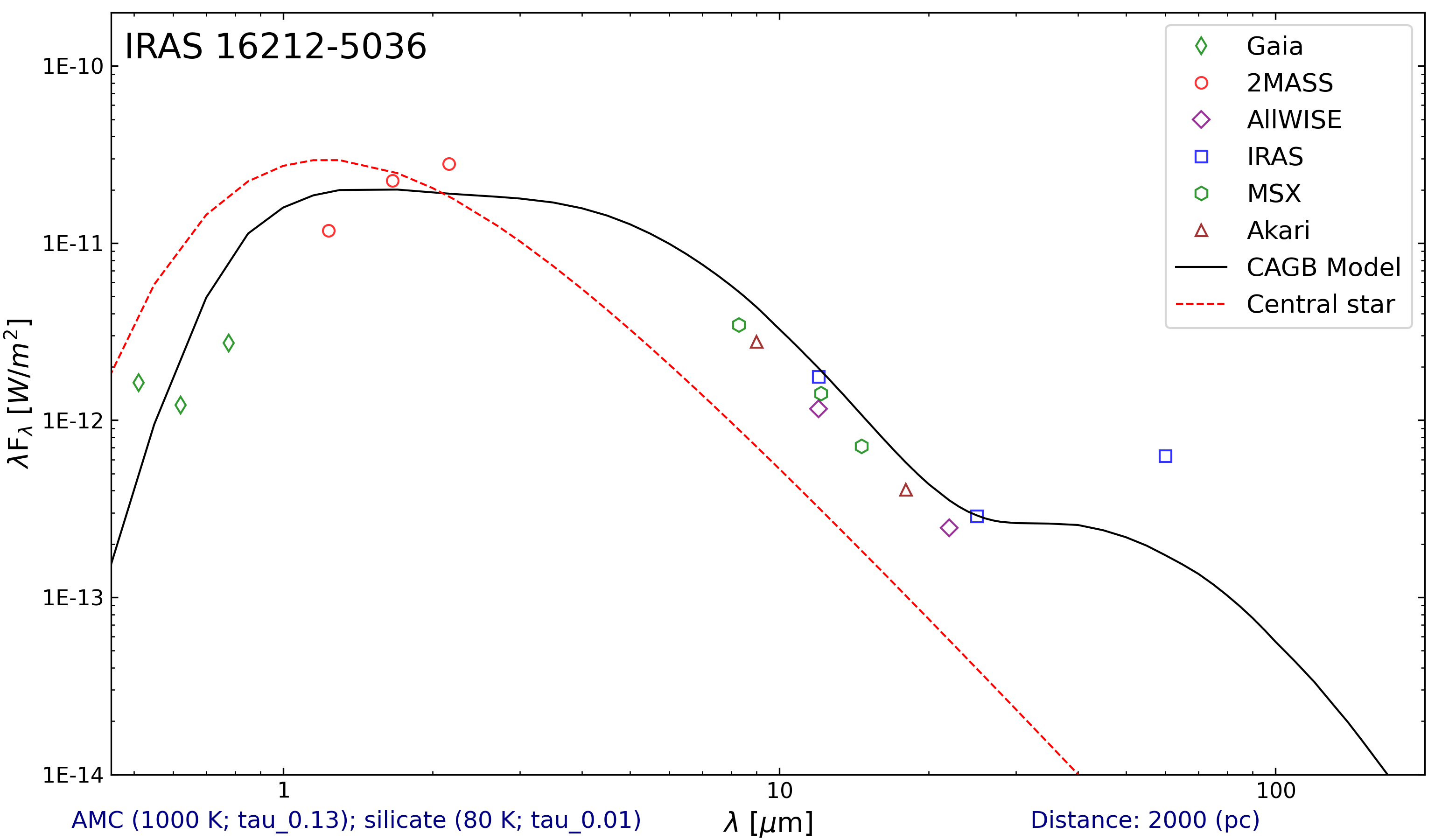}{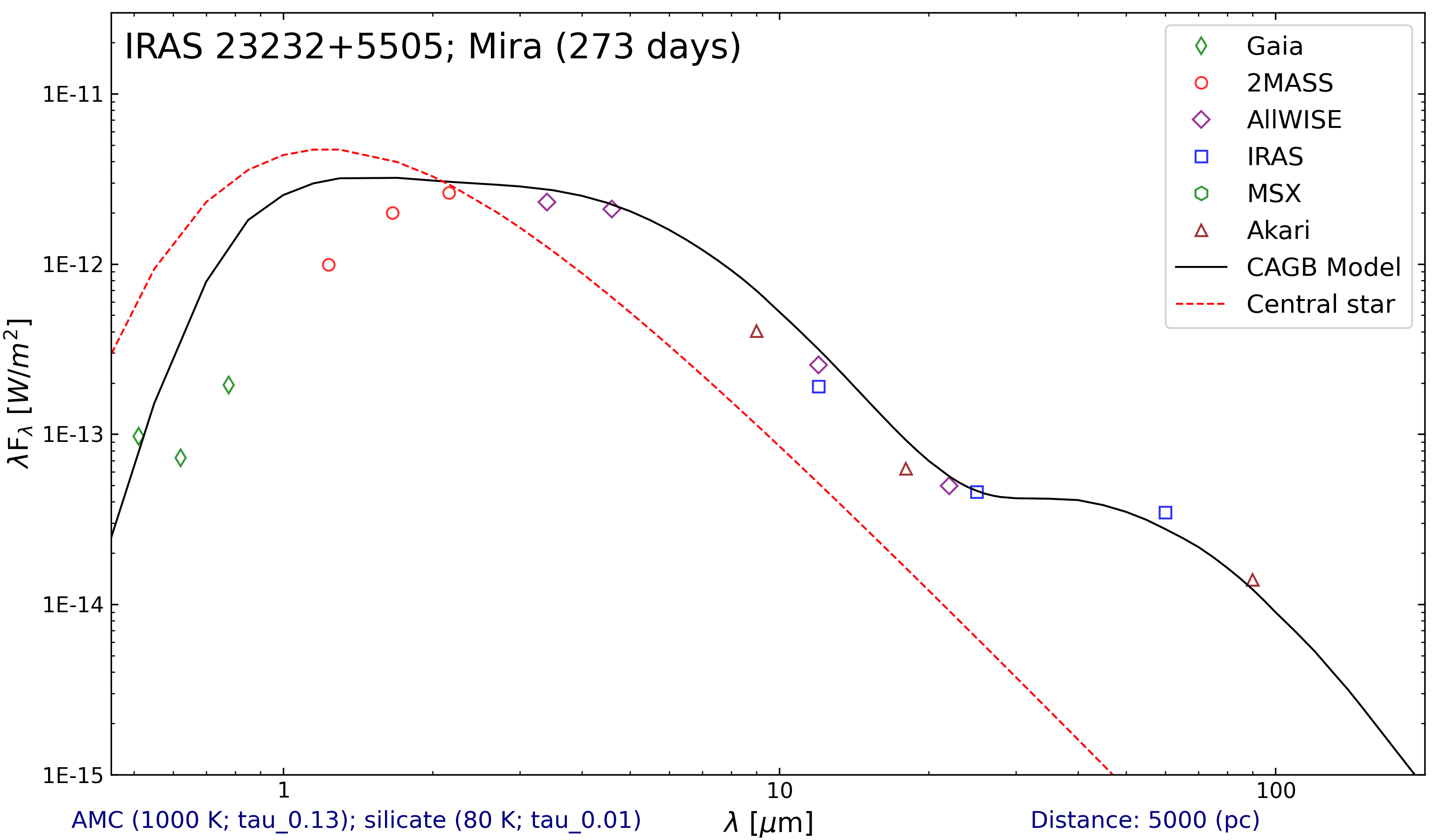}
\caption{Observed SEDs compared with more complex models (detached or double dust shell models):
two CAGB stars in upper panels with detached (O-rich) dust shell models
and other four CAGB stars with double (C-rich and O-rich) dust shell models.
See Section~\ref{sec:cmodel}.} \label{f11}
\end{figure*}

\subsection{Model SEDs for CAGB stars\label{sec:modelsed}}

From the sample of 4909 CAGB stars with IRAS counterparts (referred to as 
CAGB\_IC objects; see Table~\ref{tab:tab1}), we have selected a subset comprising 
relatively well known 152 CAGB objects (e.g., \citealt{suh2000}; 
\citealt{groenewegen2002}; \citealt{chen2012}; \citealt{groenewegen2022}). Within 
this subset, encompassing 51 visual-CAGB\_IC objects characterized by thin dust 
shells ($\tau_{10}$ $<$ 0.5) and 101 IR-CAGB\_IC objects featuring thick dust 
shells ($\tau_{10}$ $\geq$ 0.5), we have meticulously computed detailed 
theoretical model SEDs (see Section~\ref{sec:agbmodels}) that demonstrate a 
commendable ability to replicate the observations. 

The photometric data utilized for this analysis encompass observations from IRAS, 
AKARI, MSX, 2MASS, WISE, and Gaia DR3. Sub-millimeter observational data from 
JCMT (\citealt{groenewegen1993}) are also incorporated where available. And IR 
spectral data from IRAS LRS (\citealt{kwok1997}) and ISO SWS 
(\citealt{sloan2003}) are used when accessible. 

It is important to note that the theoretical dust shell model used in this work 
does not consider gas-phase radiation processes (see 
Section~\ref{sec:agbmodels}). Consequently, the CAGB model SEDs may not 
accurately reproduce observed SEDs at visual and NIR bands in detail. In many 
cases, distinguishing whether the extinction (or emission) at visual and NIR 
bands is of circumstellar or interstellar origin proves challenging. Therefore, 
Galactic extinction is not considered in SED plots.  

Figures~\ref{f9} and \ref{f10} showcase the SEDs for sixteen selected objects 
from the CAGB star sample. These are juxtaposed with theoretical model SEDs 
arranged in ascending order of dust optical depths ($\tau_{10}$). $\tau_{10}$ is 
indicated in the lower-right corner of the respective SED diagram. The 
theoretical models employ a simple mixture of AMC and SiC (20-30 \% by mass) as 
well as pure AMC for dust (refer to Section~\ref{sec:agbmodels}). Specifically, 
Figure~\ref{f7} displays SEDs of eight visual CAGB stars (chosen from 51 
visual-CAGB\_IC objects with $\tau_{10}$ $<$ 0.5). On the other hand, 
Figure~\ref{f8} illustrates SEDs of eight IR CAGB stars (selected from 101 
IR-CAGB\_IC objects with $\tau_{10}$ $\geq$ 0.5). 

For objects with available ISO SWS spectra (\citealt{sloan2003}), the target 
dedicated time number (TDT) is indicated in parentheses next to each object's 
name. Moreover, for objects identified as Mira variables by AAVSO, it is 
explicitly mentioned that they are Miras, accompanied by the pulsation period in 
parentheses.

For the 152 selected CAGB stars, a simple single dust shell model incorporating 
AMC and SiC dust effectively reproduces the observed SEDs for the majority of 
objects. However, some objects necessitate more complex models, such as detached 
or double shells, or non-spherical dust envelopes (refer to 
Section~\ref{sec:cmodel}).

\subsection{Objects that show complex SEDs\label{sec:cmodel}}

Among the 152 CAGB stars, whose detailed SEDs are compared with theoretical 
models (see Section~\ref{sec:modelsed}), twelve stars exhibit complex SEDs that 
may necessitate more sophisticated models, including detached or multiple shells 
or non-spherical dust envelopes. 

Figure~\ref{f11} showcases the SEDs for six objects selected from the 
aforementioned twelve stars, for which relatively simple detached or double dust 
shell models can provide a rough reproduction of the SEDs. In the top panels, for 
the two displayed objects, we employ detached ($T_c$ = 130 K) silicate dust 
shells to reproduce the excessive emission at 30-200 $\mu$m. For the next four 
objects displayed in the middle and bottom panels, we utilize double shell 
models: AMC dust shell and detached ($T_c$ = 80 - 116 K) silicate dust shell, 
replicating the excessive emission at 30-200 $\mu$m as well as in other 
wavelength bands. However, achieving a more detailed replication would 
necessitate more advanced models, such as those incorporating multiple 
non-spherical dust envelopes. Nevertheless, exploring such intricacies is beyond 
the scope of this work. 

In Section~\ref{sec:ircd}, we discussed visual carbon stars (objects in region R4 
on the IRAS 2CD; see Figure~\ref{f1}) that exhibit excessive flux at 60 $\mu$m 
due to remnants from an earlier phase when the stars were OAGB stars. This 
suggests the existence of CAGB stars with detached silicate dust shells, 
potentially not displaying silicate dust features due to significant detachments. 
These objects may share characteristics with single-star type silicate carbon 
stars that do exhibit silicate dust features (\citealt{ks2014}), although it is 
worth noting that most known silicate carbon stars are believed to be binary 
systems.

\begin{figure}
\centering
\smallplot{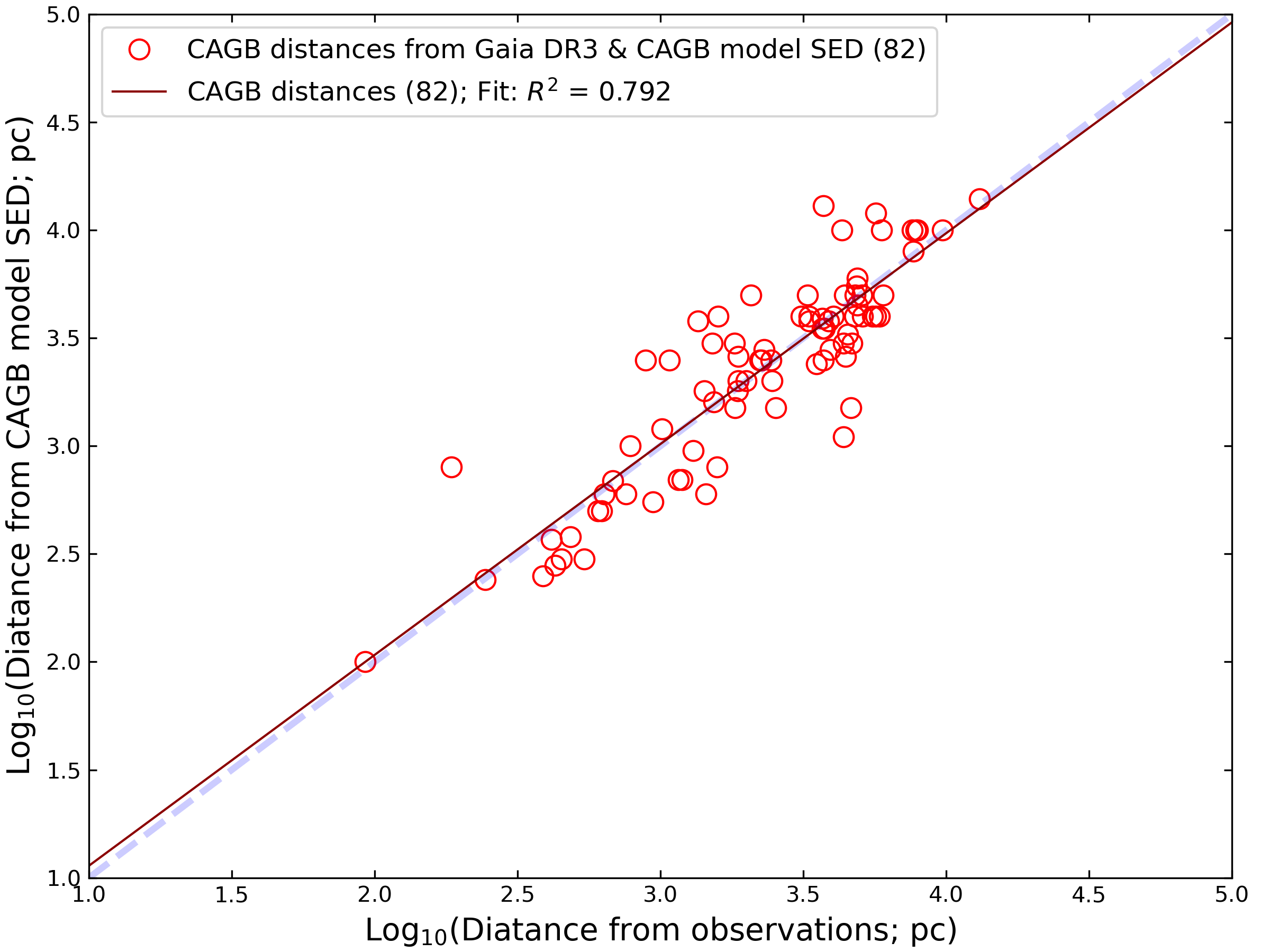}\caption{Comparison of distances obtained from observations (Gaia DR3 and Hipparcos)
and the ones obtained from theoretical model SEDs (See Section~\ref{sec:modeldistance}).}
\label{f12}
\end{figure}

\begin{figure*}
\centering
\largeplotone{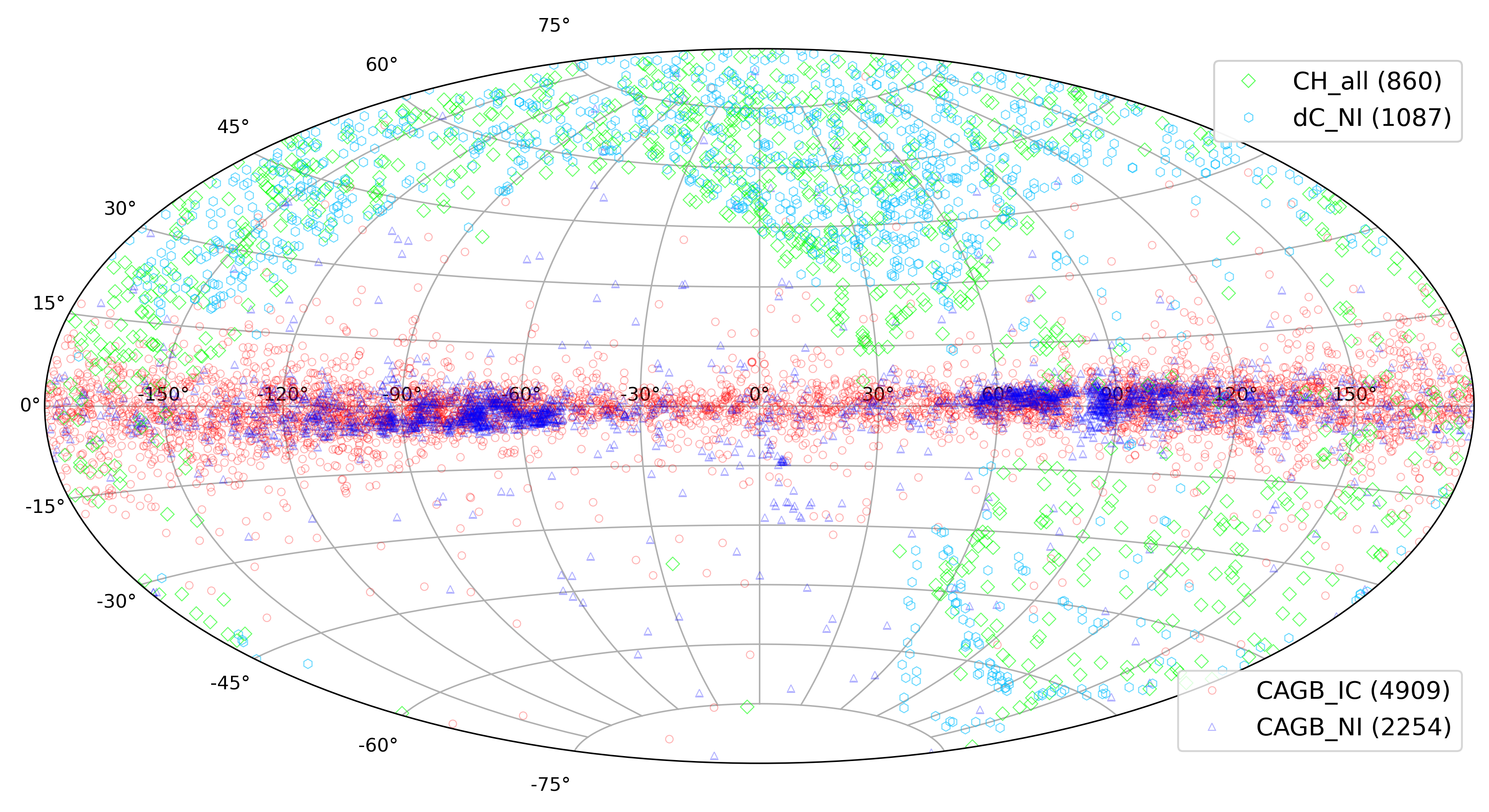}\caption{Spacial distributions AGB stars (CAGB\_IC and CAGB\_NI) in Galactic coordinate.}
\label{f13}
\end{figure*}

\begin{figure*}
\centering
\smallplottwo{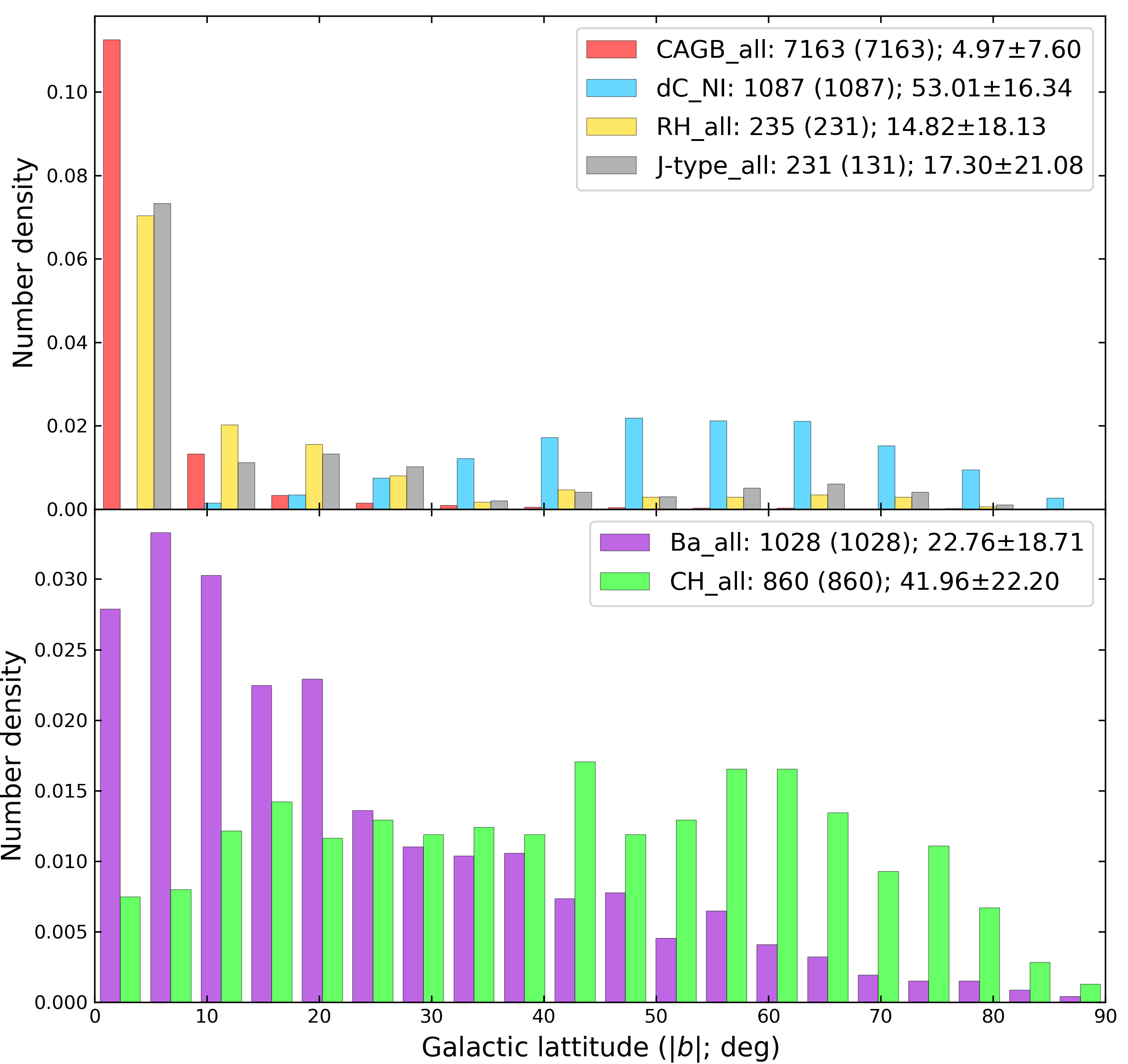}{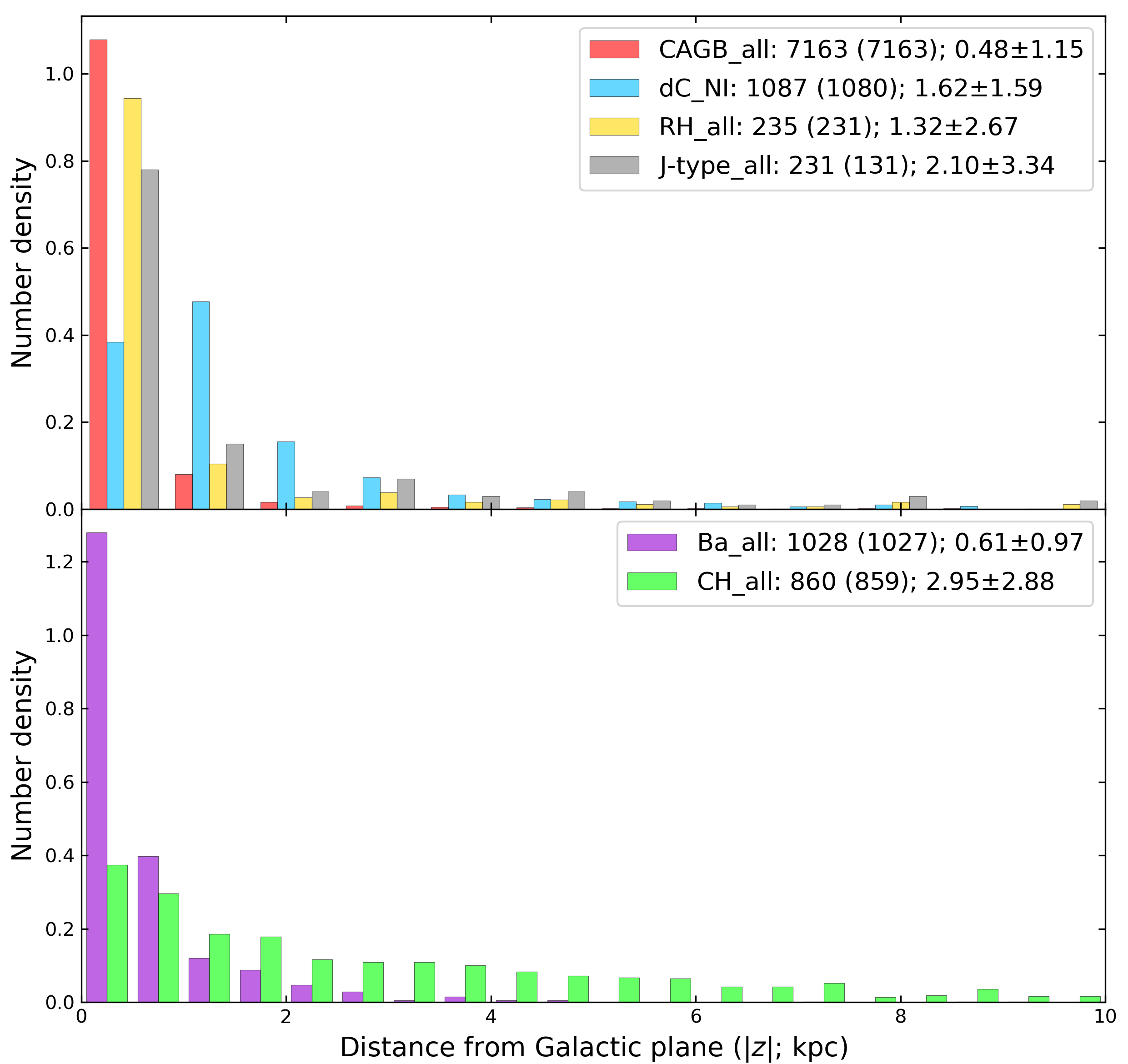}
\caption{Distances from the Galactic plane for various subclasses of carbon stares.
 (See Section~\ref{sec:spacial}).}
\label{f14}
\end{figure*}

\subsection{Distances obtained from CAGB Model SEDs\label{sec:modeldistance}}

When comparing the observed SED with the corresponding theoretical model SED for 
CAGB stars, we can determine the distance that best fits the observed SED. The 
theoretical distance for each object is indicated in the lower-right corner of 
the respective SED diagrams in Figures~\ref{f9} through~\ref{f11}. Subsequently, 
these theoretical distances can be compared with the observed distances obtained 
from Gaia DR3 (\citealt{bailer-jones2021}). 

Out of 4909 CAGB\_IC objects, distances for 4839 objects are available from 
observations (4834 from Gaia DR3 and 5 objects from Hipparcos), and distances for 
152 objects are determined using theoretical model SEDs. Among the 152 objects 
with theoretical distances, 82 objects have both theoretical and observed 
distances (80 from Gaia and 2 from Hipparcos), while 70 objects have only the 
distance derived from model SEDs. 

In Figure~\ref{f12}, a comparison is presented between distances obtained from 
observations and those obtained from theoretical model SEDs for the 82 objects 
with both sets of distances. The diagram reveals a notably strong correlation, 
the coefficients of determination ($R^2$) is larger than 0.79, indicating the 
reliability of the theoretical models. 

Therefore, we employed the mean values of distances obtained from observations 
(Gaia DR3 or Hipparcos) and distances derived from model SEDs for the 82 objects 
possessing both sets of distances. For the remaining 70 objects lacking observed 
distances, we utilized the distances derived from model SEDs. 

Out of the 2254 CAGB\_NI objects, distances for all objects are accessible from 
Gaia DR3, and 117 Mira variables have been identified through AAVSO. Among these 
117 Miras, the absolute magnitudes at the W3[12] band were found to be dimmer 
than -6 for three objects based on Gaia DR3 distances. Since Miras can be 
considered AGB stars, we compared the observed SEDs with the CAGB model SEDs, 
despite their simplicity due to fewer observed points. 

Our analysis revealed that Gaia DR3 distances for the three Miras were 
underestimated compared to the distances obtained from the CAGB model SEDs. 
Utilizing the mean values of distances from Gaia DR3 and theoretical model 
distances, the absolute magnitudes for the three Mira variables became brighter 
than -6. Consequently, we have chosen to use the mean distances for these three 
Mira variables.

\begin{figure*}
\centering
\smallplottwo{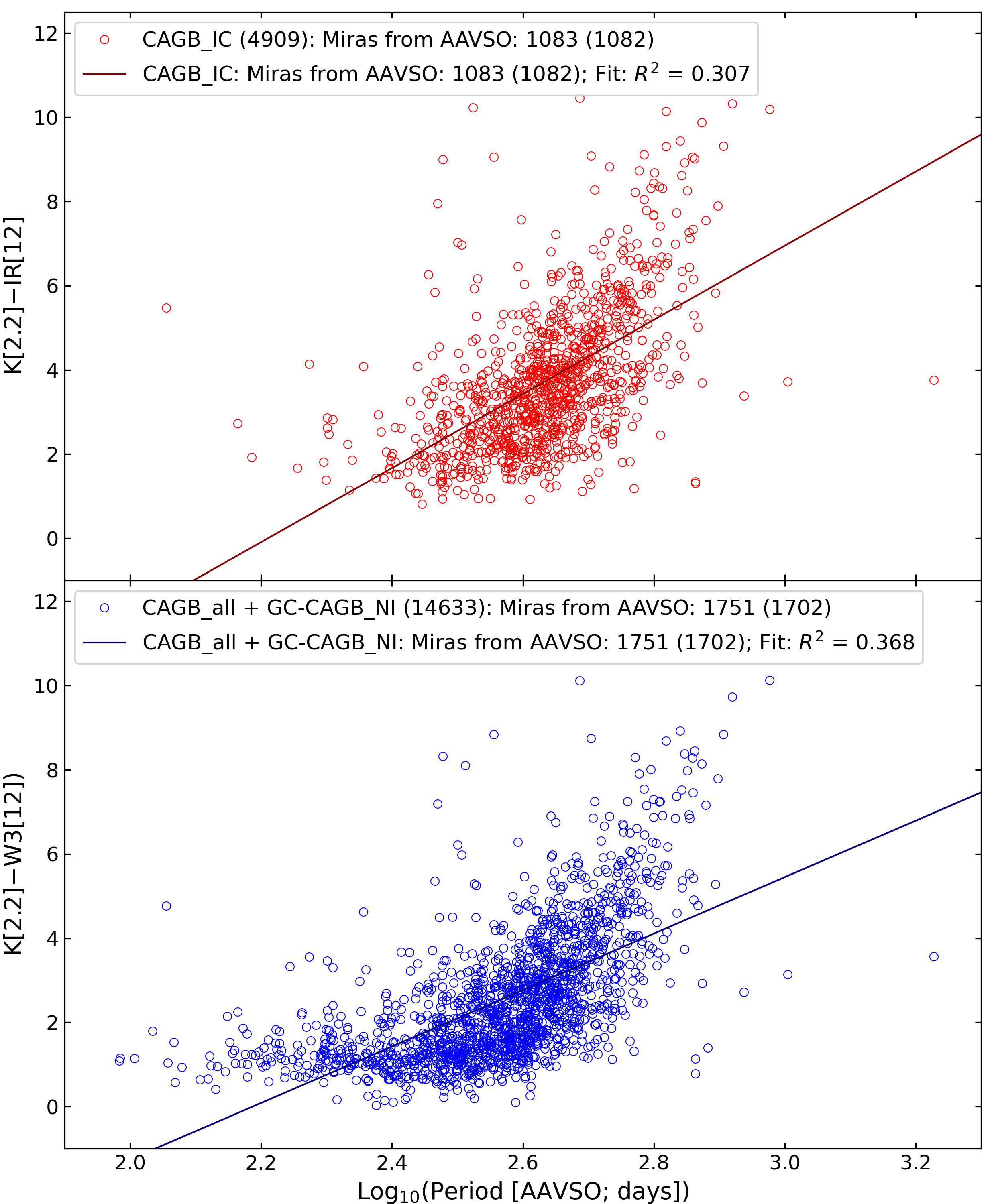}{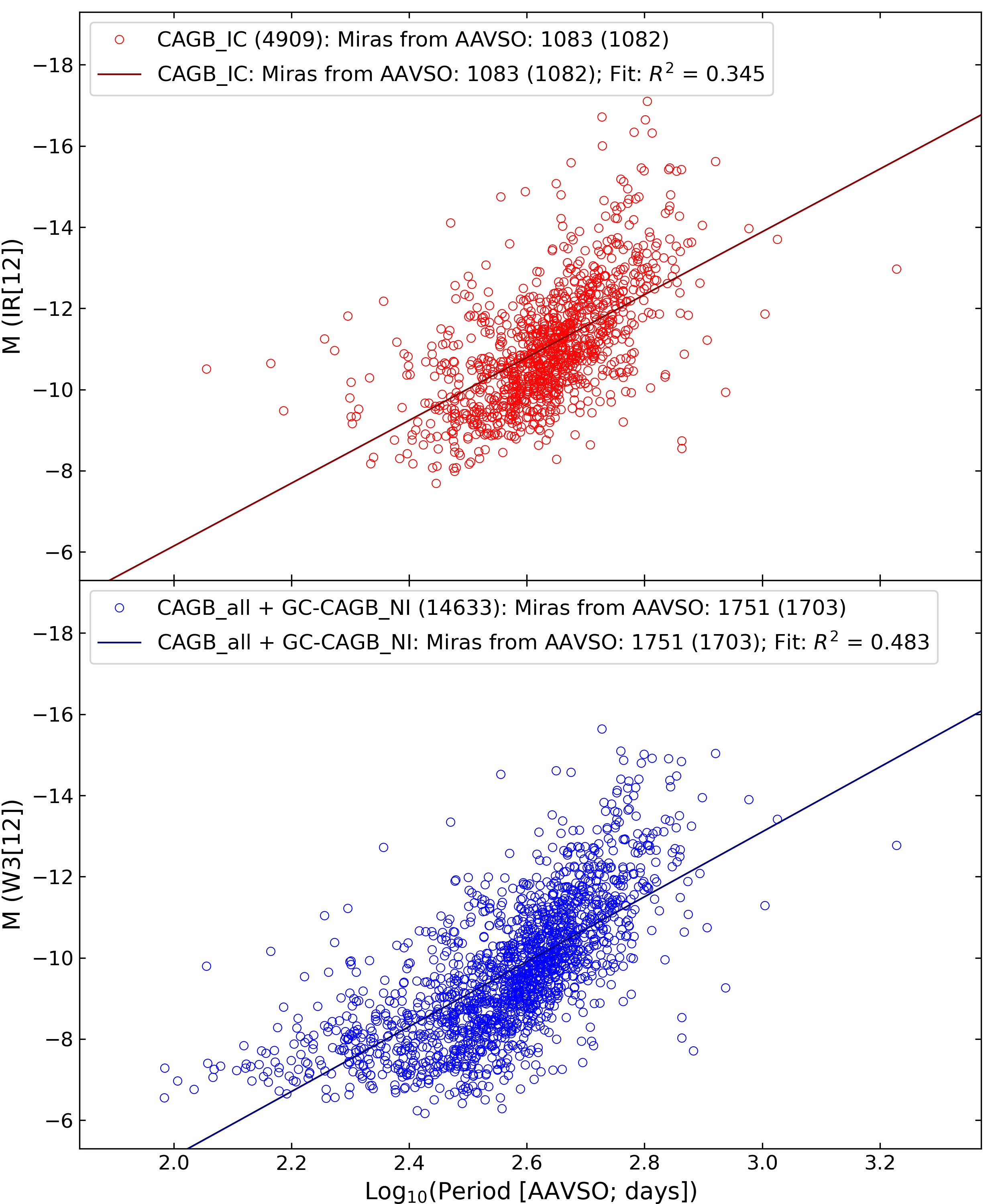}\caption{Period-Color and period-magnitude relations for Mira variables in the CAGB sample.
For each subclass, the number of objects is shown.
The number in parenthesis denotes the number of the plotted objects with good quality observed data.
See Section~\ref{sec:pmr}.} \label{f15}
\end{figure*}

\begin{figure*}
\centering
\smallplottwo{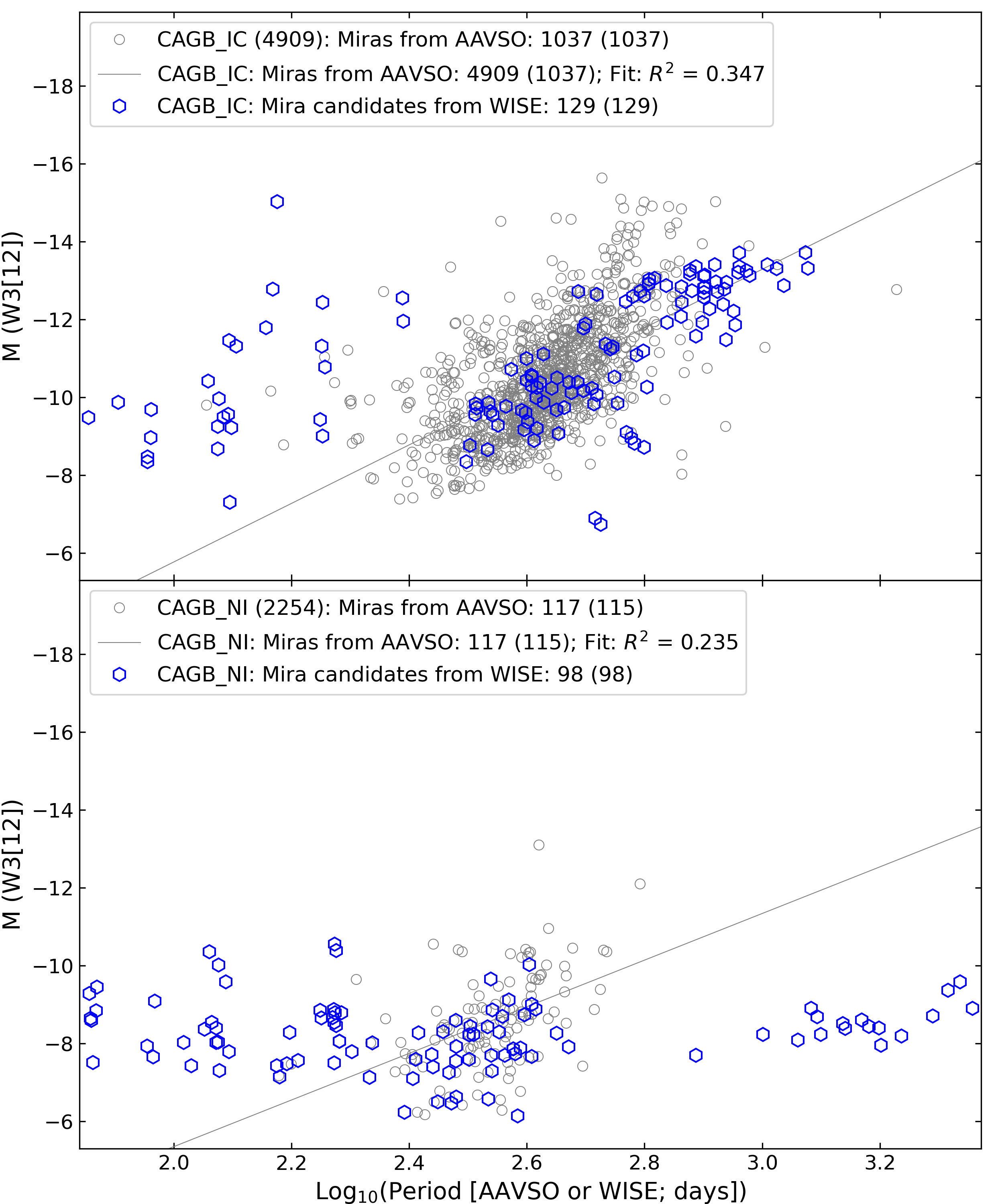}{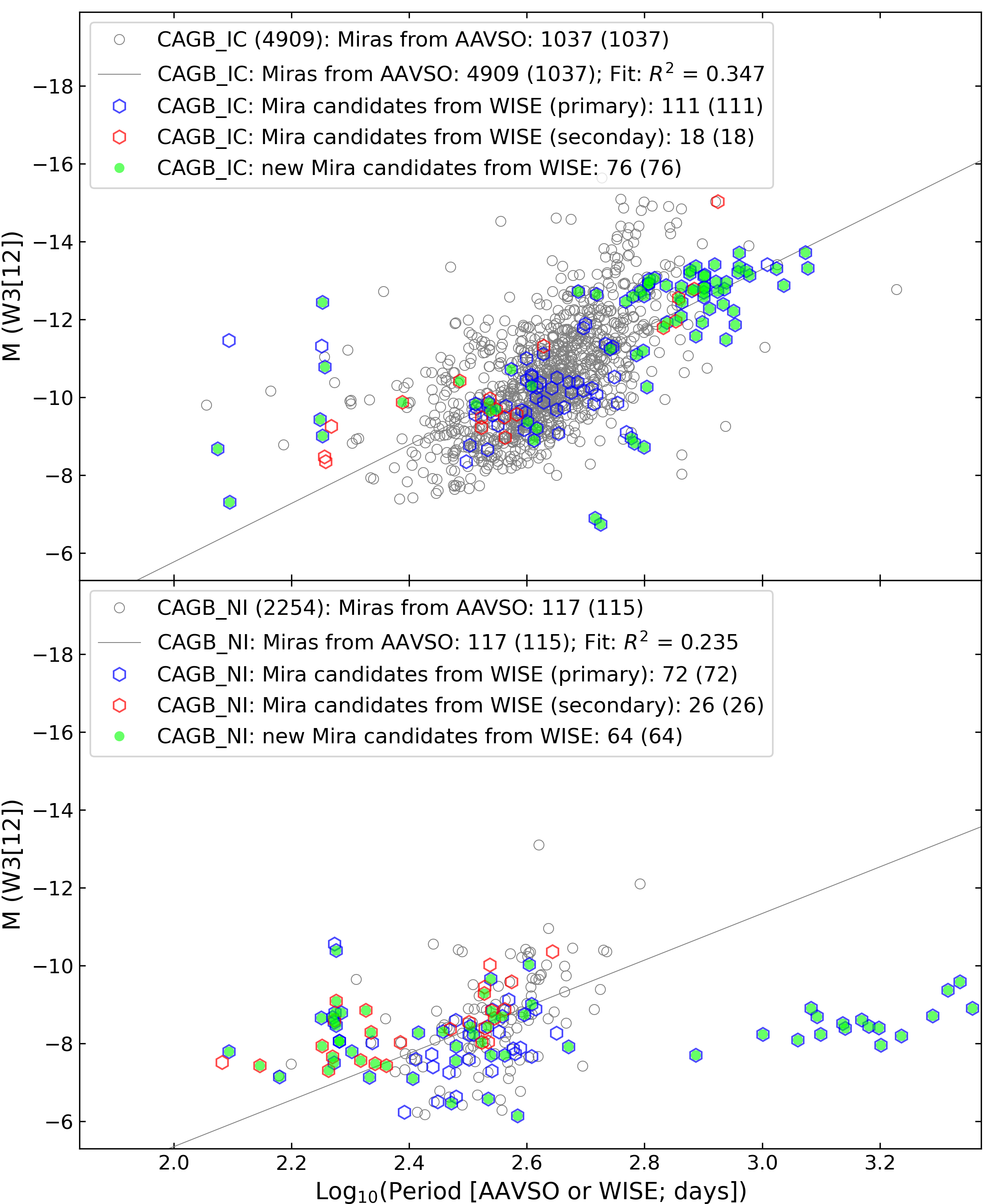}\caption{Period-magnitude relations for Mira variables and
candidates for Mira variable found from WISE data.
See Sections~\ref{sec:neo} and \ref{sec:neo-mira}.} \label{f16}
\end{figure*}

\section{Spacial distribution of carbon stars\label{sec:spacial}}

Figure~\ref{f13} illustrates the spatial distributions of carbon stars in 
Galactic coordinates. Notably, we observe a higher concentration of CAGB stars 
toward the Galactic plane compared to most extrinsic carbon stars (only CAGB 
stars, CH stars, and dC stars are depicted in this diagram). 

In Figure~\ref{f14}, the number distributions of Galactic latitudes and distances 
from the Galactic plane are presented for various subclasses of carbon stars. 
Mean values and standard deviations are provided alongside each subclass name. 
Once again, our analysis reveals that CAGB stars exhibit a pronounced 
concentration toward the Galactic plane compared to the majority of extrinsic 
carbon stars. 

As previously discussed in Section~\ref{sec:intro}, CH stars are regarded as the 
older, metal-poor analogs to Ba stars (\citealt{escorza2017}). The histograms 
presented in the lower panels of Figure~\ref{f14} illustrate the distribution of 
distances from the Galactic plane for both Ba stars and CH stars. It is evident 
that the distances for CH stars are notably greater than those for Ba stars, 
providing clear confirmation that CH stars indeed serve as older, metal-poor 
analogs to Ba stars.

\section{Infrared properties of pulsating variables\label{sec:pul}}

Most AGB stars are commonly considered to be LPVs featuring outer dust envelopes 
(e.g., \citealt{suh2021}). LPVs are categorized into small-amplitude red giants 
(SARGs), semiregular variables (SRVs), and Mira variables (e.g., 
\citealt{swu13}). While majority of SRVs and a significant proportion of SARGs 
may also be in the AGB phase, it is established that all Mira variables are in 
the AGB phase (e.g., \citealt{hofner2018}). 

The consensus in the scientific community holds that more evolved (or more 
massive) AGB stars tend to display larger pulsation amplitudes, longer pulsation 
periods, and higher mass-loss rates (\citealt{debeck2010}; \citealt{sk2013b}; 
\citealt{groenewegen2022}).

\subsection{Period-color and Period-magnitude relations\label{sec:pmr}}

Within the sample of 7163 CAGB stars (4909 CAGB\_IC and 2254 CAGB\_NI objects; 
see Section~\ref{sec:gcarbon}), 1154 Mira variables have been identified through 
AAVSO. Additionally, among the 7470 candidate objects for new CAGB stars, solely 
identified from Gaia DR3 spectra and absolute magnitudes (GC-CAGB\_NI objects), 
597 Miras have been identified through AAVSO. 

The left panels of Figure~\ref{f15} depict period-color relations for the Mira 
variables. The plot illustrates the K[2.2]-W3[12] color versus pulsation periods 
for CAGB stars. Despite noticeable scatters, a meaningful relationship between IR 
color and pulsation periods is discernible among Mira variables. 

On the right panels of Figure~\ref{f15}, period-magnitude relations (PMRs) for 
the Mira variables are presented. The plot illustrates M (W3[12]) versus 
pulsation periods for CAGB stars. The primary cause of the scatters appears to be 
uncertainties in distances. 

In contrast to Miras in our Galaxy, those in the Magellanic clouds exhibit robust 
linear-like PMRs with minimal scatters when the wavelength exceeds about 3 $\mu$m 
(\citealt{suh2020}). The study finds that Mira variables in the Magellanic clouds 
demonstrate relatively large coefficients of determination ($R^2$=0.6-0.85) in 
the linear relationship across wavelength bands in the range of 3-24 $\mu$m. This 
phenomenon is attributed to the objects being at similar distances and 
experiencing less extinctions.

\begin{figure*}
\centering
\smallploteight{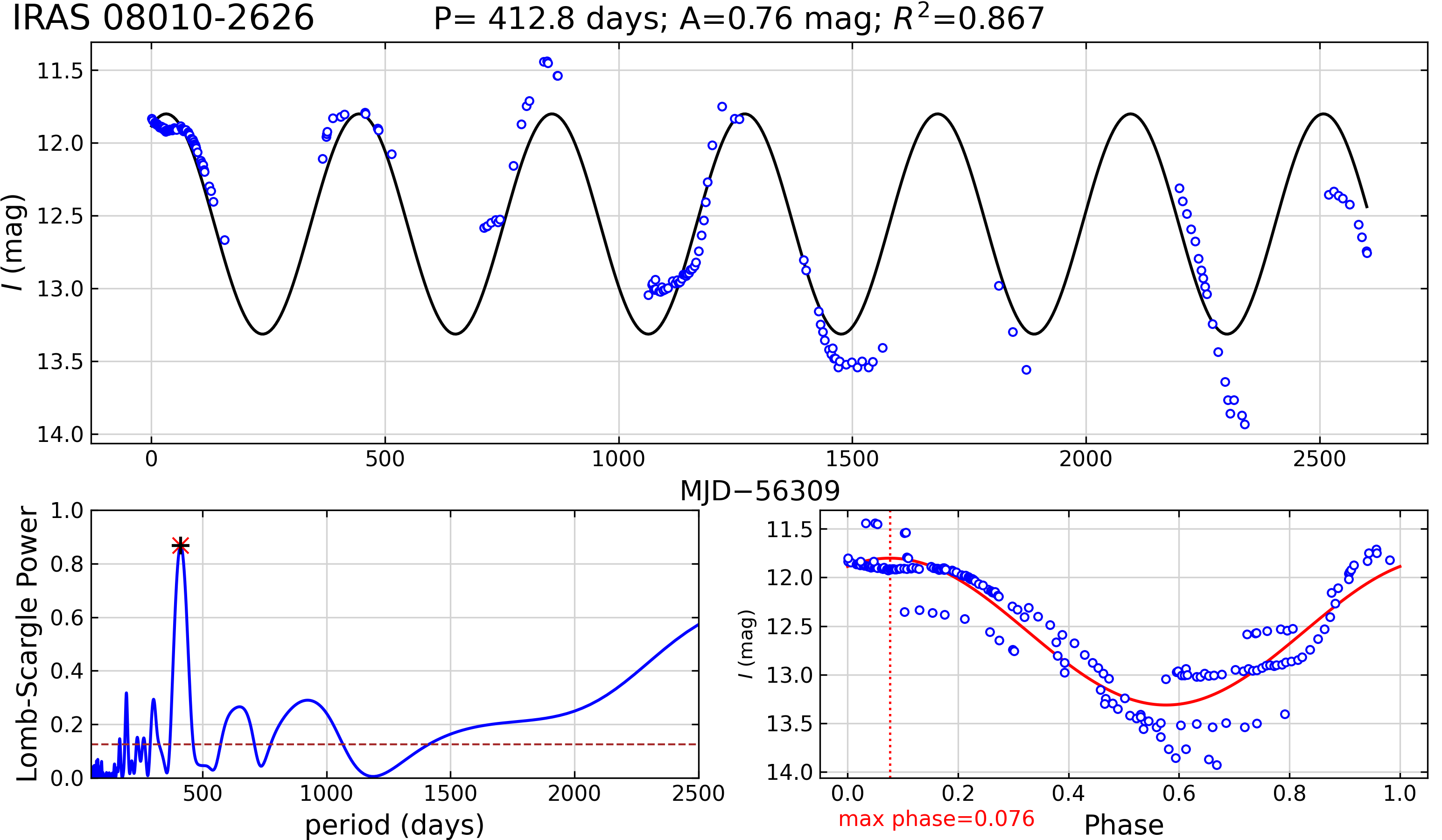}{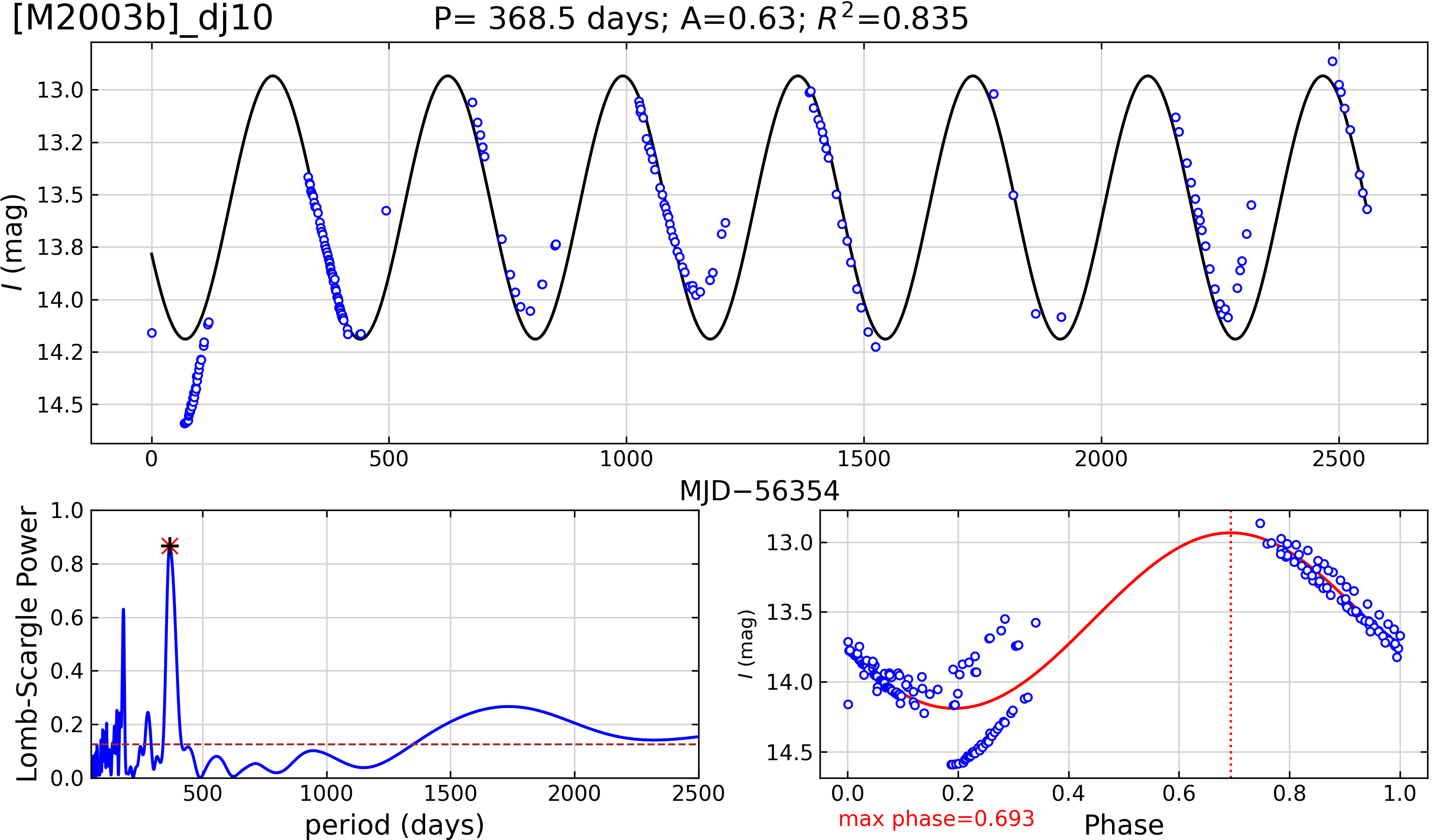}{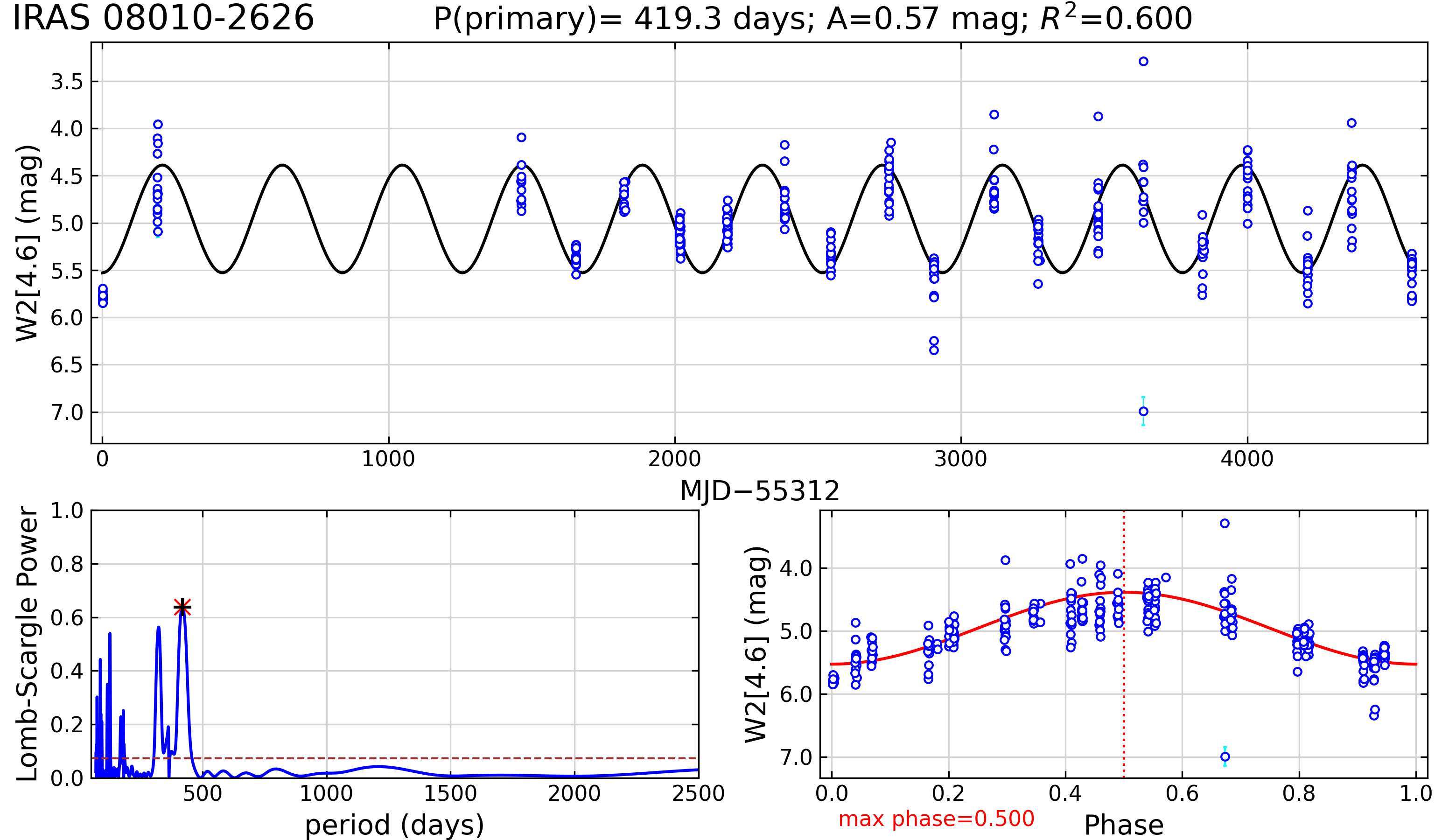}{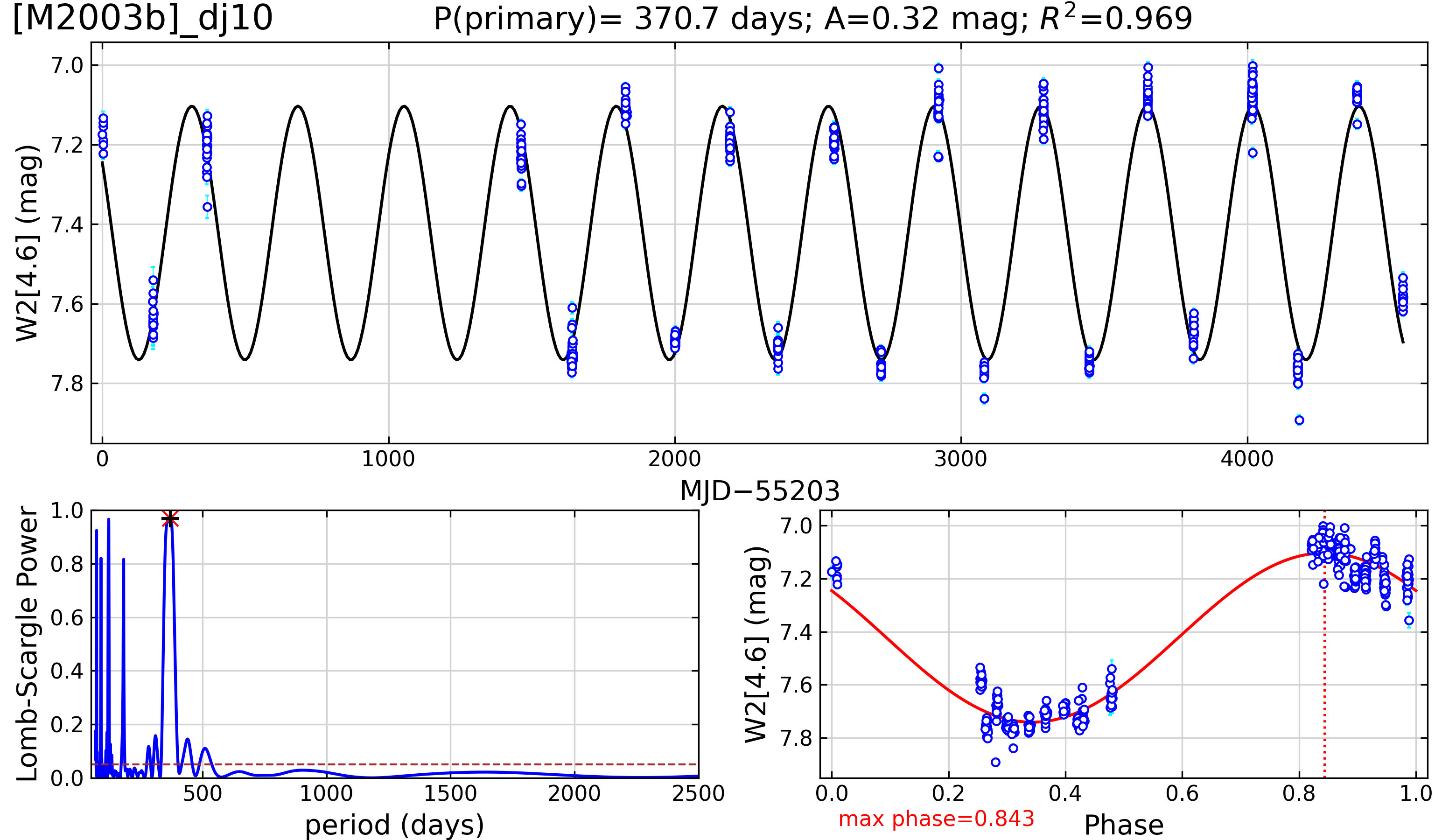}{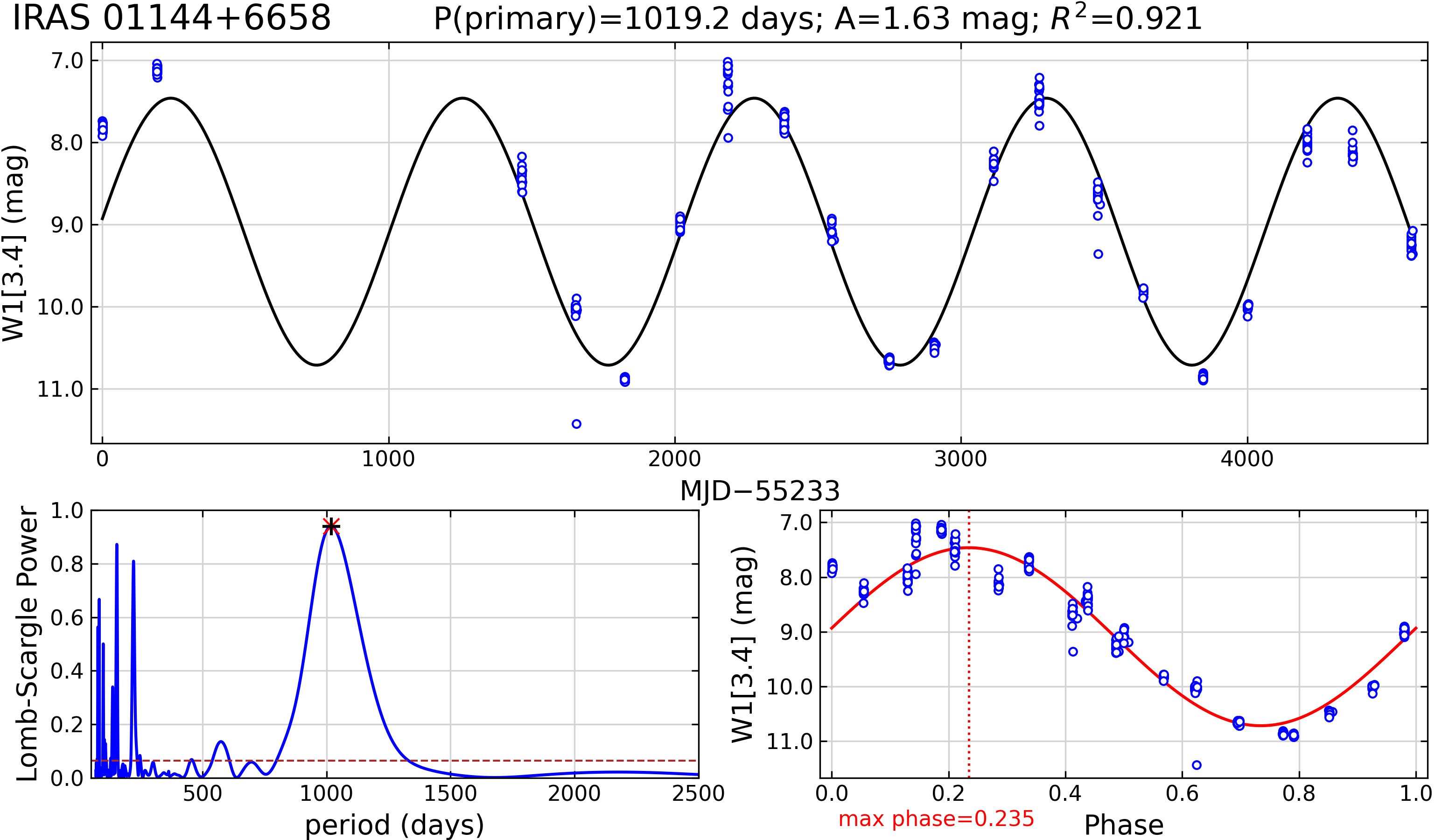}{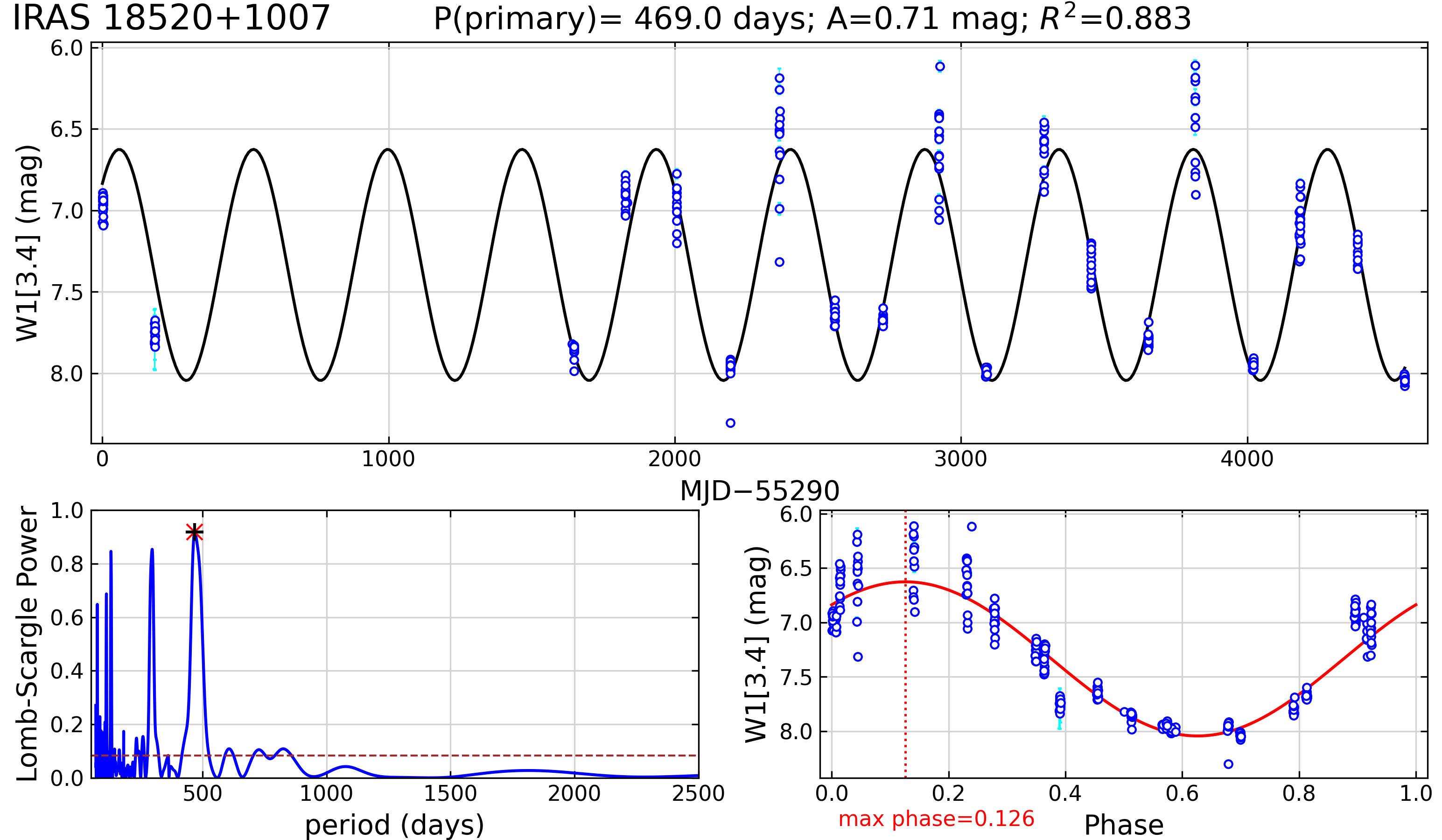}{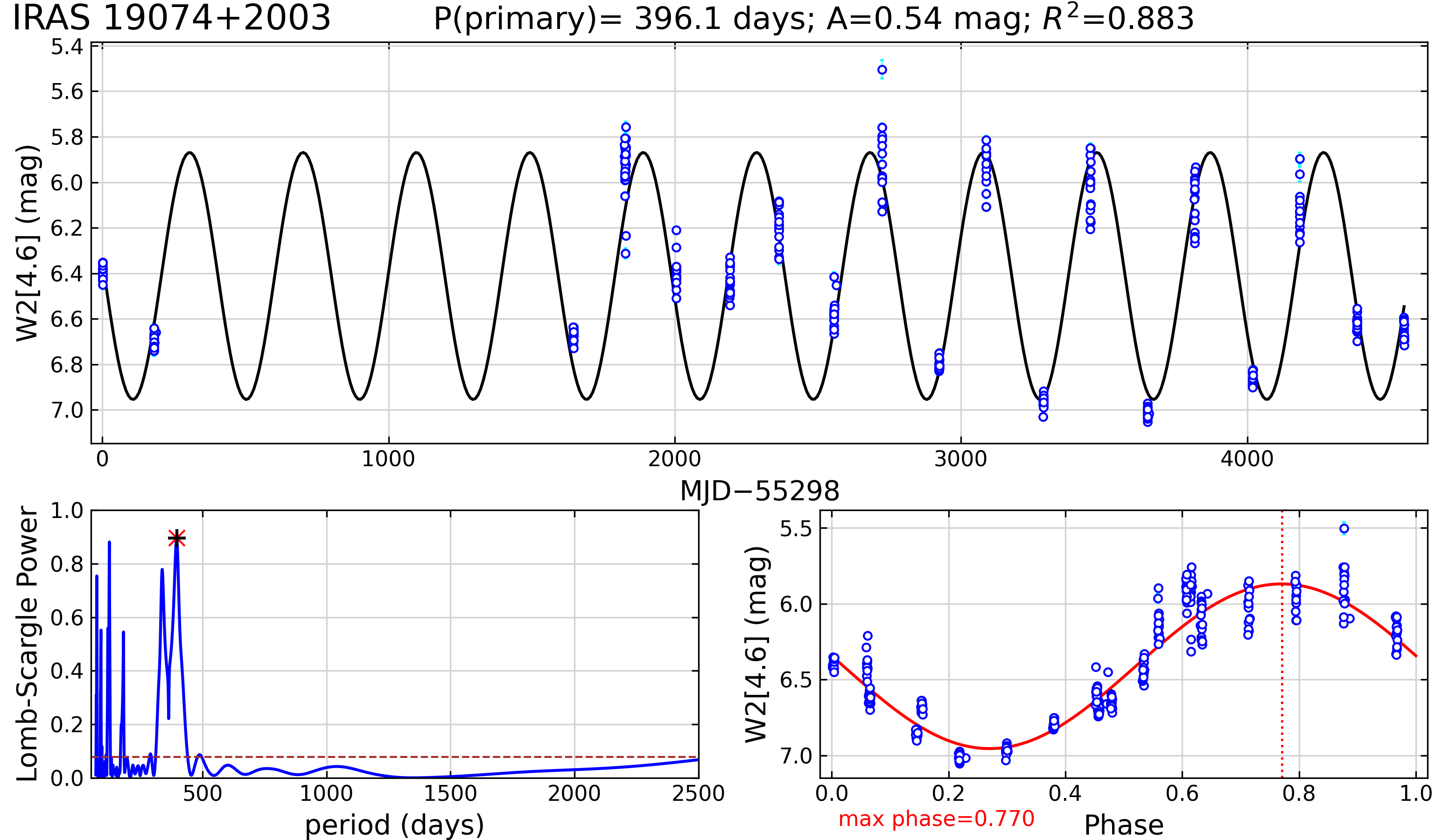}{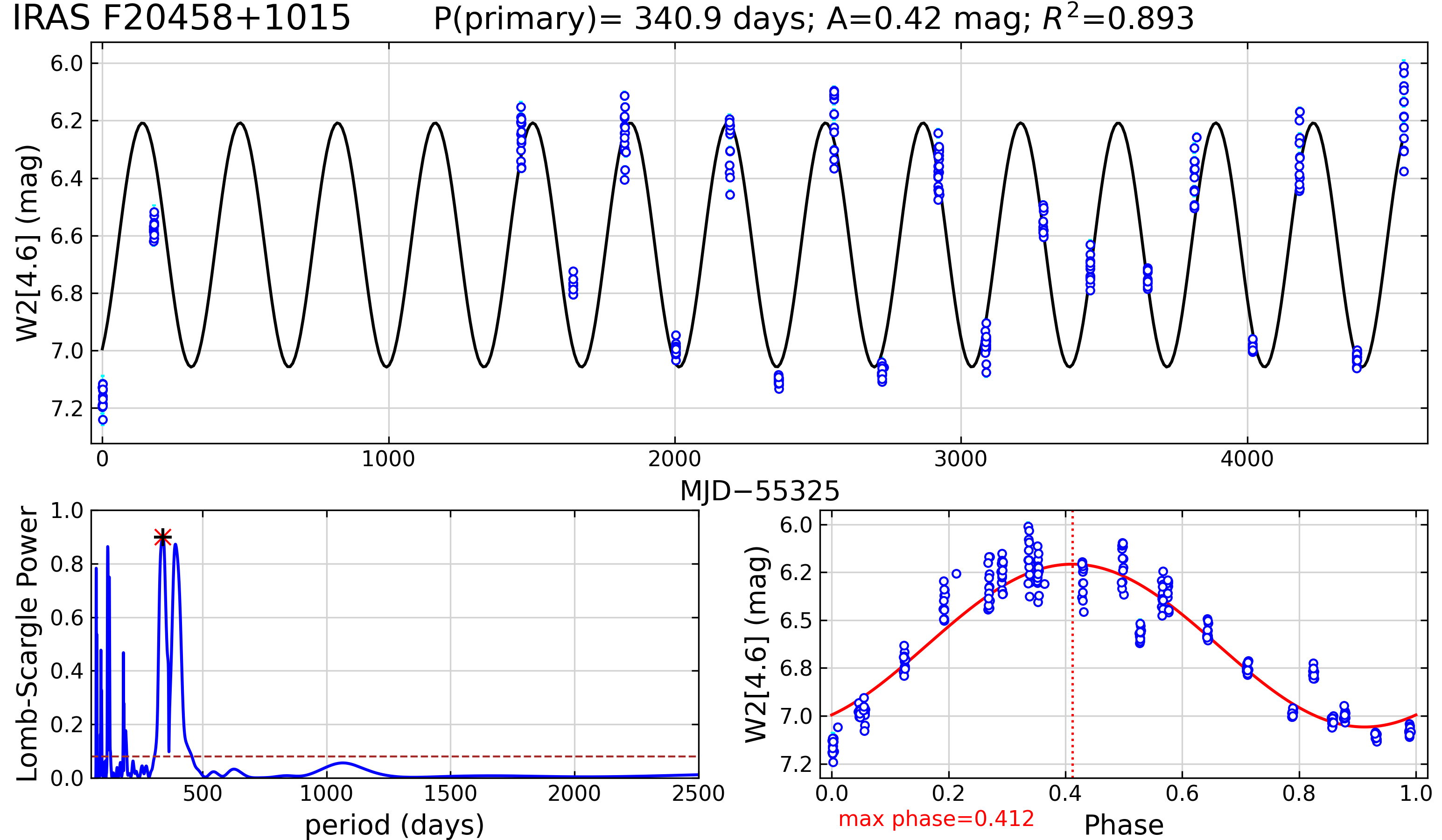}
\caption{Lomb-Scargle periodograms for six CAGB stars with known periods (AAVSO Miras).
The top panels display OGLE4 ($I$ band) light curves for two selected CAGB stars. 
Subsequently, the following six panels present WISE light curves for all six CAGB stars. 
The corresponding AAVSO periods for these six CAGB stars are as follows: 413, 365, 1060, 445, 442, and 302 days.
See Sections~\ref{sec:neo} and \ref{sec:neo-mira}.} \label{f17}
\end{figure*}

\begin{figure*}
\centering
\smallploteight{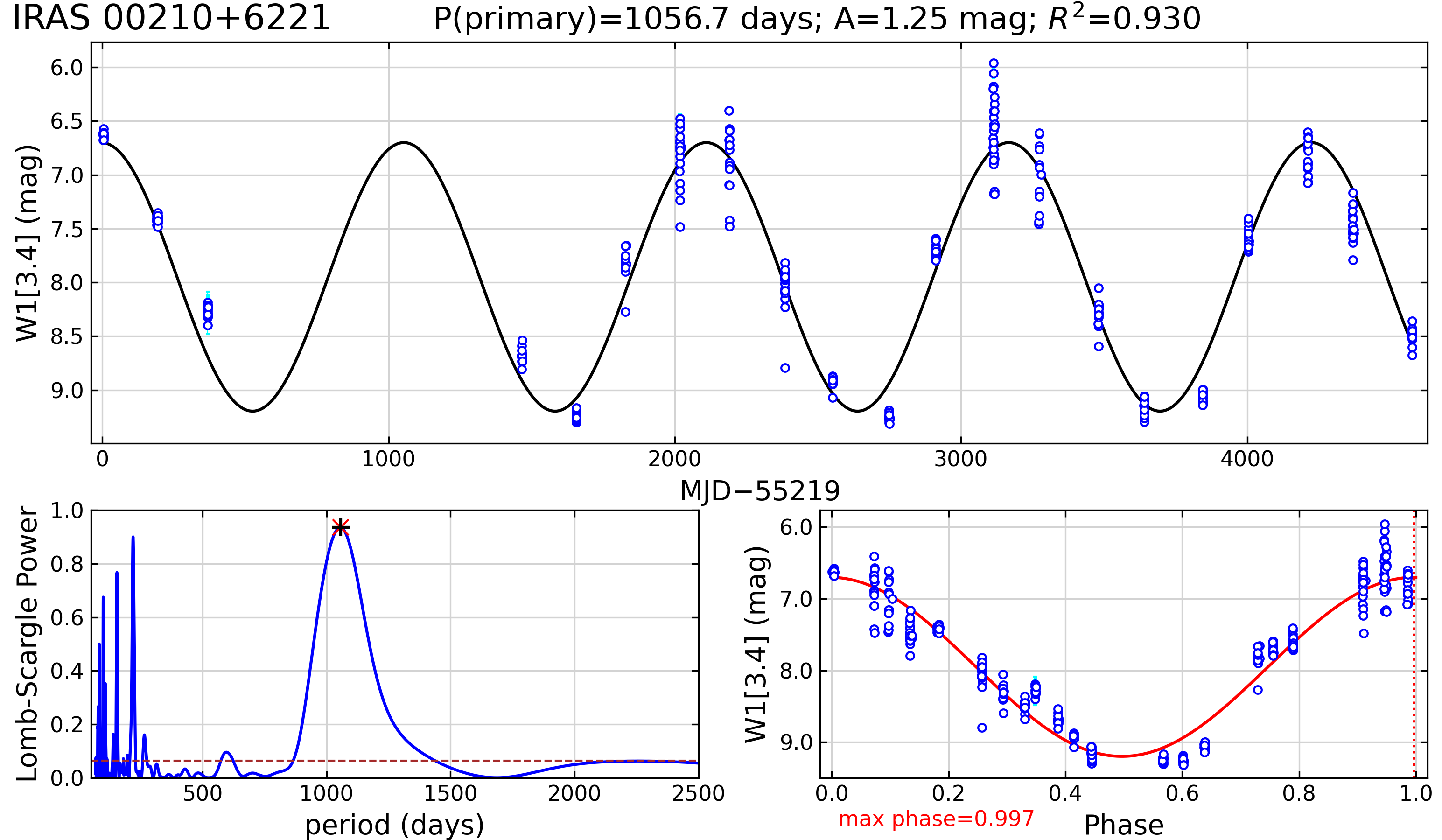}{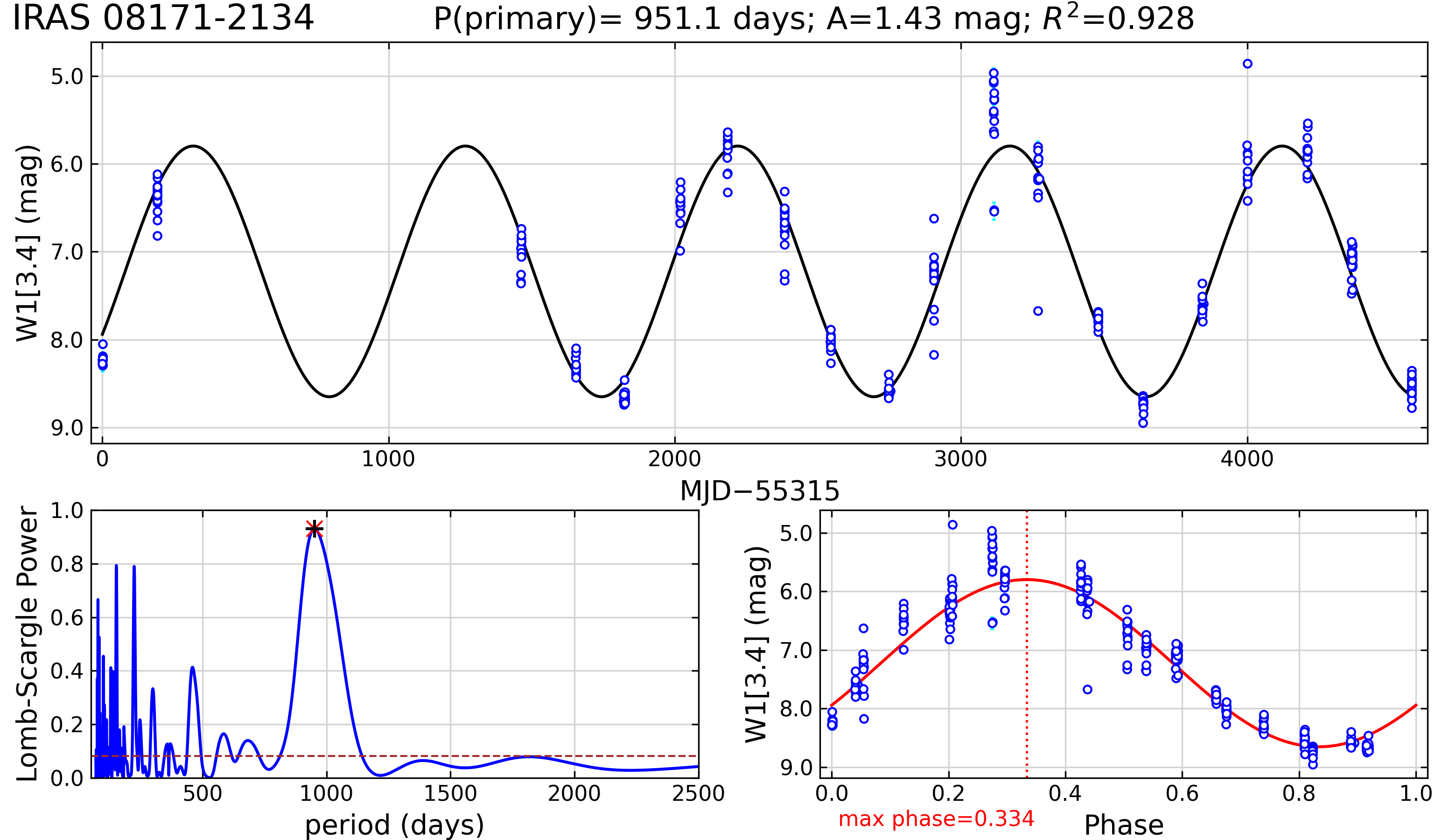}{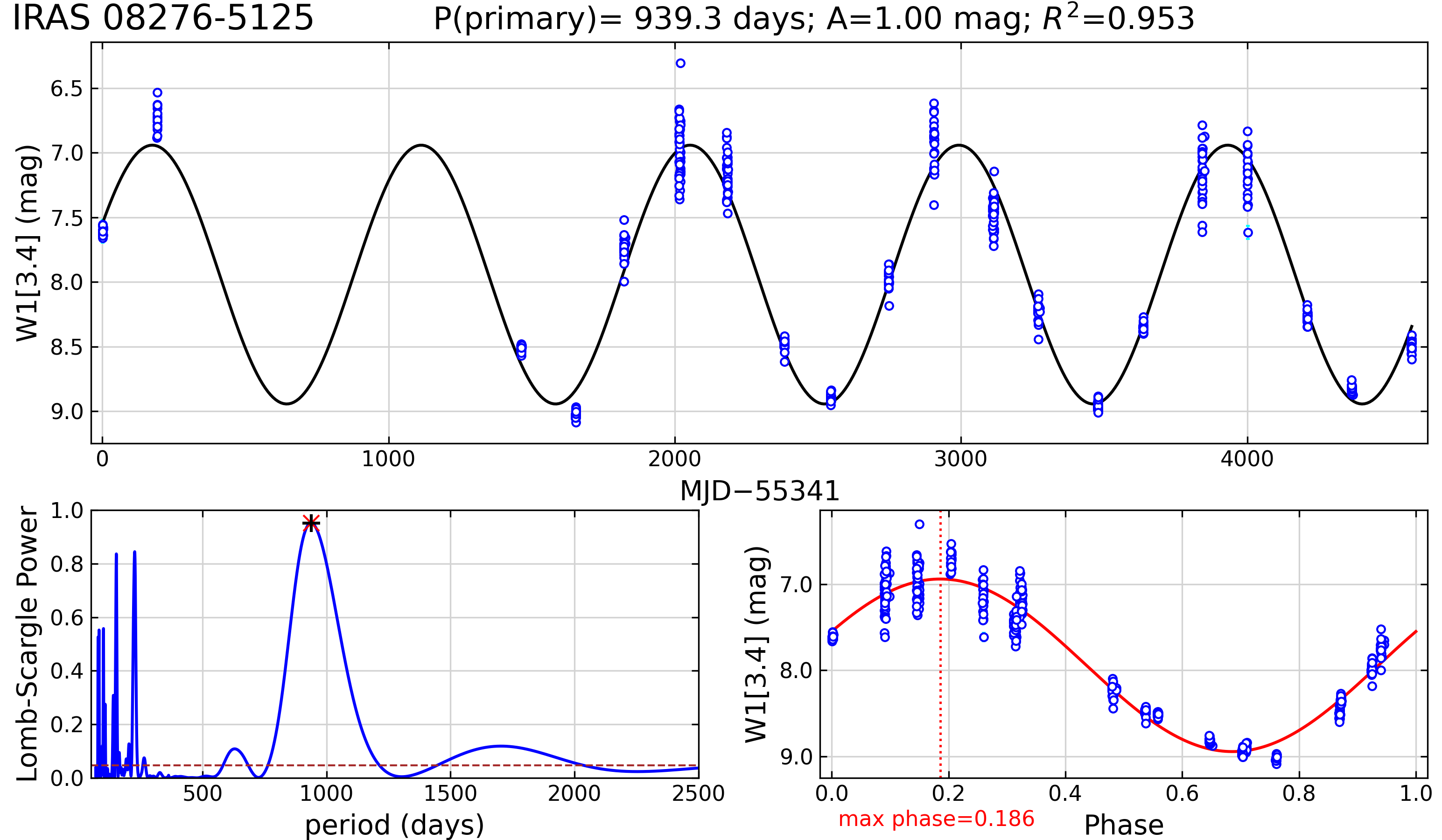}{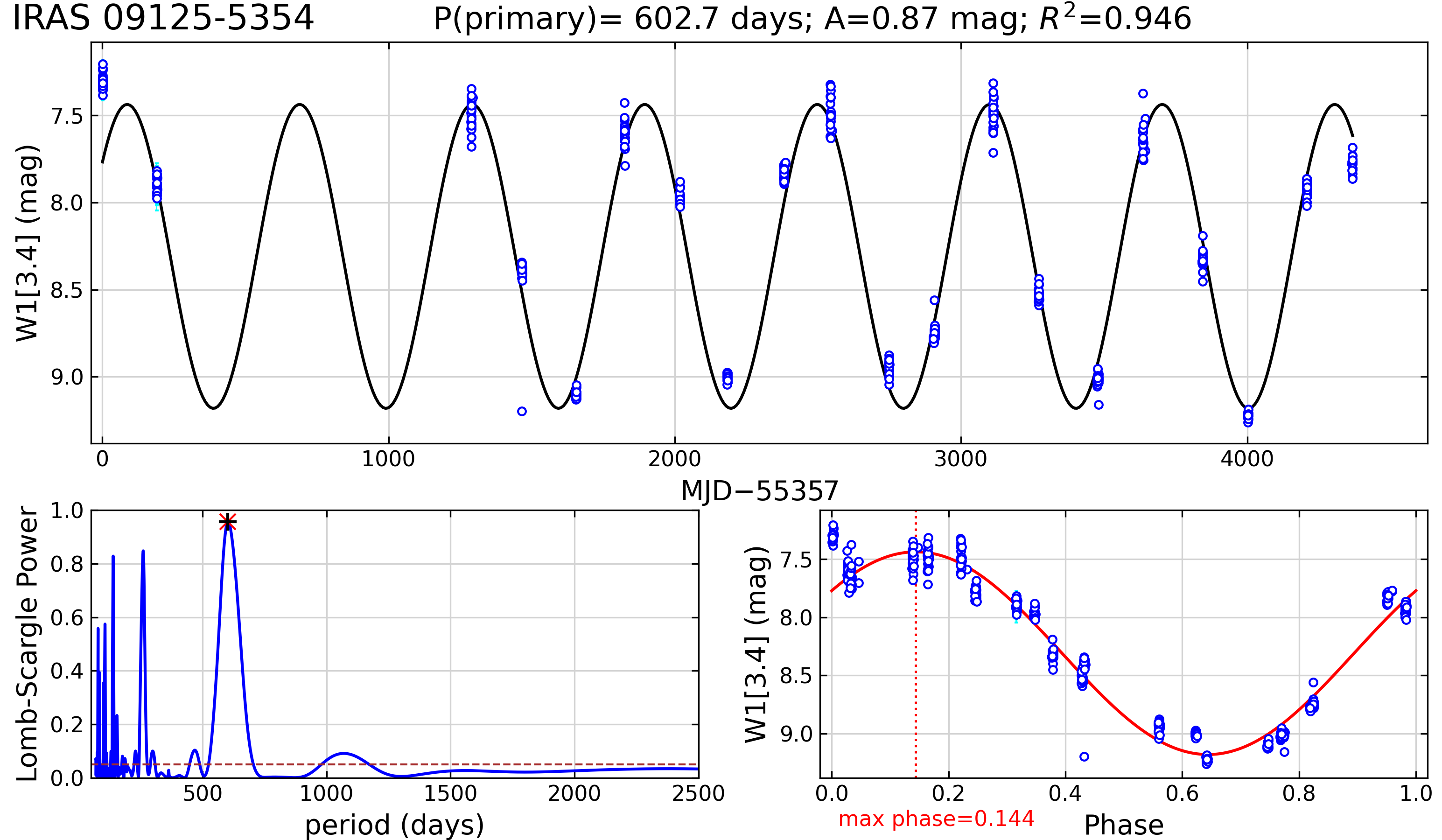}{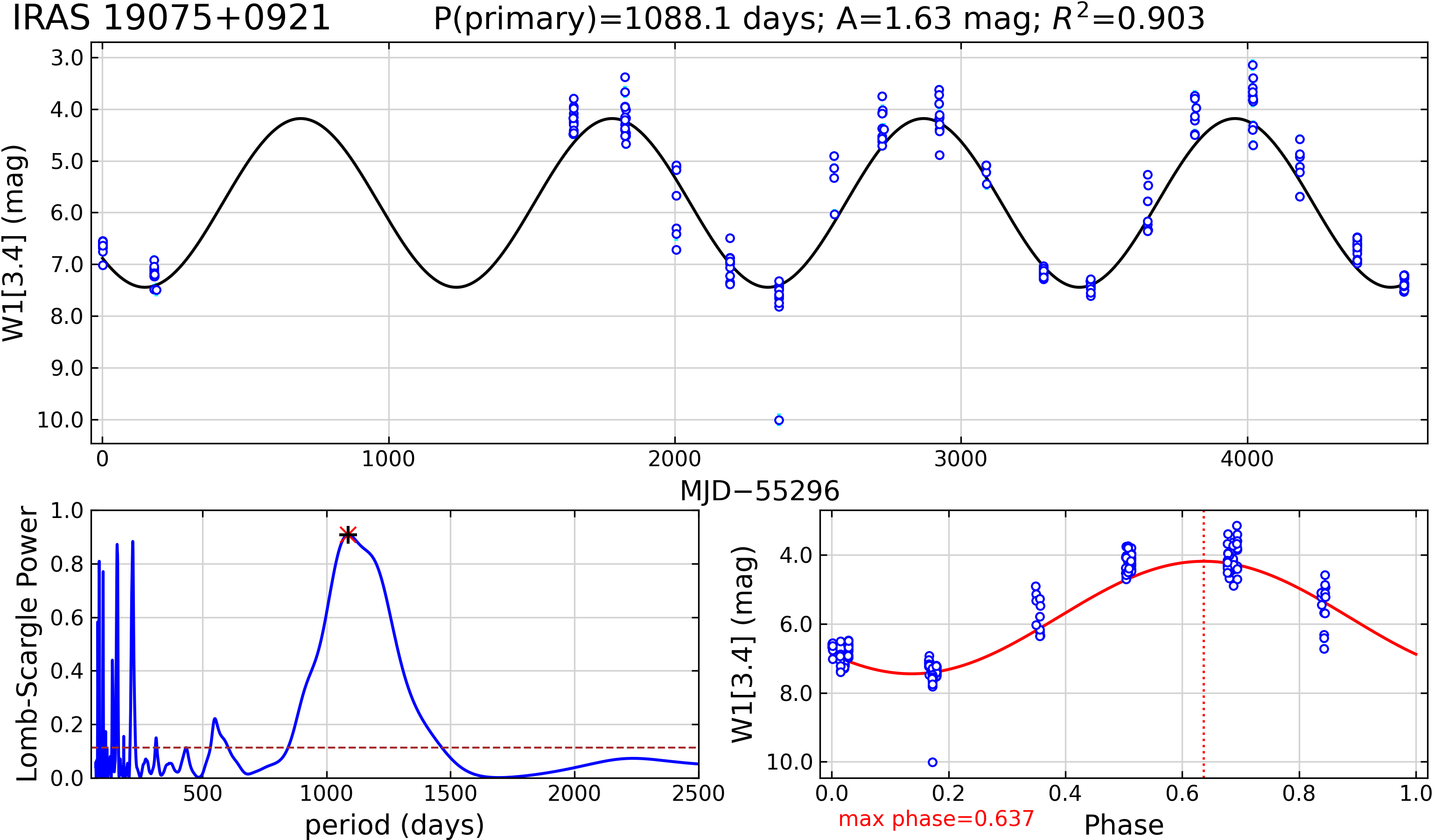}{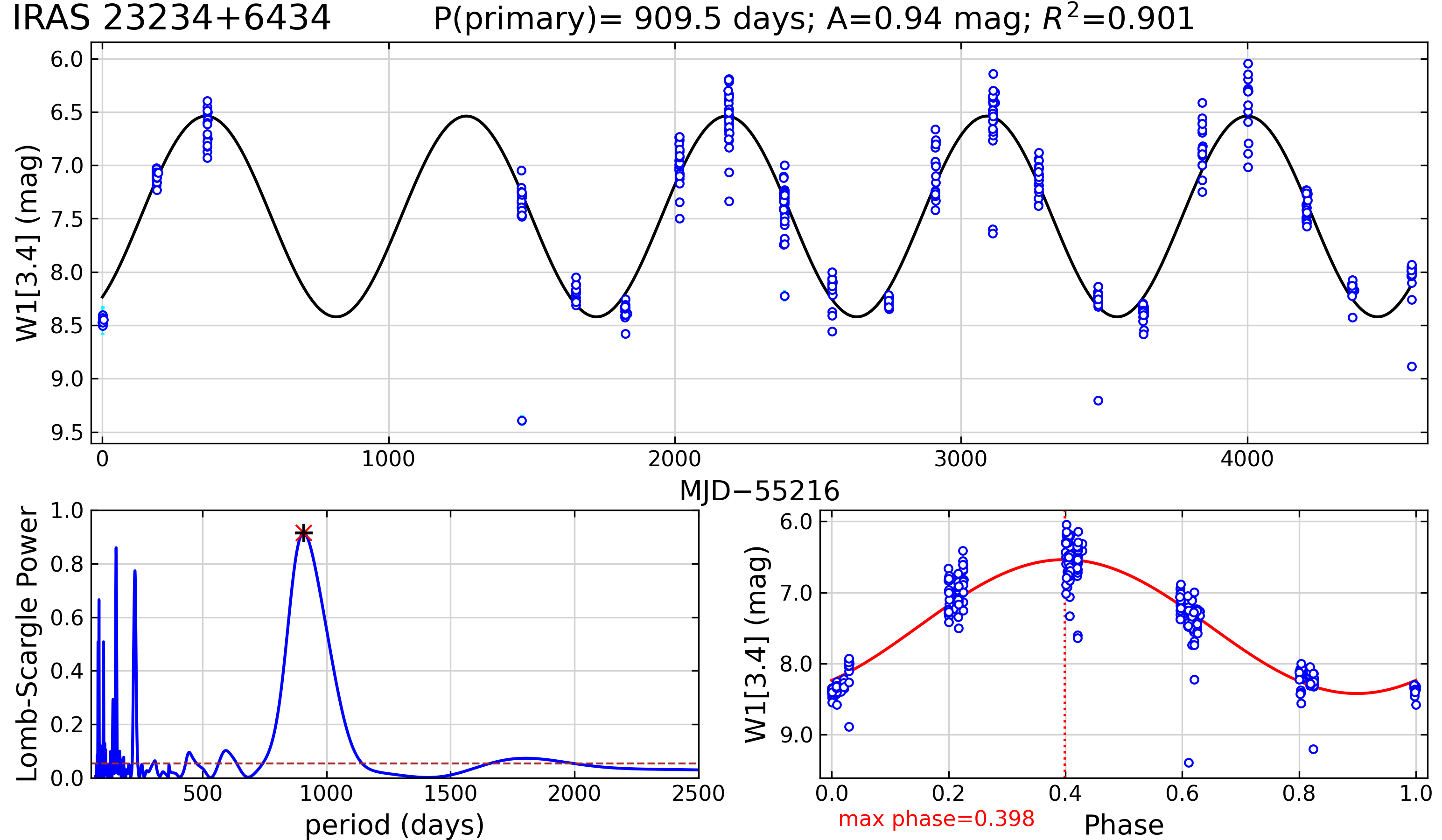}{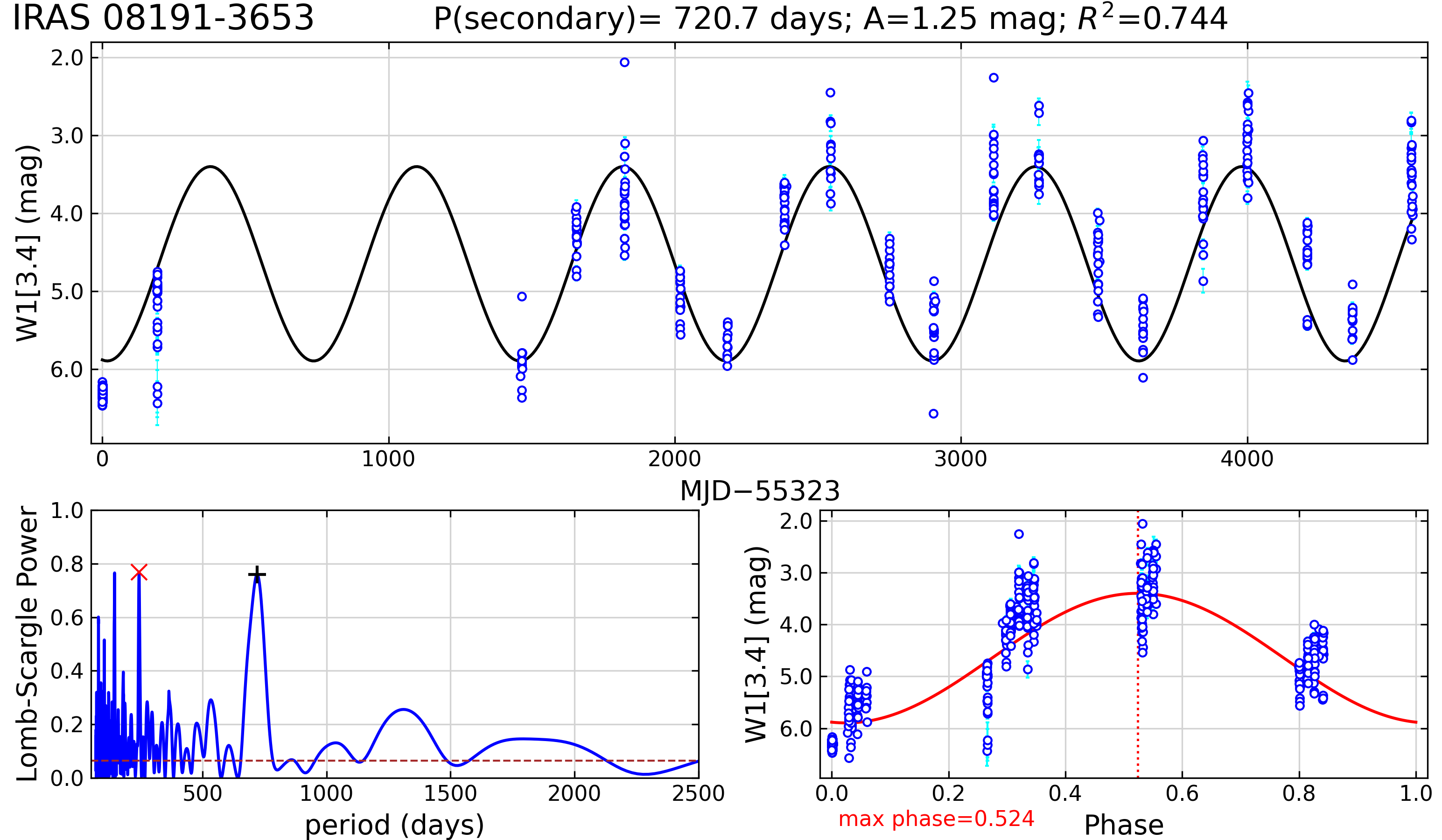}{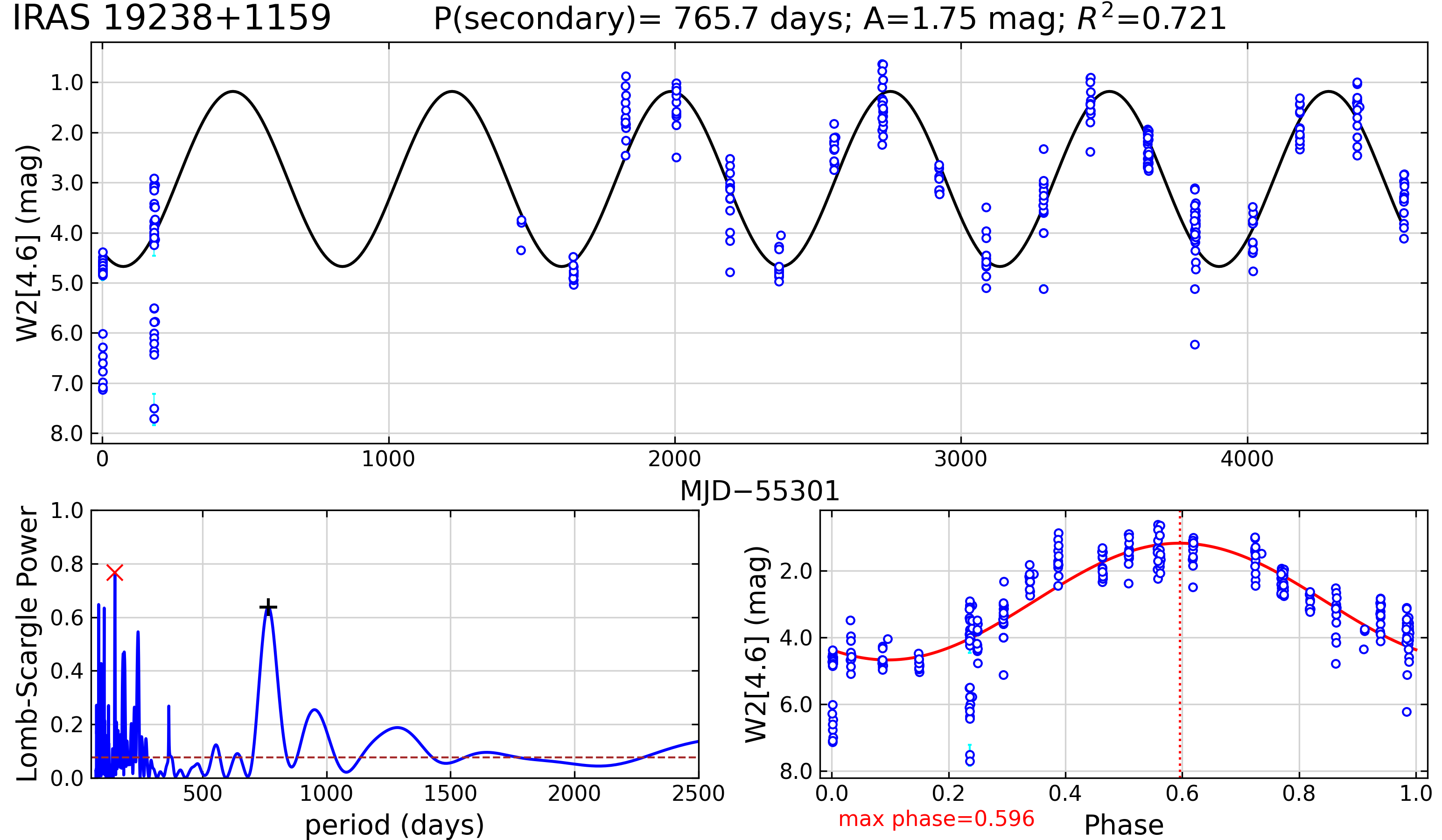}
\caption{Lomb-Scargle periodograms using WISE light curves for eight CAGB stars with unknown periods (AAVSO).
They could be candidates for new Mira variables. Note that the secondary period was used for the last two objects.
See Section~\ref{sec:neo-mira}.} \label{f18}
\end{figure*}

\subsection{Finding IR variations of CAGB stars from WISE data\label{sec:neo}}

\citet{suh2021} investigated the variability of AGB stars at W1[3.4] and W2[4.6] 
bands using WISE data. Additionally, \citet{groenewegen2022} studied the 
variation of a sample of AGB stars in our Galaxy and the Magellanic clouds using 
WISE data. This paper further explores the variability of carbon stars with a new 
sample stars and updated NEOWISE-R data. 

Our focus is on all known extrinsic carbon stars (refer to Tables~\ref{tab:tab1} 
and \ref{tab:tab2}) and 7163 CAGB stars (CAGB\_IC and CAGB\_NI objects, excluding 
GC-AGB\_NI objects). We investigate their variability at W1[3.4] and W2[4.6] 
bands over the past 14 yr. For this analysis, we use the AllWISE multiepoch 
photometry table acquired in 2009–2010 and the NEOWISE-R data (the 2023 data 
release), providing 18 epochs with two observations annually between 2014 and 
2022. 

Our analysis aims to identify Mira-like variations in the WISE light curves using 
a simple sinusoidal light curve model. The Lomb-Scargle periodogram, a 
widely-used statistical algorithm for detecting periodic signals 
(\citealt{zechmeister2009}; \citealt{vanderPlas2018}), is employed. The 
periodograms are computed using the AstroPy implementation. Additional details 
about the algorithm can be found in \citet{vanderPlas2018} and \citet{suh2021}. 

For all the sample stars, we generated light curves using WISE data and produced 
Lomb-Scargle periodograms. As emphasized by \citet{suh2021}, WISE data for a 
substantial portion of AGB stars in our Galaxy, primarily bright stars, are 
either saturated or exhibit significant scatters, rendering variation parameters 
unreliable. To ensure robust results, we selected objects with over 200 observed 
points, $R^2$ larger than 0.6, and Lomb-Scargle power exceeding 0.6. 

Only for 227 objects, pulsation parameters are reliably obtained from WISE light 
curves. As anticipated, no extrinsic carbon stars exhibit Mira-like variations in 
the WISE light curves. All 227 objects are identified as CAGB stars, comprising 
129 CAGB\_IC objects and 98 CAGB\_NI objects. Among these CAGB stars, 87 are 
recognized AAVSO Mira variables with identified periods, the remaining 140 
objects are not identified as Mira variables by AAVSO. 

In the left panels of Figure~\ref{f16}, we present PMRs for known Mira variables 
from AAVSO and the 227 objects with obtained pulsation periods from WISE data. 
PMRs derived from WISE light curves exhibit significantly larger scatters 
compared to those obtained from AAVSO data. 

The upper two panels of Figure~\ref{f17} display Lomb-Scargle periodograms using 
OGLE4 light curves at the $I$ (0.8 $\mu$m) band (\citealt{iwanek2022}) and WISE 
light curves at the W2[4.6] band (this work) for two CAGB stars, namely Mira 
variables IRAS 18010-2626 and [M2003b] dj10, according to AAVSO. While the light 
curves at $I$ (0.8 $\mu$m) and W2[4.6] bands share similar characteristics, the 
W2[4.6] band shows a smaller amplitude. In contrast to the OGLE4 periodogram, the 
Lomb-Scargle periodogram from the WISE light curve exhibits multiple peaks with 
similar power values due to the regular interval of WISE data acquisition (every 
six months). 

Due to the existence of multiple peaks with similar Lomb-Scargle power values, 
the precise determination of periods using only WISE data is challenging. As 
emphasized by \citet{suh2021}, secondary or tertiary periods might represent the 
true periodicity for specific objects, potentially overshadowing the primary 
period.

\subsection{Candidates for new Mira variables found from WISE data\label{sec:neo-mira}}

In the right panels of Figure~\ref{f16}, we again present the PMRs, but for those 
227 objects based on WISE light curves, we prioritize secondary (or tertiary) 
periods for 44 objects (red symbols) deviating significantly from the general 
trends of Miras. The revised PMRs based on WISE light curves display considerably 
fewer scatters. Out of the 227 CAGB stars with determined pulsation periods from 
WISE light curves, 140 objects (lime symbols) are not identified as Mira 
variables by AAVSO. 

Figure~\ref{f17} presents Lomb-Scargle periodograms for six selected CAGB stars 
from the group of 87 objects identified as Mira variables by AAVSO. The obtained 
pulsation periods from WISE data are similar to the periods from AAVSO. 

Figure \ref{f18} showcases Lomb-Scargle periodograms for eight selected CAGB 
stars from the group of 140 objects not identified as Mira variables by AAVSO. It 
is important to note that for the last two objects, the secondary period was 
selected. These 140 objects could potentially be candidates for newly discovered 
Mira variables.

\section{A new catalog of CAGB stars in our Galaxy\label{sec:catalog-cagb}}

In this study, we have assembled a novel sample of CAGB stars drawn from known 
carbon stars, exploring various properties associated with them. We introduce a 
new catalog comprising 7,163 CAGB stars (4909 CAGB\_IC and 2254 CAGB\_NI objects; 
see Tables~\ref{tab:tab1} and \ref{tab:tab2}) in our Galaxy. 

For the 4909 CAGB\_IC objects (see Table~\ref{tab:tab1}), Table~\ref{tab:tab5} 
lists the CAGB\_IC number, IRAS identifier, CDS SIMBAD main identifier, Gaia 
identifier, position (RA and DEC; right ascension and declination J2000), source 
of the position (SP), AllWISE identifier, variable type from AAVSO, period (in 
days) from AAVSO, period (in days) from WISE data (see Section~\ref{sec:neo}), 
distance in pc (dist\_pc), and source of the distance (SD). The sources of the 
position (SP) are categorized into two groups: 1) G3: Gaia DR3 2) AP: Akari PSC. 

For the 2254 CAGB\_NI objects (see Table~\ref{tab:tab2}), Table~\ref{tab:tab6} 
lists the CAGB\_NI number, CDS SIMBAD main identifier, Gaia identifier, position 
(RA and DEC; right ascension and declination J2000 from Gaia DR3), AllWISE 
identifier, variable type from AAVSO, period (in days) from AAVSO, period (in 
days) from WISE data (see Section~\ref{sec:neo}), distance in pc (dist\_pc), and 
source of the distance (SD). 

The source of the distance information (SD; see Section~\ref{sec:modeldistance}) 
in Tables~\ref{tab:tab5} and \ref{tab:tab6} is categorized into five groups: 1) 
G: Gaia DR3 2) GS: Gaia DR3 and CAGB model SED (mean value) 3) H: Hipparcos 4) 
HS: Hipparcos and CAGB model SED (mean value) 5) S: CAGB model SED. 

In Tables~\ref{tab:tab5} and \ref{tab:tab6}, only 21 rows of the catalogs are 
shown. The full catalogs are available in machine-readable format.

\begin{longrotatetable}
\begin{table*}
\centering
\scriptsize
\caption{CAGB\_IC objects$^a$ (13 columns; 4909 rows)\label{tab:tab5}}
\begin{tabular}{lllllllllllll}
\hline \hline               
No.  & IRAS	&	SIMBAD\_id	&	Gaia\_id	& RA	& DEC	&	SP$^b$	&	AllWISE\_id	 &	A\_Type	&	A\_Period	&	WISE\_period	&	dist\_pc	&	SD$^c$	\\
\hline   
1	&	19118+1020	&	IRAS 19118+1020                    	&	DR3 4309357708491415552	&	288.54703101 	&	10.43097188 	&	G3	&	J191411.33+102551.5	&		&		&		&	3911.8 	&	G	\\
2	&	21458+4539	&	Case 662                           	&	DR3 1974364515721462656	&	326.93394427 	&	45.89813167 	&	G3	&	J214744.15+455353.1	&	SRA	&	514.0 	&		&	2101.7 	&	G	\\
3	&	21458+5244	&	[ABC90] cyc 27                     	&	DR3 2173236138784903296	&	326.89962901 	&	52.97894764 	&	G3	&	J214735.90+525844.1	&	SR	&	386.0 	&		&	5046.6 	&	G	\\
4	&	00533+5503	&	IRAS 00533+5503                    	&	DR3 423450072786302208	&	14.05721350 	&	55.33304974 	&	G3	&	J005613.76+551958.9	&	M	&	344.5 	&		&	8051.5 	&	G	\\
5	&	00535+5923	&	V* V721 Cas                        	&	DR3 426094260814792576	&	14.13762143 	&	59.66223355 	&	G3	&	J005633.03+593943.9	&	M	&	585.9 	&		&	2769.3 	&	G	\\
6	&	00538+6410	&	V* V881 Cas                        	&	DR3 524221378626364160	&	14.23383979 	&	64.45034263 	&	G3	&	J005656.16+642701.2	&	SR:	&		&		&	5456.5 	&	G	\\
7	&	00539+5621	&	IRAS 00539+5621                    	&	DR3 423994365404423040	&	14.22248683 	&	56.62959076 	&	G3	&	J005653.41+563746.1	&	SR	&	306.0 	&		&	4954.7 	&	G	\\
8	&	17552-2814	&	PN H 1-45                          	&	DR3 4062646712567004416	&	269.59108375 	&	-28.24784676 	&	G3	&	J175821.90-281452.7	&	ZAND+M	&	416.2 	&		&	1770.1 	&	G	\\
9	&	13078-5828	&	IRAS 13078-5828                    	&	DR3 6062228557472426880	&	197.72661782 	&	-58.74670061 	&	G3	&	J131054.41-584447.8	&	M	&	444.3 	&	460.3 	&	8262.7 	&	G	\\
10	&	13078-6012	&	IRAS 13078-6012                    	&	DR3 6055840669782220544	&	197.75262618 	&	-60.47458390 	&	G3	&	J131100.64-602828.2	&	M	&	433.9 	&		&	8036.3 	&	G	\\
11	&	06344-1656	&	V* V368 CMa                        	&	DR3 2946796410619430144	&	99.16446215 	&	-16.99285239 	&	G3	&	J063639.49-165934.1	&	SRB	&	194.0 	&		&	4091.6 	&	G	\\
12	&	06344-0124	&	IRAS 06344-0124                    	&		&	99.25662000 	&	-1.45002000 	&	AP	&	J063701.62-012701.7	&		&		&	839.8 	&	4000.0 	&	S	\\
13	&	06347-1203	&	IRC -10133                         	&	DR3 2953571120234692736	&	99.26478306 	&	-12.09544908 	&	G3	&	J063703.56-120543.5	&	MISC	&	160.0 	&		&	1574.7 	&	G	\\
14	&	06348+3114	&	IRAS 06348+3114                    	&	DR3 3435862859142788352	&	99.52239512 	&	31.19535495 	&	G3	&	J063805.33+311143.6	&	L	&		&		&	2212.8 	&	G	\\
15	&	06349+2250	&	C* 3277                            	&	DR3 3379346655806590208	&	99.49454391 	&	22.80064937 	&	G3	&	J063758.71+224802.3	&	LB	&	172.5 	&		&	3033.4 	&	G	\\
16	&	20176-1458	&	IRAS 20176-1458                    	&	DR3 6875984878440208512	&	305.11527656 	&	-14.82426347 	&	G3	&	J202027.67-144926.9	&	SR	&	354.0 	&	374.2 	&	13527.4 	&	GS	\\
17	&	05185+0718	&	LEE 18                             	&	DR3 3241334169578902656	&	80.30572294 	&	7.35537774 	&	G3	&	J052113.38+072119.4	&	SRB	&	277.2 	&		&	957.1 	&	G	\\
18	&	05186+1551	&	IRAS 05186+1551                    	&	DR3 3394034241365927936	&	80.36858711 	&	15.90941185 	&	G3	&	J052128.46+155433.8	&	M	&	298.7 	&		&	9326.0 	&	G	\\
19	&	19122+1830	&	Case 185                           	&	DR3 4516021020719021440	&	288.60342816 	&	18.59832942 	&	G3	&		&	MISC	&	215.0 	&		&	2152.6 	&	G	\\
20	&	19122+2318	&	IRAS 19122+2318                    	&	DR3 2022495667287340800	&	288.59025215 	&	23.39755057 	&	G3	&		&	SR	&	176.0 	&		&	2414.6 	&	G	\\
21	&	21468+5747	&	V* V493 Cep                        	&	DR3 2202334748365238144	&	327.10479972 	&	58.01428062 	&	G3	&	J214825.15+580051.2	&	SRB	&	390.0 	&		&	6170.8 	&	G	\\
\hline
\end{tabular}
\begin{flushleft}
$^a$See Table~\ref{tab:tab1} and Section~\ref{sec:catalog-cagb}. The full catalog is available in machine-readable format.
$^b$The source of the position: 1) G3: Gaia DR3 2) AP: Akari PSC.
$^c$The source of the distance: 1) G: Gaia DR3 2) GS: Gaia DR3 and model SED (mean value) 3) H: Hipparcos 4) HS: Hipparcos and CAGB model SED (mean value) 5) S: model SED.
\end{flushleft}
\end{table*}
\end{longrotatetable}

\begin{longrotatetable}
\begin{table*}
\centering
\scriptsize
\caption{CAGB\_NI objects$^a$ (11 columns; 2254 rows)\label{tab:tab6}}
\begin{tabular}{lllllllllll}
\hline \hline
No.	& SIMBAD\_id	&	Gaia\_id	& RA	& DEC	&	AllWISE\_id	&	A\_Type	&	A\_Period	&	WISE\_period	&	dist\_pc	&	SD$^b$	\\
\hline
1	&	Kiso C1-38	&	DR3 432154116070047232	&	0.37424664 	&	64.91885902 	&	J000129.81+645507.8	&	SR	&	253.6 	&		&	8629.5 	&	G	\\
2	&	[TI98] 0000+3021	&	DR3 2861332746177405696	&	0.75359455 	&	30.63984305 	&	J000300.86+303823.4	&	M	&	253.2 	&		&	11740.0 	&	G	\\
3	&	NIKC 1-28	&	DR3 423280885436848128	&	0.92486743 	&	59.73694096 	&	J000341.96+594413.0	&	M	&	259.0 	&	246.3 	&	4809.5 	&	G	\\
4	&	Kiso C1-46	&	DR3 432237129193547392	&	1.43538322 	&	65.51587342 	&	J000544.48+653057.1	&	SR	&	121.6 	&		&	4350.8 	&	G	\\
5	&	Kiso C1-47	&	DR3 423227524761428480	&	1.43800744 	&	59.49667112 	&	J000545.12+592948.0	&	SR	&	274.4 	&		&	5385.3 	&	G	\\
6	&	CGCS 11	&	DR3 432115255205278208	&	2.29989983 	&	65.02399643 	&	J000911.97+650126.3	&		&		&		&	6730.9 	&	G	\\
7	&	Kiso C1-53	&	DR3 431355664475998976	&	2.89769938 	&	63.30855466 	&	J001135.45+631830.8	&	SR	&	310.7 	&		&	5892.5 	&	G	\\
8	&	[I81] C 3	&	DR3 429650768615042560	&	3.01956677 	&	61.54713277 	&	J001204.70+613249.6	&	SR	&	192.3 	&		&	5332.3 	&	G	\\
9	&	[I81] C 5	&	DR3 431338617735738240	&	3.31931379 	&	63.03340946 	&	J001316.63+630200.3	&	SR	&	255.5 	&		&	5554.2 	&	G	\\
10	&	BC 110	&	DR3 528876019357498112	&	3.44344370 	&	68.29170688 	&	J001346.40+681730.2	&	SR:	&	130.0 	&		&	4655.8 	&	G	\\
11	&	[I81] C 4	&	DR3 429556004454036480	&	3.62955399 	&	60.92529709 	&	J001431.09+605531.0	&	SR	&	152.9 	&		&	8384.2 	&	G	\\
12	&	[I81] M 67	&	DR3 429135887932964992	&	3.77838467 	&	60.14894424 	&	J001506.82+600856.1	&	LPV	&	278.0 	&		&	5998.7 	&	G	\\
13	&	Kiso C1-65	&	DR3 431374871565656320	&	4.46342775 	&	63.06836065 	&	J001751.22+630406.1	&	SR	&	103.9 	&		&	6861.2 	&	G	\\
14	&	CGCS 38	&	DR3 430469973497669632	&	5.02739175 	&	61.98879143 	&	J002006.57+615919.6	&	SR	&	209.4 	&		&	8237.9 	&	G	\\
15	&	Kiso C1-70	&	DR3 431035294271577088	&	5.50380093 	&	63.49398185 	&	J002200.91+632938.3	&		&		&		&	4568.3 	&	G	\\
16	&	Kiso C1-75	&	DR3 430400498104743808	&	6.02907819 	&	61.70314714 	&	J002406.97+614211.3	&	SR	&	364.3 	&		&	5600.7 	&	G	\\
17	&	BC 79	&	DR3 421752628690357632	&	6.07930817 	&	56.88220974 	&	J002419.04+565255.9	&	L	&		&		&	7893.4 	&	G	\\
18	&	Kiso C1-78	&	DR3 431011345534262784	&	6.16820049 	&	63.23646455 	&	J002440.36+631411.2	&	SR	&	55.6 	&		&	6756.6 	&	G	\\
19	&	Kiso C1-81	&	DR3 428878499133379840	&	6.59766229 	&	61.30082116 	&	J002623.44+611803.0	&	SR	&	153.0 	&		&	7797.5 	&	G	\\
20	&	C* 3222	&	DR3 2793301598079856512	&	6.72477116 	&	17.55777751 	&	J002653.94+173328.0	&	LB	&		&		&	7080.7 	&	G	\\
21	&	Kiso C1-83	&	DR3 527881893347020672	&	6.84269872 	&	66.76993732 	&	J002722.25+664611.8	&	SR	&	269.5 	&		&	7180.3 	&	G	\\
\hline
\end{tabular}
\begin{flushleft}
$^a$See Table~\ref{tab:tab2} and Section~\ref{sec:catalog-cagb}. The full catalog is available in machine-readable format.
$^b$The source of the distance: 1) G: Gaia DR3 2) GS: Gaia DR3 and model SED (mean value). 
\end{flushleft}
\end{table*}
\end{longrotatetable}

\section{summary\label{sec:sum}}

We have made an in-depth investigation into the infrared properties of carbon 
stars within our Galaxy, employing a comprehensive analysis of observational data 
across visual and IR bands. Utilizing datasets from IRAS, Akari, MSX, 2MASS, 
WISE, and Gaia DR3, we conductd an extensive comparison between observational 
data and theoretical models. The comparison involve the use of various 2CDs, 
CMDs, and SEDs. 

We have found that the CMDs, which utilize the latest distance and extinction 
data from Gaia DR3 for a substantial number of carbon stars, are very useful to 
distinguish CAGB stars from extrinsic carbon stars that are not in the AGB phase. 
In the case of all CAGB stars, the absolute magnitudes at the IR[12] or W3[12] 
bands are brighter than -6. In contrast, extrinsic carbon stars of Ba, CH, dC, 
and RH types are notably distinguishable from CAGB stars on various CMDs, 
displaying dimmer absolute magnitudes at the MIR bands. However, J-type stars, 
constituting another subclass of extrinsic carbon stars, demonstrate properties 
shared with both CAGB stars and other extrinsic carbon stars. 

We have performed radiative transfer model calculations for CAGB stars, exploring 
various parameters of central stars and spherically symmetric dust shells. A 
comprehensive comparison of theoretical models with observations on various IR 
2CDs, CMDs, and SEDs reveals a notable agreement. 

For selected CAGB stars, the observed SEDs are compared with theoretical models, 
with a simple single dust shell model of AMC and SiC dust reproducing SEDs 
effectively for most objects. Some objects require more complex models, such as 
detached or double shells or non-spherical dust envelopes. 

We have established theoretical distances for 152 CAGB stars, optimizing the fit 
to the observed SED by comparing it with the corresponding CAGB model SED. A 
comparison between distances derived from observations and those obtained from 
theoretical model SEDs reveals a significantly strong correlation, affirming the 
reliability of the theoretical models. 

We have studied the IR properties of known variable stars, revealing that for 
Mira variables, IR colors intensify in reddish hues, and absolute magnitudes 
brighten with longer pulsation periods. Furthermore, we investigated the IR 
variability of carbon stars by analyzing WISE photometric data spanning the last 
14 yr. Through a detailed examination of WISE light curves for all sampled stars, 
we have identified useful variation parameters for 227 CAGB stars. 

We have presented a novel catalog of CAGB stars, offering enhanced reliability 
and a wealth of additional information, including distances obtained from Gaia 
DR3 and model SEDs, and variational parameters from AAVSO and WISE light curves. 
We have identified CAGB stars from a substantial sample of carbon stars spanning 
various subclasses, leveraging information from diverse literature sources, the 
SIMBAD database, and examining their IR properties through 2CDs, CMDs, SEDs, and 
variations. The resulting catalog introduces 7163 CAGB stars, categorized as 4909 
CAGB\_IC and 2254 CAGB\_NI objects within our Galaxy.


\begin{acknowledgments}
I thank the anonymous referee for constructive comments and suggestions. This 
work was supported by a funding for the academic research program of Chungbuk 
National University in 2023. This research was supported by Basic Science 
Research Program through the National Research Foundation of Korea (NRF) funded 
by the Ministry of Education (2022R1I1A3055131). This research has made use of 
the SIMBAD database and VizieR catalogue access tool, operated at CDS, 
Strasbourg, France. This research has made use of the NASA/IPAC Infrared Science 
Archive, which is operated by the Jet Propulsion Laboratory, California Institute 
of Technology, under contract with the National Aeronautics and Space 
Administration. 
\end{acknowledgments}

\bibliographystyle{aasjournal}
\bibliography{reference.bib}



\end{document}